\RequirePackage{lineno}
\documentclass[biblatex]{cernrep2018}

\usepackage[detect-all]{siunitx}  %
\usepackage{enumitem}%
\usepackage{authblk}
\usepackage{hyperref}
\usepackage[utf8]{inputenc}
\usepackage{color}
\usepackage{lineno}

\begin{document}

\title{
\vspace*{-2.4cm}\hfill{{\small{CERN-PBC-REPORT-2018-003}}}\\
{\vspace*{-0.5cm}\includegraphics[width = 23mm]{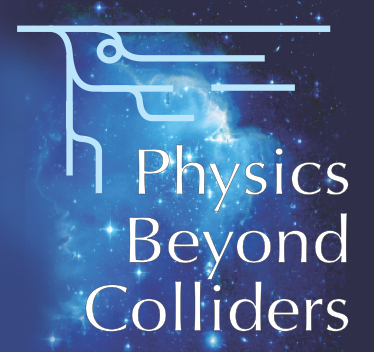}}\\[1.0cm]
Summary Report of Physics Beyond Colliders at CERN
}

\begin{abstract}
Physics Beyond Colliders is an exploratory study aimed at exploiting the full scientific potential of CERN's accelerator complex and its scientific infrastructure in the next two decades through projects complementary to the LHC, HL-LHC and other possible future colliders. These projects should target fundamental physics questions that are similar in spirit to those addressed by high-energy colliders, but that require different types of beams and experiments.

A kick-off workshop held in September 2016 identified a number of areas of interest and working groups have been set-up to study and develop these directions.  All projects currently under consideration are presented including physics motivation, a brief outline of the experimental set-up and the status of the corresponding beam and detector technological studies. The proposals are also put in context of the worldwide landscape and their implementation issues are discussed. 
\end{abstract}

\author[1]{R.~Alemany}
\author[2]{C.~Burrage}
\author[1]{H.~Bartosik}
\author[1]{J.~Bernhard}
\author[1]{J.~Boyd}
\author[1]{M.~Brugger}
\author[1]{M.~Calviani}
\author[1]{C.~Carli}
\author[1]{N.~Charitonidis}
\author[23]{D.~Curtin}
\author[34]{A.~Dainese}
\author[1]{A.~de Roeck}
\author[3]{M.~Diehl}
\author[1]{B.~ D\"obrich}
\author[1]{L.~Evans}
\author[24]{J.L.~Feng}
\author[1]{M.~Ferro-Luzzi}
\author[1]{L.~Gatignon}
\author[1]{S.~Gilardoni}
\author[19]{S.~Gninenko}
\author[32]{G.~Graziani}
\author[1]{E.~Gschwendtner}
\author[1]{B.~Goddard}
\author[16]{A.~Hartin}
\author[20]{I.~Irastorza}
\author[4]{J.~Jaeckel$^{*}$}
\author[1]{R.~Jacobsson}
\author[5]{K.~Jungmann}
\author[6]{K.~Kirch}
\author[24]{F.~Kling}
\author[13]{W.~Krasny}
\author[1]{M.~Lamont$^{*}$}
\author[7]{G.~Lanfranchi}
\author[27]{J-P.~Lansberg}
\author[3]{A.~Lindner}
\author[12]{K.~Long}
\author[1]{A.~Magnon}
\author[1]{G.~Mallot}
\author[21]{F.~Martinez Vidal}
\author[7]{M.~Moulson}
\author[1]{M.~Papucci}
\author[4]{J.~M.~Pawlowski}
\author[25]{I.~Pedraza}
\author[18]{K.~Petridis}
\author[8]{M.~Pospelov}
\author[31]{S.~Pulawski}
\author[1]{S.~Redaelli}
\author[9]{S.~Rozanov}
\author[1]{G.~Rumolo}
\author[10]{G.~Ruoso}
\author[29]{J.~Schacher}
\author[11]{G.~Schnell}
\author[22]{P.~Schuster}
\author[14]{Y.~Semertzidis}
\author[1]{A.~Siemko}
\author[7]{T.~Spadaro}
\author[1]{S.~Stapnes}
\author[28]{A.~Stocchi}
\author[15]{H.~Str\"oher}
\author[30]{G.~Usai}
\author[9]{C.~Vall\'ee$^{*}$}
\author[26]{G.~Venanzoni}
\author[33]{G.~Wilkinson}
\author[16]{M.~Wing}

\affil[1]{European Organization for Nuclear Research (CERN), Geneva, Switzerland}
\affil[2]{University of Nottingham\\Nottingham, United Kingdom}
\affil[3]{DESY\\Hamburg, Germany}
\affil[4]{Institute for Theoretical Physics, Heidelberg University \\Heidelberg, Germany}
\affil[5]{VSI (Van Swinderen Institute) University of Groningen\\Groningen, Netherlands}
\affil[6]{ETH Zurich and Paul Scherrer Institute\\ Villigen, Switzerland}
\affil[7]{LNF-INFN\\Frascati, Italy}

\affil[8]{Perimeter Institute, Waterloo and University of Victoria\\Victoria, Canada}
\affil[9]{CPPM, CNRS-IN2P3 and Aix-Marseille University\\Marseille, France}
\affil[10]{LNL-INFN\\Legnaro, Italy}
\affil[11]{University of the Basque Country UPV/EHU and Ikerbasque\\Bilbao, Spain}
\affil[12]{Imperial College\\London, United Kingdom}
\affil[13]{LPNHE, CNRS-IN2P3, Sorbonne University and University Paris Diderot\\Paris, France}
\affil[14]{KAIST/IBS\\Daejeon, Korea}
\affil[15]{FZJ\\Juelich, Germany}
\affil[16]{University College London\\London, United Kingdom}
\affil[18]{University of Bristol\\Bristol, United Kingdom}
\affil[19]{Institute for Nuclear Research\\Moskow, Russia}
\affil[20]{University of Saragossa\\Saragossa, Spain}
\affil[21]{IFIC/CSIC University of Valencia\\Valencia, Spain}
\affil[22]{SLAC\\Menlo Park, USA}
\affil[23]{University of Toronto\\Toronto, Canada}
\affil[24]{University of California\\Irvine, USA}
\affil[25]{Autonomous University of Puebla\\Mexico, Mexico}
\affil[26]{INFN\\Pisa, Italy}
\affil[27]{IPNO, CNRS-IN2P3, Universities of Paris-Saclay and Paris-Sud\\Orsay, France}
\affil[28]{LAL, CNRS-IN2P3, Universities of Paris-Saclay and Paris-Sud\\Orsay, France}
\affil[29]{University of Bern\\Bern, Switzerland}
\affil[30]{University of Cagliari\\Cagliari, Italy}
\affil[31]{University of Silesia\\Katowice, Poland}
\affil[32]{INFN\\Florence, Italy}
\affil[33]{University of Oxford\\Oxford, United Kingdom}
\affil[34]{INFN\\Padova, Italy}
\affil[ ]{    } %
\affil[ ]{    } %
\affil[ ]{    } %
\affil[ ]{-------------------------------------------------------------------------------   }
\affil[*]{PBC co-coordinator and co-editor of this summary report}

\maketitle

\tableofcontents

\vspace{2.0cm}

\section*{Introduction}
\vspace{0.5cm}
The Physics Beyond Colliders (PBC) study was mandated by the CERN Management in 2016 in order to prepare the next update of the European Strategy for Particle Physics. The PBC activities involve around a hundred contributors (appendix 3) including CERN accelerator experts, project proponents and independent physicists. The activities can be monitored on the public web site pbc.web.cern.ch. The studies of the past two years are reported in detailed documents of the PBC working groups 
\cite{
PBC:QCD,
PBC:BSM,
PBC:BDF,
PBC:CONV,
PBC:LHCFT,
PBC:LIU,
PBC:EDM,
PBC:TECHNO,
PBC:eSPS,
PBC:AWAKE,
PBC:NUSTORM,
PBC:GAMMA,
}
and summarized in the present report.

\newpage
\section{Executive summary}

\vspace{0.3cm}

CERN operates the only high-energy frontier collider worldwide, the LHC, and hosts a vigorous particle physics program on the LHC injector complex with a thousand of physicists involved. Community surveys performed to prepare the future of this program have identified a wealth of projects covering a broad range of physics domains. The PBC studies address the suitability of the CERN complex to host the projects, their general relevance in the physics landscape, their competitiveness versus other projects worldwide, as well as their implementation issues.

\subsection*{CERN complex specificities and opportunities}
\vspace{0.3cm}

The LHC injectors (including, e.g. SPS, PS) form a very flexible complex serving diverse communities with both low- and high-energy beams to address a broad physics spectrum. 

\subparagraph{Low energy facilities:}

The low energy part (< 26 GeV) of the complex has a lower power and duty cycle than high intensity frontier machines such as FNAL or J-PARC, but hosts three unique facilities worldwide: the ISOLDE radioactive ion beams, the n\_TOF pulsed neutron beam, and the very low energy antiproton beam of the Antimatter Factory. These facilities were not considered within PBC because recent upgrades (HIE-ISOLDE, n\_TOF's 2\textsuperscript{nd} experimental area, ELENA) secure their future for at least the next decade. The hosted experiments have a unique fundamental physics reach and their short evolution time scales may generate promising new ideas on short notice. It is therefore important to maintain adequate flexibility in the usage of these facilities.

A low-energy storage ring had previously been proposed to improve the measurements of the proton and deuteron EDM. 
CERN joined with the proponents within a new CPEDM collaboration in order to consolidate the ring design.
Systematic effects received close attention and results suggest that a prototype ring be exploited before the full-size ring can be built. The prototyping could be done in a facility associated to a CPEDM institute, e.g. COSY in J\"ulich.

\subparagraph{High energy Fixed Target:}

The work-horse of high energy Fixed Target physics at CERN is the SPS, which provides a worldwide unique combination of high energy up to 400 GeV, high intensity and high duty cycle. The SPS can serve many users in parallel with slow or fast extraction and high flexibility. All kinds of particles including ions are available. 

Completion of the CNGS neutrino beam program in 2012, together with injector upgrades for HL-LHC, leaves room for a new high intensity facility at CERN as regards proton yield. This opportunity motivated the proposal of the Beam Dump Facility (BDF) to exploit unique possibilities to enter a new era of beam-dump/fixed-target physics experiments at the intensity frontier.
The design of the BDF has been consolidated within PBC and is now ready for further studies towards a technical design report. The technically driven baseline schedule would allow start of operation after LS3.

The ongoing CERN accelerator R\&D offers opportunities for other new facilities at the SPS on the long term. Two options were proposed and studied within PBC: a new primary high intensity e-beam (eSPS) based on a CLIC linac, SPS acceleration up to 16 GeV and slow extraction; and a pulsed e-beam of up to 50 GeV produced in the former CNGS target area with the AWAKE plasma acceleration method.

\subparagraph{Other opportunities:}
Promising feasibility studies of a gamma factory, a novel concept exploiting the unique LHC high energy to produce a high intensity $\gamma$-ray beam, have been performed. The possible layout of a nuSTORM implementation at CERN has also been studied.

\subsection*{CERN physics potential beyond colliders}
\vspace{0.3cm}

There is growing interest for BSM models involving either new particles with low masses or new forces with very low effective couplings. This stems from the non-observation of new high-mass states at the LHC up to now and, more importantly, from the realization that these new forms of BSM physics are well motivated by both theory and phenomenology. The PBC projects cover the full range of alternatives to high-energy frontier direct searches: precision and rare decay experiments, beam-dump experiments and non-accelerator low-mass axion searches. In addition a number of PBC projects are devoted to specific QCD measurements. 

\subparagraph{Precision measurements and rare decays:}
Precision measurements and rare decays probe higher masses than accessible with LHC direct searches, through the contribution of loop diagrams. The rare $K$ decays to be investigated by NA62 ($K^+\to \pi^+\nu\bar{\nu}$) and KLEVER ($K^{0}_{L}\to \pi^{0}\nu\bar{\nu}$) are complementary to each other and to $B$ decays for BSM searches, and are accessible in a novel in-flight technique thanks to unique high-intensity/high-energy CERN hadron beams. TauFV@BDF has also a leading potential for $3^{rd}$-generation LFV decays ($\tau\to 3\mu$) thanks to the unique characteristics of the BDF beam. REDTOP devoted to $\eta$ rare decays could have access to complementary BSM models provided sufficient high luminosity can be collected. The unique high energy $\mu$-beam also offers an opportunity for a CERN contribution to the exploitation of the (g-2)$_{\mu}$ experiments, with a direct measurement by MUonE of the term responsible for the dominant theory uncertainty.

\subparagraph{High energy beam dumps:}
Beam dump like experiments probe a specific MeV-GeV mass range of the hidden sector parameter space of special interest to solve open questions in cosmology. The PBC projects make full use of the CERN opportunities with e-beams (NA64++(e) and LDMX@eSPS), $\mu$-beams (NA64++($\mu$)) and proton beams (NA62++ and SHiP@BDF). The BDF high energy beam provides extended access to the high-mass range of the targeted region. Using both appearance and disappearance signatures, as well as different types of particle beam, maximizes sensitivity in couplings and masses. Hidden sector benchmark models have been defined to compare the reach of the PBC projects with each other and with the worldwide competition including dedicated long-lived particle projects at the LHC.  

\subparagraph{Low energy hidden sector:}
In line with past CERN non-accelerator experiments, new PBC projects propose to further explore axion models with a unique reach in the sub-eV range (IAXO helioscope and JURA light regeneration experiment), motivated by the QCD axion as well as astrophysical hints. The EDM ring could also potentially probe the very low mass region with oscillating EDMs.

\subparagraph{QCD measurements:}
The growing number of QCD-oriented facilities across the world, either in operation (JLab, RHIC, J-PARC Hadron Experimental Facility), in construction (NICA, FAIR), or in discussion (EIC) reduces the windows of opportunities for competitive measurements at CERN in the future, but several unique cases were identified within the PBC projects for both hadron and QGP measurements.

\textit{At the SPS:} The CERN high-energy $\mu$-beam provides an opportunity for a meaningful contribution to the proton radius puzzle by the COMPASS++($R_p$) program. Chiral QCD could be tested in the full SU(3) sector with DIRAC++ mesonic $\pi$$K$ atoms and a COMPASS++ $K$ polarisability measurement. COMPASS++ could also perform unique pion structure measurements with DY and revisit strange spectroscopy with high statistics. The SPS kinematic domain gives a unique access to the QGP phase transition region where a Critical Point is expected. This is proposed to be revisited by NA61++ with open charm and by NA60++ with low-E dimuons. 

\textit{At the LHC:} There is a unique kinematic domain that can be explored by LHC Fixed Target measurements. Precision quark/gluon high-x PDFs, a crucial input to exploit HL-LHC for high-mass new particle searches, as well as spin asymmetries, could be extracted. The LHC-FT kinematics also corresponds to the QGP cross-over between the low-$\mu_B$ LHC/RHIC region and the SPS Critical Point region, where high statistics measurements would be important to consolidate predictions for high-$\mu_B$.

\subsection*{PBC projects competitiveness and implementation}
\vspace{0.3cm}

\subparagraph{Short-term opportunities:}
There are significant discovery potentials of NA64++(e, $\mu$) and of NA62++ in beam dump mode which are unrivaled on the Run 3 timescale. The NA64++(e) efficiency should be maximized to cover as much as possible of the DM-favored region. The NA62++ beam dump operation will also provide valuable input to finalize the SHiP design.

MUonE, COMPASS++($R_p$) and NA64++($\mu$) all have unique motivations but are in competition on the same $\mu$-beam during Run 3. The feasibility and achievable precision of the projects still need quantification. Anticipating positive outcomes, optimal use of the high-intensity $\mu$-beam requires to strengthen the studies to fit the programs together. The open charm measurement planned by NA61++ in the QCD Critical Point region is unique and has no new competition expected on the beam line.

\subparagraph{Long-term opportunities:}
The PBC benchmark models comparisons show that SHiP@BDF has a unique discovery potential compared to the worldwide competition in the next two decades. Including a small upstream experimental hall in the BDF baseline design, as proposed by TauFV, would increase the BDF potential by paving the way to a long term rare decay facility with a reach exceeding BELLE-II for the 3\textsuperscript{rd} flavor generation.

AWAKE++ and eSPS offer attractive options of new high intensity e-beams to go much beyond the reach of current CERN e-beams for dark photon searches in visible and invisible modes, respectively. LDMX proposed on eSPS has a shorter term implementation opportunity at SLAC, pending approval of LCLS-II beam extraction. The planned DESY XFEL upgrades may also offer unique opportunities for such searches at unrivaled intensity.

The KLEVER detector and high intensity $K^0$-beam can in practice only be implemented in the NA62 hall. KLEVER phasing after NA62 will depend on NA62 $K^+$ results, on the progress of the KOTO competitor at J-PARC, as well as on the evolution of the $B$ anomalies and their possible explanations.

COMPASS++ as a long term QCD factory needs to prioritize its physics goals. Measurements such as the $\pi$-structure can make use of existing beams, but others are pending on significant beam upgrades, e.g. a RF-separated $K$-beam for strange spectroscopy. Competition in the experimental hall will depend on the future of MUonE and NA64++($\mu$).

LHC Fixed Target measurements have already started with LHCb. Contributions from ALICE would extend the potential thanks to the different acceptances, strengths and operation modes of the two detectors. High statistics is essential for many measurements, and the overall physics reach will depend strongly on the possibility to combine LHC Fixed Target and collision operation in an efficient way.

A few projects with unique physics reach face specific difficulties for implementation at CERN: DIRAC++ and NA60++ can reasonably fit only in the current NA62 hall and could therefore start implementation only when the hall is freed; CERN beams are not optimal for REDTOP and would have difficulties to produce the required luminosity for a discovery. Considering the time needed to finalize the detector design, a REDTOP implementation at FNAL sounds more efficient.  

CERN participation in the design of non-accelerator experiments such as IAXO proves decisive to these projects. Several technological domains (high field magnets, RF-cavities, cryogenics, vacuum, optics) have been identified as particularly suited for future exchange of expertise around non-accelerator projects, for the mutual benefit of CERN and outside laboratories.

\newpage
\section{Setting the landscape}

\bigskip

An overview of the particle physics landscape is given, including the theoretical situation and main open questions, a description of the CERN complex and its non-LHC experimental program, a short presentation of the PBC projects proposed by the community for the future and a summary of the expected worldwide competition. 

\bigskip
\subsection{Physics landscape}
\bigskip
Over the last few years fundamental physics has made tremendous progress.
With the discovery of the Higgs and the confirmation of its scalar nature the Standard Model has been completed.
Significant progress has also been made in understanding the behaviour of the Standard Model itself, and in particular its strongly coupled sector.

Yet, at the same time more questions than ever remain open. Cosmological measurements have revealed that 95\% of the energy content of the Universe is made up from dark matter and dark energy. While the latter can be described to good accuracy by a cosmological constant, its magnitude cannot be determined within the Standard Model and appears very peculiar in its context, e.g. the corresponding energy density  is about 50 orders of magnitude smaller than the energy released during the electroweak phase transition when the Higgs field settled into its ground state and acquired its vacuum expectation value.
Dark Matter on the other hand exhibits all the features expected of a particle. But no suitable candidate exists within the Standard Model.
Moreover, the allowed mass range for such a particle is enormous. Particle masses between $\sim 10^{-22}$~eV and the Planck scale are possible; even macroscopic objects such as black holes of masses of up to $\sim 10$ solar masses are still not ruled out.
Perhaps equally pressing, while the Standard Model describes ordinary matter to an incredible precision, it does not provide a mechanism that generates the observed excess of baryons over antibaryons that is responsible for the remaining 5\% of the energy content of the Universe.

While there is thus exceedingly convincing evidence that there must be Physics Beyond the Standard Model (BSM Physics), its nature remains unknown.
This is exacerbated by the fact that so far no significant (i.e. $\geq 5\sigma$) and convincing (i.e. demonstrably unaffected by possible systematics) deviation from the Standard Model has been found in experiments on Earth.
A question is therefore why BSM physics so far has not been found in experiments. One reason could be that the new particles are heavier than the few TeV energy scale currently explored at LHC. This frontier can and will be further pushed forward by the high luminosity phase of the LHC as well as in potential, even higher energy, colliders such as the proposed FCC.

Yet over the past 10 years another option has received increasing attention. It could be that the new physics is only weakly coupled to the Standard Model and resides in an aptly named ``dark'' or ``hidden sector''. This is particularly suggestive since both dark matter and dark energy must be very weakly coupled to the Standard Model (at least effectively) in order to be in line with observations and experimental attempts to detect these substances.
On the theoretical side particles weakly interacting with the Standard Model are also very appealing. They can be straightforwardly realized in theoretical models in the form of new fields that are uncharged under the SM gauge groups and which can then interact only via a relatively small set of ``portals'' to the SM particles. These models can in many cases also be embedded in more fundamental theories such as string theory.

There is an additional important feature motivated both from theory as well as phenomenology: particles (very) weakly coupled to the SM could even be light and have masses well below the electroweak scale, possibly much smaller than an eV. Phenomenologically it is clear that sufficiently weakly coupled particles evade detection. On the theory side symmetries can ensure a small mass, the classic example being pseudo-Goldstone bosons of slightly broken global symmetries. 
Many phenomenological applications have been discussed. Examples are dark matter, messenger particles to dark matter, explanations of the $(g-2)_{\mu}$ anomaly, the proton radius anomaly, stellar cooling anomalies and many more.

Detecting light and moderate mass new particles (very) weakly coupled to the Standard Model provides both new challenges as well as opportunities and the last few years have seen an exciting development of many new ideas in a lively and fruitful exchange between theorists and experimentalists. Typical features of the experiments involved are high intensities and/or high precision and accordingly this new direction has been dubbed the Intensity/Precision Frontier.

Importantly the Intensity and Precision frontier is directly related to the High Energy Frontier explored at colliders such as the LHC.
Very weak interactions often arise by integrating out heavier particles and the weakness of the interactions is related to the underlying very large energy scale. This is most explicit when the interactions feature a dimensionful coupling constant related to a power of an inverse mass scale $\sim 1/M^{k}$. The mass scale is then related to the masses of new particles in the underlying theory. Examples where this is explicit are, e.g. axions. However, such small extra interactions do not necessarily need to involve new BSM particles, instead they can also contribute (weak) extra interactions for the SM particles. Again this can be explored in precision experiments. A prominent example of this strategy at work is the study of rare meson decays. Typically these decays are suppressed (e.g. by approximate symmetries) in the Standard Model and even small new physics effects could be observed. Of course to see these decays we need a sufficient number of initial mesons, requiring high intensities. Then this allows these experiments to indirectly explore energy scales often way above the TeV scale. Similarly, small effects can be observed in deviations of very precisely calculable quantities such as $(g-2)_{\mu}$, or in violations of symmetries with e.g. CP violating electric dipole moments of particles such as neutrons, protons and electrons.

\bigskip
While the Standard Model has proven to be an excellent description of ordinary matter, our understanding of it is far from complete.
This is most evident in the strongly interacting QCD sector, despite enormous progress such as the discovery of the quark gluon plasma. Most well controlled calculations in quantum field theory are performed using perturbation theory in a small coupling constant. At low energy in QCD the coupling constant is large and this approach fundamentally breaks down. In the last years non-perturbative methods such as lattice and renormalization group methods have made significant progress. Nevertheless a full first principles quantitative description of the phase diagram and of strong coupling dynamics is still missing. Moreover, in interesting situations QCD lattice and renormalization group methods face problems, e.g. when the density (as featured for example by neutron stars) is high lattice calculations have a sign problem preventing convergence with a reasonable amount of computational time.
To further develop our methods and a quantitative physical picture requires a healthy interplay between theory and experiment, in particular those that achieve extreme conditions of high temperature and high density. An important way to achieve these conditions is high and moderate energy collisions of heavy ions.

In addition to furthering the understanding of QCD itself, strong interactions play a part in experiments exploring physics beyond the SM: measurements of QCD quantities are often required as input parameters for new physics searches.
An important example are measurements of the parton distribution functions that enter cross section calculations in proton-proton  collisions. They enter precision measurements at LHC but they can also increase the reliable reach for heavy particles in the high luminosity phase of the LHC. Another crucial quantity is the hadronic contribution to the $(g-2)_{\mu}$ that at present can not be calculated with the required precision and has to be obtained from measurements of suitable quantities. More indirectly, but also importantly, indirect detection of dark matter in astrophysical observations often requires inputs on nuclear cross sections.
All in all there is a crucial interplay between QCD measurements and searches for new physics.

\subsection{CERN complex landscape}

\vspace{0.5cm}

This section provides an overview of the CERN complex together with details on the low energy and high energy facilities associated to the LHC injectors.

\subsubsection{Complex overview}

\begin{figure}[!htb]

   \centering
   \includegraphics*[width=450pt]{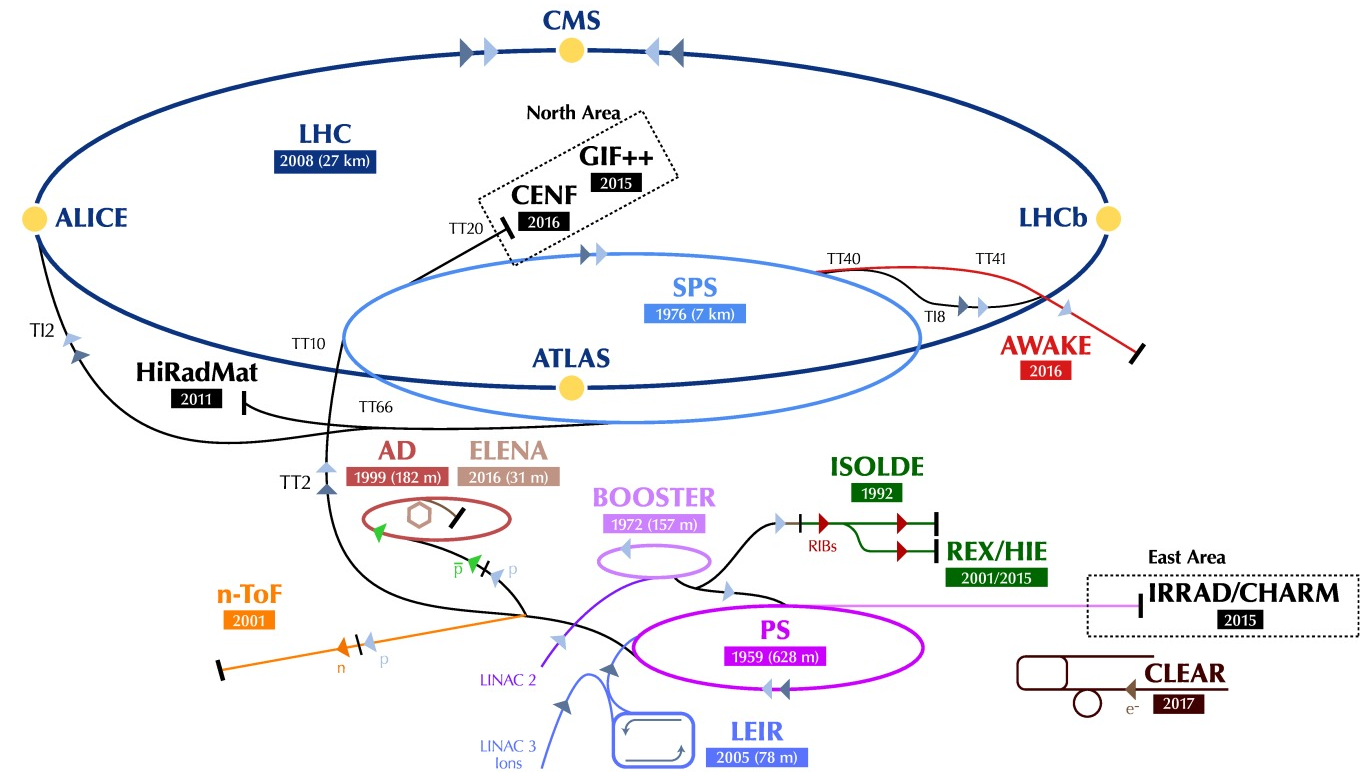}
   \caption{Overview of the CERN accelerator complex circa 2018. Figure taken from~\cite{Mobs}.}
   \label{fig:complex}

\end{figure}

CERN's accelerator complex has a long history of evolution and adaptation with each of the four operating synchrotrons
now serving a diverse physics community. An overview of the complex is shown in figure \ref{fig:complex} and the current parameters of the main facilities summarized in table \ref{tab:complex}.

\begin{table}[h]
\begin{center}
\caption{Proton beams delivered by the CERN complex to the main facilities/users. (SE -- slow extraction; FE -- fast extraction)}
\label{tab:complex}
\begin{tabular}{p{3cm}lll}
\hline\hline
\textbf{Machine}  & \textbf{Facility}           & \textbf{Energy at extraction}
                                                & \textbf{$\sim$Protons per cycle}\\
\hline
Linac4  &        & 160 MeV          &  user dependent \\
Booster  &  ISOLDE  & 1.4 GeV             & \num{3e13} \\
PS    &  n\_TOF &      20 GeV         &   \num{9e12} \\
PS    &  East Area &      24 GeV        &   \num{6e11}  \\
PS    &  AD/ELENA &     26 GeV          &   \num{1.5e13} \\
SPS (SE)   &  North Area &    400 GeV           &  \num{2.8e13}  \\
SPS (FE)   &  LHC & 450 GeV              & \num{3.4e13}  \\
SPS (FE)   &  HiRadMat & 440 GeV   &  \num{4e13} \\
SPS (FE)   &  AWAKE & 400 GeV       & \num{3e11}  \\
SPS (FE)   &  CNGS (2005 -- 2012) &  400 GeV             &  \num{4.5e13}  \\

\hline\hline

\end{tabular}
\end{center}
\end{table}

The complex provides multi-user facilities with a low duty cycle from Linac4, with the defining cycle time of 1.2 s being given by the Booster. 
Despite the low duty cycle, impressively high intensity per user per cycle are delivered to the lower energy users: ISOLDE, n\_TOF, Antimatter Factory, and the experiments and test facilities of the East Area. These users profit from the operational configuration which delivers low rep rate but high energy and high pulse intensity.
The complex's operational model is in contrast to high intensity CW delivery which characterises operation in, for example, PSI, which makes CERN noncompetitive at the so-called Intensity Frontier characterised by high mean power on target at low energy.

The possibility for CERN to move to a more rapid cycling regime had been investigated in the 2000's with the well-developed concept of a superconducting proton linac (SPL) injecting into a higher energy version of the PS (PS2). The SPL/PS2 goal was to act as a proton driver for a European based neutrino program in the first instance in parallel with delivering beam to the established users of the complex. A technical design for a rapid cycling synchrotron to replace the Booster was also produced.

\begin{figure}[!htb]

   \centering
   \includegraphics*[width=400pt]{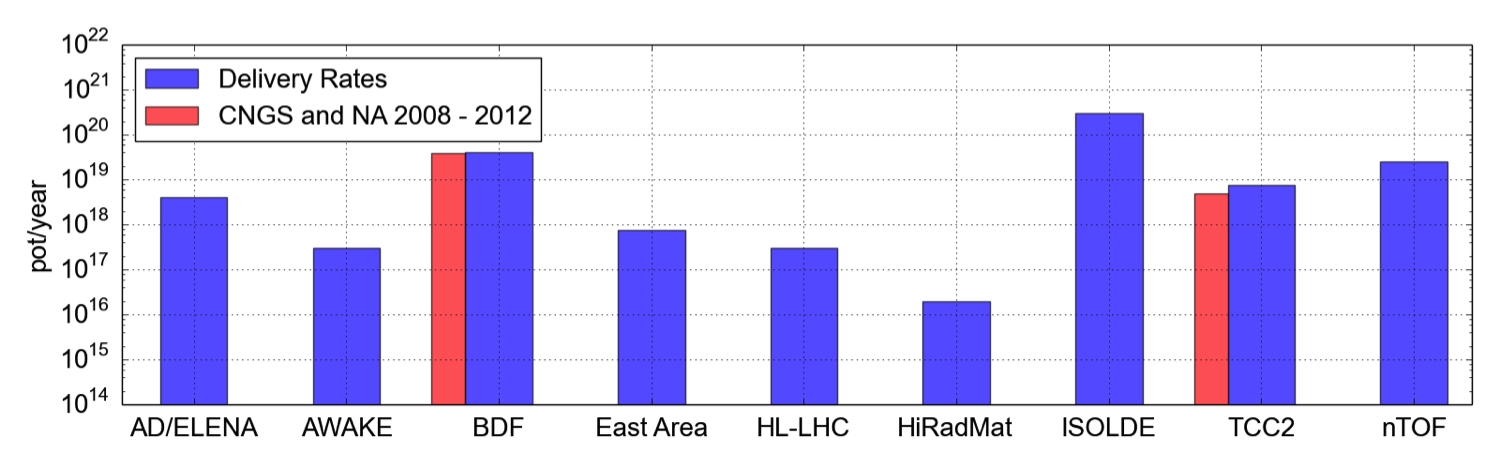}
   \caption{Possible proton delivery per year in the complex following the LIU upgrade. BDF, discussed below, represents a major additional user at the North Area, with a PoT requirement similar to former CNGS. TCC2 are the North Area targets serving the Conventional Beam experiments, discussed below. Figure taken from~\cite{rumolo}.}
   \label{fig:protons-LIU}

\end{figure}

These proposals were rejected in favour of a major upgrade of the existing LHC injector complex (LIU), with the principle aim of delivering the requisite high bunch population and low transverse emittance beams for the HL-LHC. The HL-LHC is due to be commissioned in 2026 and aims to deliver around 250 fb\textsuperscript{-1} per year and a total of 3000 fb\textsuperscript{-1} over the following 10 years of operation. The HL-LHC upgrade program impacts all of the injector chain. Its potential to increase beam intensity and energy to existing clients has been studied in detail under the auspices of Physics Beyond Colliders \cite{PBC:LIU}. A summary of the possible proton use in the post-LIU era is shown in figure \ref{fig:protons-LIU}.
The demands of the lower energy facilities can be seen.
The ability of the SPS to deliver beam to an additional major user while maintaining regular service to the North Area users is also indicated.
In this respect, the delivery of beam to CNGS, \num{1.8e20} protons on target (PoT) total in the years 2008 - 2012, is shown for reference.

Comparison with other laboratories worldwide (appendix 2) shows that with their high energy and high luminosity both the SPS and LHC provide unique capabilities. The SPS offers high intensity and high energy protons via either slow or fast extraction.
The LHC is, in the main, delivering high luminosity, high energy proton-proton collisions and is clearly the defining energy frontier machine in the medium term.

The SPS and LHC also operate with heavy ions. The SPS delivers ions to experiments in the North Area at a variety of energies.
The LHC has a vigorous and successful ion program serving ATLAS, CMS, ALICE and LHCb. The provision of high intensity ion beams from the injectors has received attention under the auspices of the LIU project and beam parameters commensurate with those foreseen for the HL-LHC era have been realized in the 2018 run.

\begin{figure}[!htb]
  \centering
  \includegraphics*[width=450pt]{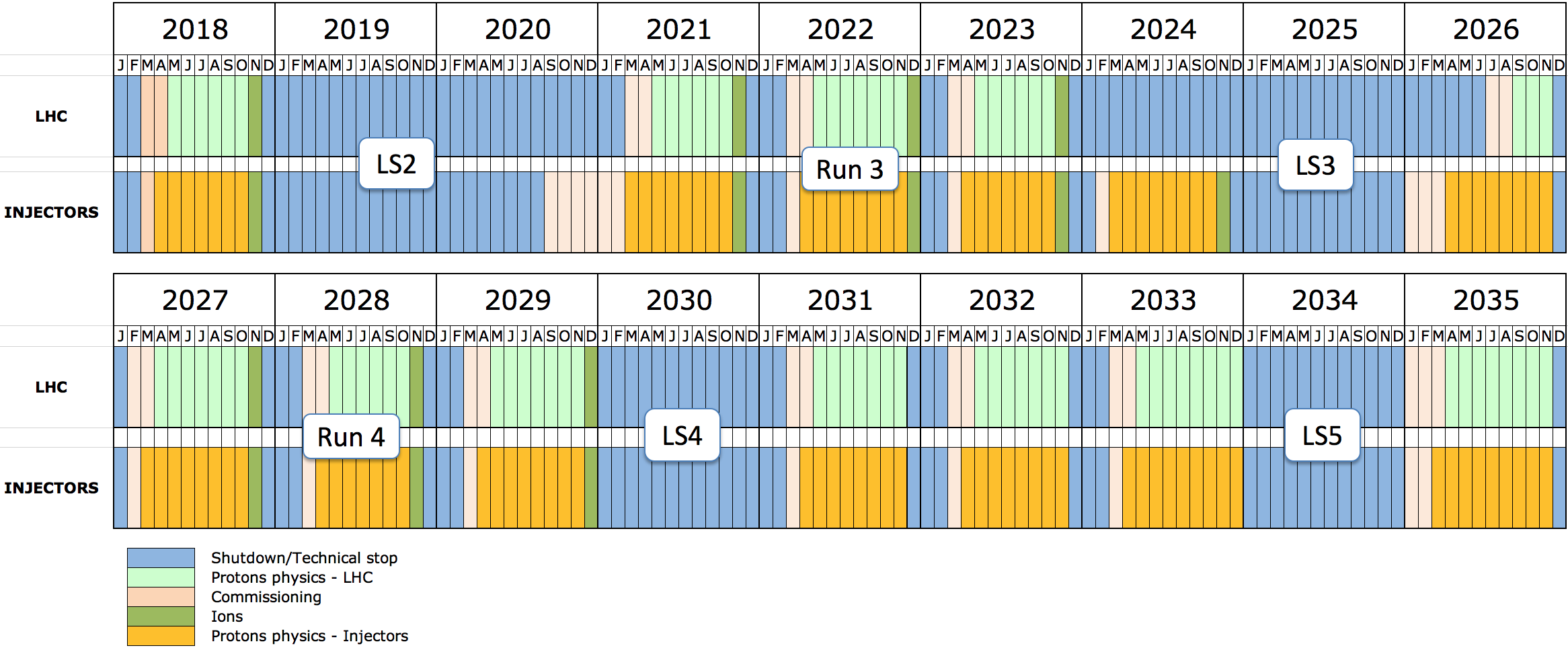}
  \caption{Long-term LHC and injectors schedule}
  \label{fig:LHCschedule}
\end{figure}

The LHC, subsequently the HL-LHC, will remain the CERN flagship in the medium term. The injectors long term operational planning is driven by the LHC's rhythm of 3 to 4 year runs interspersed with long shutdowns of 1 to 2 years, with one exception being an extra year's injector complex operation in 2024 on the lead in to LS3. The current schedules for the LHC and its injectors operation are summarized in figure \ref{fig:LHCschedule}.

\subsubsection{Low energy facilities}
\vspace{0.5cm}

The complex has some very well-established low energy facilities (ISOLDE, n\_TOF, Antimatter Factory) with passionate, vocal, and equally well-established user communities. 
Their beam requirements from the Booster and PS are clearly defined and significant, leaving little room for any additional user or facility requiring high accumulated intensity below 26 GeV.
It is assumed that recent upgrades have secured the future of these programs for at least the next decade:

\begin{itemize}
\item The Antimatter Factory was up to now based on the Antiproton Decelerator (AD), currently the world's only facility providing $\bar{p}$'s of very low (5.3 MeV) energy needed for precision measurements of antiprotons and antihydrogen ($\bar{H}$) atoms. 
The new synchrotron ELENA (Extra-Low ENergy Antiproton ring) of 30-m circumference is under commissioning to capture the 5.3-MeV $\bar{p}$ provided by AD and decelerate them to E = 100 keV.
ELENA will reduce the momentum spread of the $\bar{p}$ using electron cooling techniques, thereby achieving a low transverse beam emittance. The number of $\bar{p}$ captured in Penning traps per unit time will be increased by a factor 100 for the experiments which used the 5.3-MeV $\bar{p}$ beam up until the end of 2018. ELENA will also allow parallel operation of the experiments.

\item The on-line isotope mass separator ISOLDE is a facility dedicated to the production of a large variety of radioactive ion beams for many different experiments in the fields of nuclear and atomic physics, solid-state physics, materials science and life sciences.
HIE-ISOLDE represents important upgrades of the ISOLDE radioactive beam facility. The main focus lies in increasing the energy of the radionuclide beams from 3 MeV/u up to 10 MeV/u via a superconducting post-accelerator. The upgrade substantially enhances research opportunities in most aspects of nuclear structure and nuclear astrophysics, making ISOLDE the only facility in the world capable of accelerating medium to heavy radioactive isotopes in this energy range.

\item n\_TOF is a pulsed neutron source coupled to a 200 m flight path designed to study neutron-nucleus interactions for neutron kinetic energies ranging from a few meV to several GeV. 
The study of neutron-induced reactions is of importance in a wide variety of research fields, ranging from stellar nucleosynthesis, symmetry breaking effects in compound nuclei, and the investigation of nuclear level densities, to applications of nuclear technology, including the transmutation of nuclear waste, accelerator driven systems and nuclear fuel cycle investigations. The recent construction of a $2^{nd}$ n\_TOF experimental area has significantly enhanced the physics reach of the installation.

\end{itemize}

\subsubsection{High energy facilities}
\vspace{0.5cm}

High energy Fixed Target physics is performed at the SPS, where beam time and proton allocation is less constrained than for the above low energy facilities. The SPS can serve several high energy clients in parallel within a repetitive operation sequence (``Supercycle"), of a typical length of a few tens of seconds. The LHC makes clear but limited demands, and excellent availability and few premature dumps have led to over 50\% of the scheduled LHC physics time being spent in Stable Beams, lessening the time required from the SPS.
The most demanding users of SPS conventional beams are sited in the North Area -- detailed below -- and the standard slow extraction cycle can be used to fill any time available.
For the moment, HiRadMat and AWAKE make limited, punctual demands.
Machine development, both dedicated and parasitic, should also be factored in. 
Approximately, one month of an operational year is dedicated to the delivery of ions to both the North Area and the LHC. In general, this is not compatible with proton operation.

The North Area (NA) receives a primary proton beam at 400 GeV/c from the CERN SPS \cite{PBC:CONV}. 
The full SPS proton beam is slowly extracted to the NA during a flat top of typically 4.8
seconds, which could be modified in case the physics program requires a shorter (e.g. Beam
Dump Facility) or longer flat top (e.g. in case of cohabitation with the Beam Dump Facility) in
the same super-cycle.
A typical duty cycle (defined as the ratio between useful flat top length and super-cycle) is between 20 and 30\%.
The recent maximum extracted proton flux was about \num{3.5e13} protons-per-pulse (ppp), but the aim is to reach \num{4.0e13} ppp in
the future. This proton flux is transported and shared through two series of splitter magnets
onto three primary targets, T2, T4 and T6, from which the NA beam lines are served.

An important limitation on the intensities extracted to the North Area is the inherent losses involved during slow extraction. 
In particular these impact the electrostatic septa situated in the extraction region.
A concerted program to examine novel methods of loss reduction (e.g. shadowing with diffusers and/or crystals) is in progress.
Significant reduction of the present loss rate is a prerequisite for, say, the BDF initiative.

\begin{figure}[!htb]
   \centering
   \includegraphics*[width=420pt]{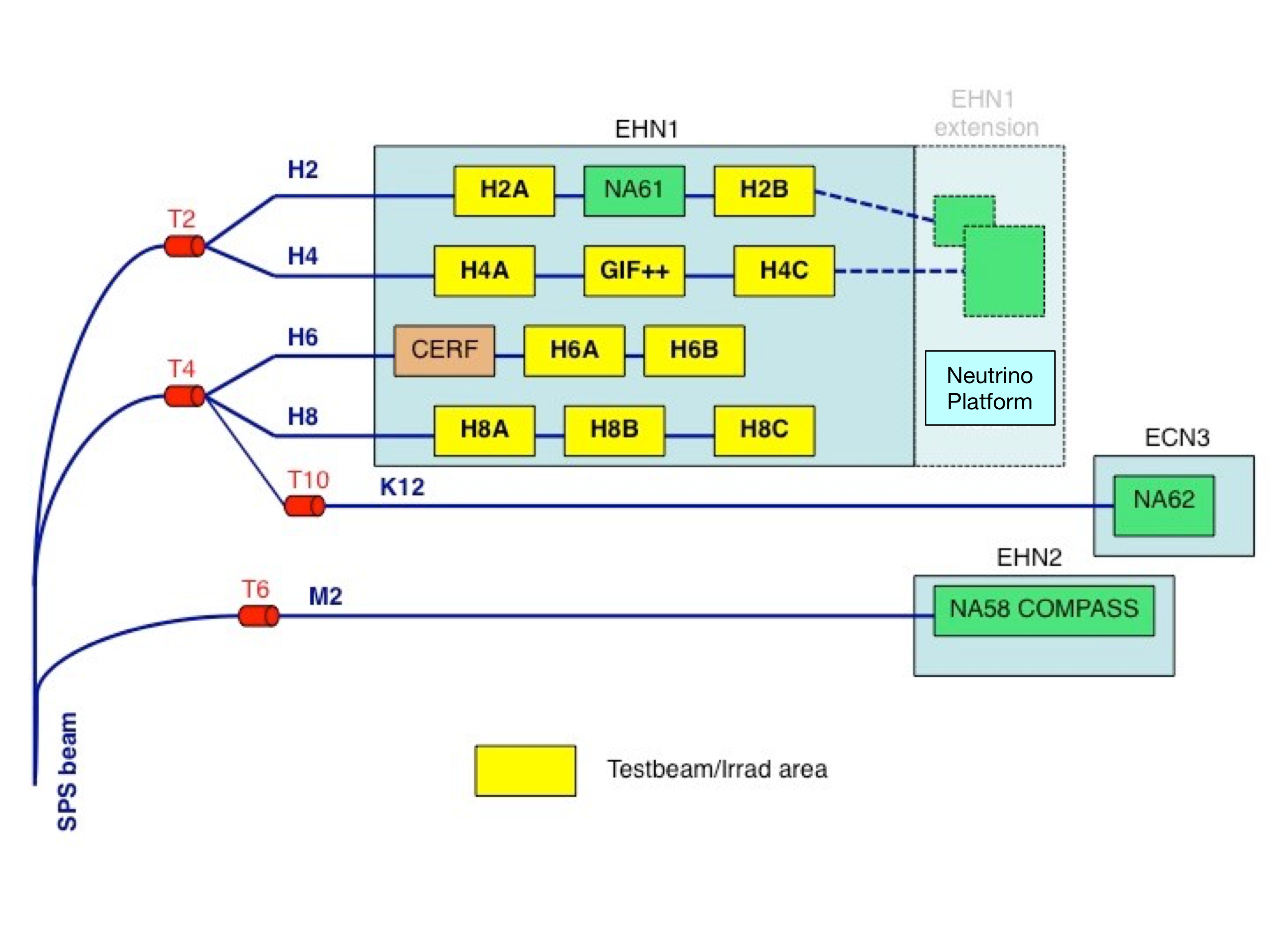}
   \caption{Schematic of North Area showing targets and beam-lines. Figure taken from~\cite{aida}. }
   \label{fig:NA}
\end{figure}

The North Area comprises two surface halls, EHN1 and EHN2, and an underground cavern, ECN3. 
EHN1 is the biggest surface hall at CERN (\num{330 x 50} \si{m^2}) and houses the H2, H4, H6 and H8 beam lines. 
EHN2 is served by the M2 beam line for muon, hadron and electron beams and serves at the moment the COMPASS experiment. 
In ECN3 the NA62 experiment for rare kaon decays is served by the K12 beam line. 
A schematic layout of the current NA complex is shown in Figure \ref{fig:NA}.

\subsubsection{Ongoing R\&D}
\vspace{0.5cm}

CERN has wide-spread R\&D across the accelerator sector. This ranges from advanced computing techniques, through material science for advanced collimation and targets, 
advanced vacuum technology, to solid state RF power generation and so forth. 
Undoubtedly, part of CERN's strength is its ability to stay abreast of, and in some cases, push, technological advance.
In most cases, novel development is requirement driven and has a medium-term objective in mind.
Some, however, target more long term objectives such as the High Energy Frontier.

Of note, in terms of scale, are:

\begin{itemize}
\item  {\bf High field magnets} is a well-funded activity backed by a well-established international collaboration.
In the medium term, there are hard deliverables for the HL-LHC in the form of wide-aperture \ensuremath{\mathrm{Nb_3Sn}} inner triplet quadrupoles and 11 T dipoles for use in dispersion suppressor collimator insertions. 
In the longer term, development targets the requirements of the FCC, in particular the 16~T dipoles required for FCC-hh, also foreseen to be based on \ensuremath{\mathrm{Nb_3Sn}} technology.

\item {\bf Superconducting RF} has a long history at CERN and indeed it has become a work-horse of the laboratory, firstly at LEP, subsequently at the LHC, and in recent novel developments such as the 
HIE-ISOLDE 1/4 wave resonator structures and the crab cavities of the HL-LHC. R\&D in this area is typically driven by requirements, and in this regard, the medium term at CERN awaits a driver.

\item { \bf X-band RF} technology has seen prolonged attention by the CLIC collaboration. Given proven 12~GHz accelerating structures, with or without drive beam, the aim in this area would appear to be application driven rather than further concerted R\&D.

\item {\bf Plasma wakefield acceleration} is developed by the AWAKE experiment (see chapter 3) including implementation of plasma cell technologies.  

\item {\bf Detector technology R\&D} largely anticipates the serious challenges of medium term upgrades (HL-LHC) and the longer term options (ILC, CLIC, FCC). 
\end{itemize}
The above are mainly project driven with dedicated lines in CERN's Medium Term Plan.

\newpage
\subsection{CERN PBC experimental landscape}
\vspace{0.5cm}
\subsubsection{Experimental evolution in the past decade} 
\vspace{0.5cm}
The CERN LHC injector complex has been hosting a vigorous particle physics program in the past decade, gathering about a thousand physicists within about 20 experiments.

The CNGS neutrino beam program was successfully conducted until 2012 with a total of \num{1.8d20} PoT delivered for the OPERA and ICARUS detectors located in the Gran Sasso underground Laboratory. OPERA established the oscillation of $\mu$-neutrinos to $\tau$-neutrinos with a significance higher than 6$\sigma$, and ICARUS paved the way to large underground liquid argon TPCs, which are now being further developed within the CERN neutrino platform for the future DUNE neutrino long baseline program in the US. Completion of CNGS operation leaves room for new high intensity/high energy programs as regards the proton yield of the CERN complex. 

Long-standing experiments devoted to QCD measurements have been steadily continued: the COMPASS experiment, a program initially mainly devoted to the parton spin content of the proton, is currently completing its COMPASS II program focused on the proton orbital angular momentum through muon Deeply Virtual Compton Scattering (DVCS), and on Transverse Momentum Dependent (TMD) measurements with polarized $\pi$-$p$ Drell-Yan interactions; the NA61 experiment, an upgrade of one of the heavy ion fixed target experiments (NA49) which established QCD deconfinement in the 90's, is continuing the exploration of the QCD phase transition by looking at the Critical Point expected in the kinematical range of the SPS; the DIRAC data taking at the PS was completed in 2012, providing the first observation of $\pi$$K$ mesonic atoms among other results.

Antimatter precision measurements at the Antimatter Factory got a strong impetus in the past years: The ATRAP, BASE and ASACUSA experiments considerably improved antiproton parameters measurements (magnetic moments, mass, etc.); magnetic trapping of anti-hydrogen atoms was established on a routine basis and first measurements of microwave and laser spectroscopy of anti-hydrogen performed by the ALPHA experiment; new detectors devoted to the gravitational interaction of anti-hydrogen atoms are being commissioned (ALPHA-G, AEGIS) or built (GBAR). The ELENA ring under commissioning will increase by 2 orders of magnitude the trapping efficiency of anti-hydrogen atoms while allowing experiments to operate in parallel. ELENA secures the long term future of the Antimatter Factory. For this reason, and also because small-size antimatter experiments evolve on short time scales, the Antimatter Factory was not explicitly considered in the PBC study. Care will however have to be taken to maintain enough flexibility in the future implementation of new ideas within the facility since the CERN Antimatter Factory is unique worldwide and may yield breakthroughs in the understanding of the matter-antimatter asymmetry.     

The new projects recently started in the North Area have focused on searches for new physics: the NA64 experiment was installed on the SPS H4 electron beam (Figure \ref{fig:NA}) and provided improved limits on the radiation of dark photons by electrons, excluding a possible explanation of the muon g-2 anomaly. After many years of construction, the NA62 kaon experiment, a successor of NA48 mainly devoted to ultra rare $K^+$ decays into $\pi^+\nu\bar{\nu}$, started data taking on the SPS and reported first candidates. The detector has a vocation to run several years and will investigate the flavour sector complementary to B factories.    

Last but not least, CERN has been hosting non-accelerator experiments exploiting the unique opportunities offered by the LHC high field magnets: the CAST experiment points a LHC prototype dipole at the sun and has provided unique limits on the production of axions, excluding part of the QCD axion domain. The OSQAR experiment has conducted a laser light shining through a wall experiment based on two  LHC dipoles and has provided, to date, the best laboratory limits on axions.   

\subsubsection{Brief summary of future options} 

\vspace{0.5cm}

The community surveys performed within the PBC study collected a wealth of suggestions on how to further exploit the CERN complex in the future. An overview of the initial ideas is presented here, with more details on the programs and the current status of their PBC studies given in the next section.

Most of the ongoing North Area experiments have proposed detector upgrades and extensions of their data taking beyond the currently approved programs. As regards QCD measurements, COMPASS wishes to further operate as a long term QCD facility devoted to hadron structure, spin and spectroscopy, with a focus on hadron beams including higher intensity antiproton and kaon RF-separated beams; NA61 plans to further study the QCD phase transition with open charm production. With regard to proposed searches, NA62 plans part-time operation in beam-dump mode to investigate the hidden sector, and NA64 wishes to employ active beam dump and missing energy techniques with electrons at higher intensities, as well as to extend the experiment to muon and hadron beams in searches for light dark bosons from a dark sector that are coupled to photons.  

New experiments have also been proposed with conventional (upgraded) beams. A new idea, MUonE, is proposed to measure the vacuum hadronic polarization (one of the two dominant sources of theory uncertainty in the muon g-2 prediction) in the t-channel with a high energy muon beam, complementary to the e\textsuperscript{+}e\textsuperscript{-} method used up to now. NA60++ plans to revive the NA60 concept to explore the QCD phase diagram through an energy scan of low-E dimuon production in Pb-Pb collisions. DIRAC++ wishes to install a DIRAC-like detector on an SPS proton beam, which would boost the production of mesonic atoms by a factor 20 compared to the previous DIRAC location at the PS. In the search domain, the KLEVER project wishes to extend the NA62 method to ultra-rare $K^0$ decays using a high intensity $K^0$ beam, whereas REDTOP is motivated to investigate ultra-rare $\eta$ decays using a high intensity CW low energy proton beam.

New Fixed Target facilities based on the SPS are considered. The Beam Dump Facility (BDF) based on a high intensity, high energy, and slow extracted SPS proton beam, is proposed as a new general purpose high intensity facility.  In the first instance, the SHiP detector could perform a comprehensive investigation of the hidden sector. In addition, a small fraction of the upstream beam is proposed to be used by the TauFV experiment to investigate forbidden $\tau$ decays. 
Complementary to the BDF, an extension of the AWAKE set-up could provide a high intensity medium energy pulsed electron beam for investigation of the hidden sector with a detector located in the former CNGS decay tunnel. The eSPS study team proposes that the SPS could deliver a primary low energy electron beam to a missing momentum dark matter experiment. The electrons would be provided  by implementing a new e-injector linac based on CLIC technology .    

Fixed Target physics is also proposed to be developed at the LHC using internal targets located near the LHCb or ALICE collider interaction points, as pioneered by LHCb with the SMOG system.  This opens a unique new kinematical domain to proton and heavy ion measurements, primarily for QCD physics. Several methods based on internal gas targets, including polarisation options, as well as on beam halo crystal extraction using UA9 developments, are being explored.  

Other novel, longer term, facilities have been proposed within the PBC study. One highlight is an electrostatic storage ring for a high precision measurement of the proton electric dipole moment. The feasibility and motivation for a "$\gamma$-factory", a very high intensity $\gamma$-beam produced by conversion of a laser beam through partially stripped ions stored in the LHC, is being studied. A CERN implementation of nuSTORM, a high intensity $\nu$ beam based on a muon storage ring, has also been proposed.   

Finally, several groups have proposed to further exploit the synergies between CERN technological expertise and projects developed in external institutes for searches and precision measurements. Some highlights include the CERN contribution to the design of the large scale magnet of IAXO, the next generation axion helioscope considered as a successor of CAST, and the potential use of high field magnets being developed for the HL-LHC and the FCC in a variety of searches for axion-like particles.  

\subsection{Worldwide landscape}
\vspace{0.5cm}

The proposed projects of PBC are part of some lively worldwide activity in particle physics. 
A short overview is presented here with a focus on programs that have a significant overlap in physics reach with PBC. 
The expected sensitivities of the experiments will be compared quantitatively in chapter 4. 

CERN currently hosts the only collider operating at the high energy frontier, the LHC, whereas other large laboratories have moved their interest to high-intensity and precision facilities (see Appendix 2 for a survey of existing beam facilities). On the LHC, aside the two general-purpose experiments ATLAS and CMS which focus on the search of new high mass states, there is growing interest to extend searches to more weakly coupled new particles thanks to detectors located further away from the collision points (FASER, MATHUSLA, CODEX-b, milliQan). Outside CERN, the main worldwide projects in operation, construction and discussion for the next two decades are listed below. For convenience they are classified as hadron, muon, and electron beam facilities, low energy colliders, and non-accelerator experiments.      

\subparagraph{Hadron beam facilities:} The two flagship long-term high-intensity multi-GeV proton facilities are FNAL in the US and J-PARC in Japan. At FNAL, the former Tevatron injector complex provides 8~GeV and 120~GeV proton beams whose intensity will be upgraded to the MW range in the next decade within the PIP-II program. The 8~GeV beam produces a short baseline (SBL) $\nu$-beam devoted to sterile $\nu$ searches. The 120~GeV beam from the Main Injector currently serves the long baseline $\nu$-beam to NOvA and will later produce the $\nu$-beam to DUNE. It also hosts a fixed target DY experiment (SeaQUEST). Hidden sector searches will be performed with the near detector of DUNE as well as with a possible beam-dump discussed for the SBL.
At J-PARC, the 30 GeV primary proton beam is the precursor of the T2K and future HyperK $\nu$-beams, with intensity upgrades to the MW range planned in the next decade. The J-PARC proton beam also feeds the J-PARC Hadron Experimental Facility hosting e.g. the KOTO experiment devoted to the $K^0\rightarrow\pi^0\nu\bar{\nu}$ decay. A new 20 GeV hadron beamline in construction, together with an extension of the Hadron Experimental Facility in discussion for the long term, may allow diverse QCD studies and important KOTO upgrades in the future. 

In the US the flagship hadron collider RHIC at BNL will continue operation with a short term program devoted to spin physics and investigation of the QGP phase transition at higher baryonic potential $\mu_B$ with a beam energy scan. In the mid-term significant detector upgrades will allow better characterization of the deconfined medium at low $\mu_B$.

Lower-energy high-intensity leading hadron facilities are also in operation or being built in Europe. The FAIR program in Germany will investigate both the QCD phase transition at high $\mu_B$ with high intensity ion beams (CBM experiment), and hadronic exotic states with antiproton beams (PANDA experiment). The European Spallation Source (ESS) in construction in Sweden will provide a 5 MW 2.5 GeV proton beam. A fundamental neutron physics beamline is in discussion at the ESS and would be of highest interest for many precision studies such as neutron $n\bar{n}$ oscillations and the neutron EDM.

\subparagraph{Muon beam facilities:} PSI in Switzerland has world-leading high intensity $\mu$-beams hosting the MEG II and future Mu3e experiments devoted to $\mu$ lepton flavor violating (LFV) decays, and other efforts including a possible search for the $\mu$-EDM in a small ring. FNAL and J-PARC also use their high intensity proton beams to produce high intensity low-energy $\mu$-beams for precision physics. They both host competing experiments, in preparation, to search for LFV $\mu$-decays into electrons (Mu2e@FNAL and COMET@J-PARC) and to (re-)measure $(g-2)_\mu$ with a precision higher by a factor 4 compared to the current determination. 

\subparagraph{Electron beam facilities:} The main laboratory with a long term high intensity electron beam program for particle physics is JLab. Based on a superconducting linac, the beam was recently upgraded up to 12 GeV (JLab12) and can provide highly polarized e-pulses with unprecedented luminosity. The JLab experimental hall complex is currently being extended and a wealth of experiments are operating or in the  project stage.
The main physics goals focus on the proton structure, tomography and spin using e.g. DVCS, DIS and SIDIS processes with polarized targets, and on exotic hadronic states produced by photo-production. The high intensity beams also allow for searches of dark photons within the ongoing HPS and future APEX experiments.

The former high-energy frontier laboratories SLAC and DESY have now turned mainly into photon science facilities with the LCLS-II and XFEL free electron lasers, respectively. Their unique high-intensity e-beams, however, remain of interest for particle physics: at SLAC extraction of a fraction of the LCLS-II CW 4 GeV beam (8 GeV in the future) to a facility primarily devoted to hidden sector searches with, for example, the LDMX experiment, is under discussion. At DESY, the XFEL very high intensity 17.5 GeV pulsed e-beam is not yet available for particle physics, but upgrades under discussion involving a CW option and new experimental halls may present new opportunities.

Europe also hosts lower energy high-intensity e-beam facilities, such as MAMI/MESA in Mainz and DAFNE/BTF in Frascati, which conduct diverse precision physics programs.  

\subparagraph{Low energy colliders:} The coming years will be marked by the start of the low energy, high luminosity, $e^+e^-$ collider SuperKEKb in Japan. Within the next decade, BELLE II should collect 50 times more data than the previous BELLE and BABAR experiments, opening new avenues to precision measurements in the flavour sector as well as to searches for low mass hidden particles. 
At a lower energy the $\tau$ and charm factory BEPC/BES~III in China is undergoing upgrades for the next decade. Low energy hadron colliders are also in preparation for the future: NICA in construction in DUBNA will provide high intensity ion collisions for spin and dense matter studies. The electron-ion collider (EIC) under discussion in the US would allow high-precision studies of hadron structure in a regime dominated by gluons and sea-quarks.

\subparagraph{Non accelerator experiments:} The wealth of non-accelerator experiments worldwide performing precision measurements or searches cannot be covered extensively. A domain strongly connected to accelerator experiments is the search for new weakly interacting particles.
For WIMPs a new generation of large recoil detectors based either on liquid Xenon (LZ, Xenon-nt, DARWIN) or depleted liquid argon (Darkside-20t) is under preparation for the next decade, with the goal to reach the background level of the solar "neutrino floor". 
R\&D is ongoing to further improve using directional detection methods, and to cover the few GeV mass domain with lower density detectors (CRESST-III, SuperCDMS). For dark matter axions, haloscopes based on resonance with microwave cavities are being exploited (e.g. ADMX) and are expected to explore part of the QCD predicted parameter band in the coming decade. New experimental setups (e.g. MADMAX) are being implemented to reach higher masses. Finally, it is also worth mentioning the wealth of neutron EDM projects being pursued in Europe and elsewhere, with many improvements planned in the short- and mid-term future.

\newpage
\section{PBC proposed projects}

\vspace{0.5cm}
This section summarizes the main intrinsic features of the proposed PBC projects introduced in section 2.3.2. The physics motivation, beam requirements and detector implementation issues are presented taking into account input from the proponents and the current status of the PBC working group investigations. The presentation focuses on the main issues, with more details being available in the ancillary project and PBC working group documents 
\cite{
PBC:QCD,
PBC:BSM,
PBC:BDF,
PBC:CONV,
PBC:LHCFT,
PBC:LIU,
PBC:EDM,
PBC:TECHNO,
PBC:eSPS,
PBC:AWAKE,
PBC:NUSTORM,
PBC:GAMMA,
}.
Projects initiated from past or present collaborations are given their initial experiment name to which a generic "++" tag is added. All projects are ordered along their implementation within the CERN complex (section 2.2). Summary tables give a comprehensive overview of their main characteristics and current status.

\subsection{Upgrades of existing PS/SPS beam lines}
\vspace{0.5cm}

\subsubsection{EHN1:}
\vspace{0.5cm}

\paragraph{NA61++}
\vspace{0.5cm}
The post-LS2 NA61 plans were submitted to the SPSC as proposal addenda \cite{NA61++,NA61++b}. 
\subparagraph{Physics motivation:} 
The main fundamental motivation of NA61 after LS2 is to further investigate the onset of deconfinement with open charm in Pb-Pb collisions. This hard probe was not accessible to the first generation of heavy ion experiments at the SPS in the 90s. Its measurement could be done in the full phase space thanks to Fixed Target kinematics, and would complement the available charmonium results to constrain dynamical models of deconfinement. In addition NA61 plans to further provide hadron production measurements to better control the neutrino fluxes of the future DUNE, T2K-II and Hyper-Kamiokande programs, and to measure nuclear cross-sections for cosmic ray experiments. 
\subparagraph{Beam requirements:}  

At present, the maximum intensity of primary ions that can be transported into EHN1 is limited to around 10\textsuperscript{5} heavy ions per $\approx$17 s spill. Higher intensities should be possible \cite{PBC:CONV} but will require additional studies with particular attention to shielding requirements. Fragmented ions beams have already been delivered to NA61 and no special changes are required in the beam-line.

For the hadron production measurements at momenta below 10 GeV/c, a low energy, tertiary beam line would be required. A possible location has been identified, and a preliminary layout and optics design produced \cite{PBC:CONV}. A potential target and down-stream options have been explored, and the H2 beam line and tertiary branch as been implemented in GEANT-4.
More studies are necessary but future implementation certainly looks possible.

\subparagraph{Detector implementation:}
The NA61 detector is a Fixed Target spectrometer based on four large acceptance TPC's and superconducting coils. The detector has an old design but benefited from regular improvements and consolidations in the past years. The main upgrades needed for open charm measurements are the addition of a vertex detector to tag charmed hadron decays, and an increase of the data acquisition rate by one order of magnitude to accumulate sufficient statistics. A prototype vertex detector based on CMOS MIMOSA chips was successfully tested in the NA61 heavy ion collision environment. The final large acceptance vertex detector is intended to use the ALPIDE chip developed for ALICE. The DAQ upgrade will be implemented using the present ALICE TPC electronic boards to be decommissioned for the ALICE upgrade. Altogether including improvements for neutrino and cosmic measurements, the NA61++ upgrade costs can be kept at a level of $\approx$2 MCHF thanks to collaboration with ALICE. The NA61 collaboration gathers $\approx$150 physicists from 14 countries with a strong contribution of Eastern Europe but no formal CERN group participation.

\paragraph{NA64++(e)}
\vspace{0.5cm}
  NA64 is a recent experiment installed on the H4 beam line. It exploits the concept of missing energy to look for radiation of invisible dark photons by electrons. After LS2 NA64 plans to continue its searches with higher cumulative beam intensities \cite{NA64E++}.
\subparagraph{Physics motivation:} 
The missing energy method provides a unique opportunity to search for weakly coupled particles in the MeV-GeV range that decay dominantly into invisible particles. 
The latter are possibly candidates for dark matter.
NA64 has collected \num{3e11} Electrons on Target (EoT) before LS2, and wishes to reach  \num{5e12} EoT afterwards.
With \num{5e12} EoT NA64++ could probe a significant range of parameter space in the next 5 years. NA64 can also configure its detector to look for visible decays which probe a complementary parameter space. 

\subparagraph{Beam requirements:}  

The target of a total of \num{5e12} EoT will require around \num{2e18} protons on target. No modifications should be required in the H4 beam line. Preparation of a user zone to allow the quasi-permanent installation of NA64 on the H4 beam line began in 2018.

\subparagraph{Detector implementation:}
The NA64 expected signal consists in an electron with energy below the incident beam energy in an otherwise empty event. The key issues to avoid fake signals are the beam purity and the detector hermeticity. In NA64, the beam purity is drastically enhanced by a synchrotron radiation monitor which individually tags electrons at the nominal 100 GeV beam energy and rejects all other contributions. The detector hermeticity is favored by the 100 GeV Lorentz boost and ensured by a set of hadronic calorimeters downstream the electromagnetic calorimeter. The initial detector implementation has been done at low cost thanks to re-use of older detector components.

Data taking is currently not limited by the H4 beam intensity, but by the ability of the experiment to cope with event pileups, since the method requires to tag individual interactions in the detector. Detector improvements and data acquisition upgrades are planned after LS2 and being already tested with higher intensity operation. Taking them into account, 6 months of operation will be needed to accumulate the targeted number of EoT. The implementation of the required detector upgrades will have to be monitored carefully to avoid wasting operation time of the otherwise highly solicited H4 beam line.

\subsubsection{EHN2:}
\vspace{0.5cm}

\paragraph{COMPASS++}
\vspace{0.5cm}
The future plans of COMPASS were presented in \cite{COMPASS++}.

\subparagraph{Physics motivation:}
The short term plans in the post-LS2 Run 3 period are mainly based on muon beams: COMPASS was already approved to run in 2021 with $\mu$-scattering on a transversely polarized deuterium target, in order to improve the precision of the transversity measurement on deuterium. It is also proposed to address the proton radius ($R_p$) puzzle with $\mu$-proton elastic scattering, complementary to existing measurements based on electron elastic scattering or on spectroscopy (Lamb shift in electronic and muonic hydrogen atoms). The existing hadron beams could also allow the first collection of a high statistic $\pi^+p$ DY sample for separating valence and sea quarks in the pion, an exploratory investigation of exotic charmonium spectroscopy with low energy antiproton-proton collisions, and anti-proton production measurements in p-He scattering, of interest for astrophysical programs.

On the longer term COMPASS wishes to further operate as a general purpose QCD facility (COMPASS++ program). The work-horses would be higher intensity, higher purity $K$ and anti-proton beams produced with the RF separation technique. A wealth of measurements are proposed, including Drell-Yan measurements of the $K$ structure and proton asymmetries, a Primakov measurement of the $K$ polarisability and a comprehensive high statistics survey of strange spectroscopy using the powerful Partial Wave Analysis technique developed by COMPASS. 
A process is ongoing within the COMPASS collaboration to prioritize between the various data taking periods, which would altogether require more than 10 full years of operation.   

\subparagraph{Beam requirements:}  

The standard M2 muon beam as used in the years 2016 and 2017 has the required characteristics for the measurements foreseen by COMPASS++ with muons.
For the proton radius measurement, the collaboration requests a standard muon beam with lower intensities of
10\textsuperscript{5} to 10\textsuperscript{7} muons/sec, depending on the rate capabilities of a future trigger system and TPC.

The COMPASS++ program with kaons and antiprotons requires beams with higher purities and intensities than currently available. The only efficient method to increase the COMPASS secondary beam purity at high energy is RF-separation, which is based on the different velocities of particles of same momentum but different masses. A preliminary study of this option was performed by the PBC conventional beams WG \cite{PBC:CONV}. The maximum energy at which particles can be separated is intrinsically limited by the maximal distance between the 2 RF-kickers, and corresponds to $\approx$ 75 GeV and $\approx$ 100 GeV for kaons and antiprotons, respectively. Additional implementation constraints are likely to further reduce the maximum momenta and will limit the momentum ranges to small bands. The available intensities in the desired particles will result from the achievable purities and the maximal total rate of 10\textsuperscript{8} particles per second on flat top dictated by radio-protection in the EHN2 hall. Finalizing the design of RF-separated beams is a major endeavour. The beam implementation would require significant resources and could be done at earliest for post-LS3 operation. 

\subparagraph{Detector implementation:}
The COMPASS detector consists of a 2-stage large acceptance spectrometer equipped with many tracking devices. The main upgrade for the short term $R_p$ measurement would consist in the implementation of a hydrogen TPC target providing a precise measurement of the recoil proton in $\mu$p elastic scattering for determination of the t-transfer. Promising tests have been performed with a prototype TPC, but the determination of the required optimal $Q^2$ range for the measurement and a full simulation of the detector will be necessary to demonstrate the control of systematics. In particular, it is not clear whether the COMPASS data will allow to alleviate the current fit ambiguities which affect the extraction of $R_p$ in ep scattering. For an efficient use of the unique M2 $\mu$-beam, the TPC data taking rate will have to be significantly higher than in the prototype test where only a tiny fraction of the feasible intensity was used. Further upgrades are foreseen for the long-term program, possibly including improved precision tracking and particle identification devices for spectroscopy, an active absorber equipped with a vertex detector for DY, a liquid Helium target for anti-proton production and a recoil silicon detector for the E GPD. 

The overall cost for the full set of detector upgrades is estimated between 10 and 20 M\texteuro. In order to support its program the COMPASS collaboration should be maintained at a level of $\approx$ 250 physicists. On the short-term new groups from the Russian and German communities are expected to join the $R_p$ measurement. On the longer term a larger re-configuration of the collaboration is needed to compensate for the loss of historical groups currently retiring or moving to other QCD facilities. Ongoing investigations show good prospects from e.g. groups of the US hadron physics community.

\paragraph{MUonE}
\vspace{0.5cm}
MUonE \cite{MUONE} is a dedicated experiment designed to address the contribution of the hadronic vacuum polarization (HVP) to the muon g-2 with a new experimental method.
\subparagraph{Physics motivation:}
A significant improvement on the (g-2)$_\mu$ theory uncertainty is required by the new projects being set up at FNAL and J-PARC, expected to reduce the experimental uncertainty by a factor 4. The theory uncertainty is currently dominated by the uncertainty on the contribution of the hadronic vacuum polarization. MUonE proposes a novel method to determine this contribution, based on the measurement of the t-distribution in elastic $\mu$-e scattering. The unique high energy muon beam available at CERN allows to access the full t range of interest for the measurement. The HVP is expected to change the elastic cross section at high-t by $\mathcal{O}(10^{-3}$): it could be extracted with the required precision of a sub\% by normalizing the high-t to the low-t region, provided relative systematic effects can be kept below $\approx$ $10^{-5}$ on the whole t-range. A collaboration with the worldwide theory expert community has been set up to compute the elastic cross section at the required order in QED perturbation theory, and regular theory workshops monitor the progress.
\subparagraph{Beam requirements:}  

The MUonE experiment requires a parallel 150 GeV/c muon beam with maximum intensity, i.e. 5 x 10\textsuperscript{7} muons/sec. 
The beam must have a small divergence in both planes to match the small detector dimensions. 
Low hadron contamination of the muon beam and event by event control of its energy better than 1\% is required for an efficient use of the kinematical constraints in elastic scattering.

In 2018 MUonE ran parasitically with a test beam set-up downstream of the COMPASS experiment to check the beam conditions and provide input to the trigger strategy. The beam optics and purity are adequate for MUonE. Use of the Beam Momentum Station would allow to control the event-by-event beam energy fluctuations at the required sub-\% level. 

\subparagraph{Detector implementation:}
The MUonE detector consists in a long segmented target made of thin Be planes equipped with high precision Silicon trackers, followed by an EM calorimeter and a muon detector in the region of possible overlap of the scattered e and $\mu$. A minimum of two years of data taking are foreseen to accumulate the required high statistics event sample. The main challenge is to maintain the relative systematic uncertainties below the $10^{-5}$ level.  Beam tests of prototype targets on CERN muon beams have shown that non-Gaussian multiple scattering is manageable. More work is ongoing on the required uniformity of tracking efficiency, particle identification and beam effects. All systematics should benefit from the very high statistics event sample to be kept under control. So far no show stopper has been identified. It is however clear that a pilot run with a reduced set up will be desirable to validate the overall method before a full scale project can be conducted, in which case the pilot run should be led in conditions as close as possible to the expected final set-up.     

The total cost of the detector is estimated to less than $\approx$ 10 M\texteuro. A dozen institutes, mostly from Italy, are currently connected to the project. There is an urgent need to strengthen a core of several full time experimentalists to finalize and validate the design of the detector in close contact with CERN. On the longer term the collaboration should be able to gather the $\approx$ 50-100 physicists required to build and operate the detector. The detector could be built within a few years and be ready for operation within the run 3 period, in time for the improved (g-2)$_\mu$ determinations expected from the new (g-2)$_\mu$ projects.

\paragraph{NA64++($\mu$,h)}
\vspace{0.5cm}
After LS2 NA64 wishes to extend its method for the search of dark photons with missing energy to high energy muon and hadron beams available in EHN2 \cite{NA64MU++}.
\subparagraph{Physics motivation:} 

Most searches for new physics are currently performed with initial particles involving only the first generation of the Standard Model. However, many models of new physics feature flavor dependent interactions. It is possible that new particles interact dominantly with second or third generation particles, significantly weakening the sensitivity of most experiments. It is therefore prudent to also perform searches based on second and third generation particles. In NA64++($\mu$) this is done using the high energy muon beam available at CERN and the same missing energy technique as employed in the electron mode. An important target is a dominantly muon-coupled vector with MeV mass that could explain the anomaly in the $(g-2)_{\mu}$ measurement. The sensitivity of the proposed NA64++($\mu$) Phase I should allow to exclude such an interpretation of the anomaly. A later Phase II program at higher intensity is foreseen to improve limits on dark vector bosons coupled to muons and millicharged particles.

\subparagraph{Beam requirements:}

Two experimental phases are foreseen. 
Phase I requires of the order $10^{11}$ Muons on Target ($\mu$oTs) at an energy of 100-160 GeV, which could be collected in a few months at moderate intensity.
Phase II requires a much higher integrated intensity of \num{5e13} $\mu$oTs to be competitive. This would represent at least 12 months of data taking at the highest muon beam intensity of $10^7$ muons/s.
Both phases require small beam divergence in both planes. Beam tests conducted in the COMPASS hall have confirmed that the $\mu$-beam purity is adequate for NA64++($\mu$).  

\subparagraph{Detector implementation:}
The detector set-up proposed for searches with a muon beam follows the same philosophy as with electron beams: an active muon target (electromagnetic calorimeter) is followed by a large hadron calorimeter located in a magnetic field, which is used both to measure the outgoing muon momentum and to tag associated hadrons. The signal consists in an outgoing muon with momentum significantly lower than the incident beam energy, within an otherwise empty event. As for electrons key issues are the incident muon beam purity and momentum control, and the detector hermeticity. The Phase I detector could be built within a budget of $\approx1$ MCHF and would fit in the EHN2 hall upstream of COMPASS. Phase I data taking is scheduled after LS2 to match the expected new results of the current (g-2)$_\mu$ experiments. Phase II is not foreseen to start before LS3, with detector upgrades to be designed according to the lessons learned from the Phase I run.  

The NA64 plans with hadron beams are less advanced. Integrated luminosities of $10^{13}$ $\pi$oTs, $10^{12}$ $K$oTs and $10^{11}$ poTs are anticipated to investigate complementary dark photon models. These measurements would require mid-energy (20-50 GeV/c) hadron beams, which could also be supplied by the H4 beam line without modification.

\subsubsection{ECN3:}
\vspace{0.5cm}

\paragraph{NA62++}
\vspace{0.5cm}

NA62 is currently the CERN flagship Fixed Target search experiment, primarily designed to measure the ultra-rare  $K^{+}\to\pi^{+}\nu\bar{\nu}$ decay with a 10\% precision, comparable to the present theoretical uncertainty of the Standard Model prediction.

\subparagraph{Physics motivation:}
The measurement of the $K^{+}\to\pi^{+}\nu\bar{\nu}$ decay channel with 10\% precision will provide a precise test of the Standard Model with sensitivity to new physics. Searching for features in the missing energy spectrum NA62 also has sensitivity to other invisible kaon decays, e.g. into light dark sector particles.
Beyond the currently approved program, NA62 plans to have a dedicated beam dump run that would allow to directly search for a variety of weakly coupled particles in the MeV-GeV range that feature visible decays. With $10^{18}$ PoTs, NA62++ has significant sensitivity to new particles such as axion-like particles and heavy neutral leptons. 

\subparagraph{Beam requirements:}  

An in-depth study of the NA62++ beam dump option has been performed~\cite{PBC:CONV}.
Summarizing briefly:

\begin{itemize}
\item switching to dump mode would consist in removing the T10 target and closing the TAX collimator located $\approx$ 80m upstream of the NA62 decay volume, with appropriate optics modifications; 
\item simulations have been performed and benchmarked to address the issue of the detector background due to pions and kaons decay to muons;
\item optimisation studies with an unmodified layout have been performed: this includes proton beam transport to a more downstream location in the K12 beam line; scanning of the achromat magnet around the TAX;  scanning of the muon sweeping magnets; scanning of the scraping magnets and possible replacement using magnetized iron blocks;
\item the baseline implementation is with an unchanged layout but the possibility of an optimised layout has also been considered.
\end{itemize}

Current studies are promising and further optimization of the beamdump configuration is ongoing~\cite{PBC:CONV}. A total muon rate reduction for momenta p$>$20 GeV/c of about one order of magnitude is achievable by a different powering of the magnets, and a large fraction of the remaining muons at this momenta hit the muon veto detector at its outer acceptance. The introduction of magnetized iron blocks could further improve the situation at high momenta for one particular charge.

\subparagraph{Detector implementation:}
The NA62 detector exploits a novel high energy in-flight decay technique, combining a high rate/low material tracking with an hermetic calorimetric coverage for full particle identification of $K$ decays in a high intensity environment. A 60m-long vacuum decay vessel is preceded by an ultra fast precision tracker for the initial individual $K$ direction measurement, and followed by a spectrometer equipped with a RICH for measurement and identification of the decay products. The set-up is complemented by a hermetic electromagnetic calorimetry coverage for $\pi^0$ rejection and $\gamma$ measurement, and by a forward hadronic calorimeter and muon system.   

The detector has been operated close to its nominal rate (> 60\%) since 2017. First $\pi^+\nu\bar{\nu}$ candidates have been observed, with a sample of $\approx20$ events expected until LS2. It is anticipated that 2 more years of running will be needed after LS2 to reach the targeted precision on the $\pi^+\nu\bar{\nu}$ channel. Specific triggers will in parallel allow to investigate other $K$ rare decays. In this context NA62 is considering the option to devote up to 1 year ($10^{18}$ PoT) of the run 3 period to specific operation in beam dump mode. The main signature for production and decay of hidden particles consists in 2 tracks or 2 photons pointing to the same decay vessel region in an otherwise empty event. The present detector is already well suited for such a search: possible minor upgrades under consideration include a charged particle veto hodoscope to be placed upstream of the decay vessel. The main experimental issue for such a dedicated run is background rejection. Dedicated test runs in beam dump mode have confirmed the 0-background assumption for an integrated luminosity of $\approx$ \num{3e16} PoTs. 

NA62 is an efficient well established collaboration and there is every confidence that the program can be performed. The collaboration will have to prioritize its goals as function of the results of the ongoing approved program and the future evolution of the anomalies currently reported in the heavy flavour sector.

\paragraph{KLEVER}
\vspace{0.5cm}

The KLEVER experiment \cite{KLEVER} aims at a measurement of the ultra-rare neutral kaon decay, $K^{0}_{L}\to \pi^{0}\nu\bar{\nu}$ within a high intensity $K^0$ beam. 

\subparagraph{Physics motivation:} 
The $K^{0}_{L}\to \pi^{0}\nu\bar{\nu}$ neutral mode and the corresponding charged mode studied by NA62 are complementary in several respects. The combination of both modes is sufficient to determine the CKM unitarity triangle independent of B decays. The two modes have also different sensitivity to BSM models. In particular if NA62 found a deviation from the Standard Model prediction for the charged kaon decays, the neutral decay mode can provide crucial information to distinguish between different models and in particular to get information on the nature of the underlying flavour violation.

\subparagraph{Beam requirements:}  

For NA62 the nominal beam flux on T10 is 3x$10^{12}$ ppp and the nominal beam intensity of the 75 GeV/c secondary beam is 2.2x$10^9$ ppp. 
In an underground area like ECN3, with hermetic zone perimeter excluding any access with beam on, the secondary beam intensity for NA62 does not pose particular problems.

The required proton flux for KLEVER is at least 2x$10^{13}$ protons per 4.8s spill, around a factor 7 higher than the nominal NA62 beam. 
The secondary beam will transport all long-lived or stable neutral particles including $K^0_L$ and $K^0_S$, $\Lambda^0$'s, neutrons and $\gamma$'s. 
A first study \cite{PBC:CONV} addressed the optimum production angle based on benchmarked simulations. The baseline target material is beryllium but alternative materials could be investigated. 
The study considered the prompt muon flux outside, above, and behind the ECN3 cavern, and addressed the design and optimisation of the neutral beam, in particular background reduction, collimation, and muon sweeping. The ability of ECN3 to take the higher beam intensities requested by KLEVER is certainly an issue and has also received concerted attention.

\subparagraph{Detector implementation:}
The KLEVER project extrapolates the NA62 high energy in-flight decay technique to the neutral $K^0$ decay into $\pi^0\nu\bar{\nu}$. The signature consists of 2 photons at the $\pi^0$ mass associated to missing $P_T$, which is much more challenging than for the NA62 charged mode since only neutral particles are involved. Because of these different and somewhat stronger experimental constraints KLEVER would have to build most of its components from scratch. The detector consists mainly of a large 160m-long decay vessel hermetically instrumented with electromagnetic calorimetry. Signal events are measured with a new Shashlik calorimeter replacing the NA62 LKr calorimeter with better timing. The main experimental issue is the rejection of the copious physics background dominated by $K$ decays into $\pi^0\pi^0$ with one $\pi^0$ escaping detection. For this purpose the Shashlik calorimeter is complemented by new large angle lead-scintillator vetos distributed along the decay vessel, and by a very forward EM calorimeter located within the neutral beam. The latter is a challenging device requiring a high photon detection efficiency and a maximal transparency to the $\approx$ 500 MHz neutral hadron rate in the beam. A new technology based on crystalline tungsten absorbers is under test together with the AXIAL collaboration. A high statistics fast simulation of the setup indicates that the signal/noise ratio can be kept $\approx1$ and that 60 signal events could be accumulated in 5 years of operation (\num{5e19} PoTs). Detailed simulations are necessary to consolidate these estimations and are being implemented.   

The detector has an estimated total cost of $\approx30$ M\texteuro, and could be built at earliest for a post-LS3 operation. The KLEVER program is supported by about half of the current NA62 collaboration members. Initiatives are ongoing to reach the required level of contributors to conduct such a project. An Expression of Interest is being prepared for the SPSC Committee.

\paragraph{NA60++}
\vspace{0.5cm}

The NA60++ project aims to revive the former NA60 concept to characterize the QCD phase diagram with low energy dileptons.
Because of its high intensity beam the experiment requires an underground experimental hall for which the only existing option is the ECN3 cavern.

\subparagraph{Physics motivation:}
NA60++ proposes dilepton and hadronic precision measurements in the finite $\mu_B$ region of the QCD phase diagram where the existence of a first order phase transition with co-existence of a mixed-phase and a critical end-point were suggested but not confirmed experimentally.
Evidence for a first order phase transition will be searched by the first measurement of a caloric curve. The temperature will be measured from the thermal dimuon mass spectra, and a scan of the baryochemical potential $\mu_B$ performed varying the collision energy.
Additionally, chiral symmetry restoration, expected at the phase boundary, can be quantitatively investigated by a precision measurement of the dimuon thermal yield expected from the $\rho$-$a_1$ chiral mixing. Charmonium and open charm measurements are also foreseen.

\subparagraph{Beam requirements:}  

NA60++ will use primary ion beams over a range of energies, with minimum beam
intensities of $10^7$ ions per second, i.e. \num{5e7} per 5 second spill. In addition, they request some
periods of a few weeks each with proton beams at \num{5e8} protons per second.
Because of the lower intensities, massive front-end shielding can be avoided and the intensity requirement of NA60++ should not pose major problems.
The beam design for the beam to NA60++ could be a straight copy of the former H10 beam used by NA60.

\subparagraph{Detector implementation:}
The NA60++ detector, similar to NA60, consists in a dimuon spectrometer with a vertex detector for high precision tracking. The main new features compared to NA60 are a data taking rate higher by 2 orders of magnitude, and the acceptance variation associated to the energy scan. The experiment will also benefit from the latest technologies as developed for the LHC upgrades. It is estimated that the detector would require $\approx3$ years of R\&D and $\approx2$ years of construction, for a total cost between 15 and 25 M\texteuro. A group of physicists has expressed interest to participate in the simulations and detector design. In the longer term the project would require consolidation of a collaboration at the level of $\approx$ 100 physicists. Data taking could start at earliest after LS3.      

\paragraph{DIRAC++}
\vspace{0.5cm}

The DIRAC++ project wishes to perform the study of mesonic atoms at the SPS to improve on precision compared to what was done by DIRAC at the PS. Because of its high intensity beam the experiment requires an underground experimental hall for which the only existing option is the ECN3 cavern.

\subparagraph{Physics motivation:} 
The SPS energy is expected to increase the production rate of $\pi$$K$ and $\pi$$\pi$ mesonic atoms by a factor of 20-50 compared to the PS. This would allow significant improvements of the measurement of the QCD SU(3) chiral symmetry breaking sector: in particular the error on the $\pi$$K$ scattering length is expected to decrease from 34\% (DIRAC determination) to 5\% (present theory error). The $\pi$$\pi$ scattering length measurement is also expected to improve on the current best determination from NA48.    
\subparagraph{Beam requirements:}  

DIRAC++ is requesting a proton beam intensity of \num{.55e12} protons per spill.
These beam intensities would induce activation of equipment and of the air in the ECN3 cavern. This would have
impact on the cool-down time before access and impose ventilation in the ECN3 cavern.
However, the DIRAC++ target is almost transparent and no loss points along the beam are
foreseen. The beam can be dumped in a properly designed re-entrant dump, as was the case
for the DIRAC experiment in the East Area. Details remain to be studied.

\subparagraph{Detector implementation:}
The foreseen DIRAC++ detector is similar to DIRAC. It consists of a low energy spectrometer with efficient particle identification using RICH detectors. The total cost is roughly estimated to 3 MCHF re-using some of the DIRAC components such as the magnet. It is hoped that the required collaboration of a few ten physicists will build up once the project gets momentum, as was the case for DIRAC.   

\subsubsection{PS}
\vspace{0.5cm}

\paragraph{REDTOP}
\vspace{0.5cm}
The REDTOP project plans to study ultra-rare $\eta$ decays. It was initially introduced within the FNAL context \cite{REDTOPFNAL} but recently expressed potential interest to be hosted at CERN. A survey of the existing CERN facilities \cite{CB-REDTOP} identified the PS East hall as the most suited area to implement it. 
\subparagraph{Physics motivation:} 
REDTOP aims at studying rare $\eta$ and $\eta'$ decays with a sample of $10^{13}$ $\eta$ and $\eta'$. The $\eta$ and $\eta'$ particles are unique in that they do not carry flavor quantum numbers and therefore allow to study symmetry violations as well as new BSM particles without requiring flavour violating interactions. An example is the variation of the usual $\eta\to\gamma\gamma$ decay where one of the photons is replaced by an up to GeV mass dark photon leading to $\eta\to\gamma X\to\gamma\ell\ell$. Moreover, REDTOP could explore a sizeable number of C, CP and even CPT violating decays.

\subparagraph{Beam requirements:}  

The REDTOP program requires a proton flux of $10^{11}$p/s with high duty factor (> 75\%) at a particle kinetic energy of 2-3 GeV.
The production rate and background rejection requirements lead to an optimum incident proton kinetic energy of 1.9 GeV (even 3 GeV for the heavier $\eta'$). 

Both LEIR and the Booster were considered as options and rejected \cite{CB-REDTOP}.
It was also suggested to extract the 2 GeV beam to the PS which could then deliver it to the East Hall. 
In 2018, a 24 GeV/c proton beam is routinely slow-extracted into the CHARM and IRRAD facilities along the T8 beam line with a maximum intensity of \num{6.5e11} protons per extraction over 0.4 seconds. 
REDTOP would require a much longer flat top of e.g. 10 seconds at around 2 GeV kinetic energy (as delivered directly from the PSB post-LS2) in a cycle of 9 basic periods = 10.8 s. 
For the moment no blocking issue is identified, but many machine studies would be required. REDTOP would however remain significantly below $10^{18}$ PoT per year, and the impact on the rest of the CERN physics program would be very significant.
In addition REDTOP would probably have to be installed in place of the CHARM facility. In summary it was concluded by the team assigned to examine the possibilities that there is no easy way to integrate REDTOP efficiently in the existing CERN physics program. 
All options investigated require major machine studies and would have a significant impact on the CERN physics program.

\subparagraph{Detector implementation:}
REDTOP will intercept less than 1\% of the low energy incident proton beam with a small segmented target surrounded by a detector with $4\pi$ coverage. One key issue is the large background from inelastic interactions in the target. Since $\eta$ decays of interest for BSM searches mostly involve charged leptons, the inelastic background can be strongly reduced if the detector is hadron-blind while maintaining good lepton identification. The proposed implementation is based on an optical TPC, in which charged particles are tracked from their Cerenkov radiation, and a dual readout calorimeter based on the ADRIANO concept. Consequent R\&D would be necessary to bring both technologies to the required level: the ADRIANO technology was validated with several prototypes but no large scale detector has yet been constructed. No prototype of the foreseen optical TPC exists. Concerns on the radiation hardness and mitigation were also raised by the FNAL Program Advisory Committee. On the other hand, since beam intensities feasible at the PS are one order of magnitude smaller than the REDTOP nominal intensity, staging of the detector could be considered.  

The nominal detector cost was estimated to $\approx$50 M\$. The project is supported by a sizeable collaboration dominated by US-oriented groups. Significant progress on the detector R\&D and detailed simulations of the expected sensitivities are needed before an operational implementation can be considered.

\subsection{New SPS beam lines}
\vspace{0.5cm}

\subsubsection{Beam Dump Facility}
\vspace{0.5cm}

One potential avenue to explore the so-called Hidden Sector is to search for MeV to GeV hidden particles in fixed target experiments. To this end, the SPS Beam Dump Facility (BDF) has been proposed \cite{PBC:BDF} as a fixed-target facility foreseen to be situated at the North Area of the SPS. Beam dump in this context implies a target which aims to absorb as many of incident protons as possible and contain most of the cascade generated by primary beam interactions. Exploitation of the facility in the first instance is foreseen by the SHiP and TauFV experiments.

\paragraph{BEAM FACILITY}
\vspace{0.5cm}

The proposal is to site the BDF near the existing facilities at the North Area of the SPS. Ongoing studies are based on the assumption that the target and experimental facility will be located upstream of the existing experimental hall EHN1.

Slow extraction of 400 GeV/c protons from the SPS is foreseen using existing slow extraction hardware situated at point 2.
The extracted beam is to be transferred along the existing transfer line to the North Area (TT20) up to a splitter magnet in an existing cavern (TDC2).
This magnet sends the BDF beam into a $\sim$350 m long new beam line up to the target.
This line defines the beam geometry up to the target, and has to be equipped with a system to provide adequate spatial dilution of the beam on the target together with instrumentation and steering elements.  

The target is designed to contain most of the cascade generated by the primary beam interaction. 
It is embedded in a massive iron shielding absorbing the remaining primary protons and produced hadrons emerging from the target. 
The hadron absorber is magnetized and followed by an active muon shield and an experimental hall, expected to house the SHiP detector in the first instance.

As SHiP requires pushing the primary proton beam to a power of around 355 kW, 
radiation protection considerations strongly determine the design of the facility.
In particular high prompt and residual dose rates call for considerable shielding and remote interventions in the target area.
The risk and environmental impact from releases of radioactivity by air and water as well as the soil activation also heavily influence the design.  
Studies include expected prompt and residual dose rates in the various accessible areas of SHiP as well as the levels of stray radiation in the surrounding areas.

Exploratory studies have also been conducted to look into the possibility of installing the TauFV detector in the new beam line.
Here around 2\% of the protons are intercepted on their way to the BDF target to serve an in-line experiment.

The key beam parameters foreseen are shown in table \ref{shipP}.
\begin{table}[hbt]
   \centering
   \caption{Key beam parameters foreseen for BDF/SHiP}
   \begin{tabular}{lcc}
\hline  
Momentum [GeV/c] 	   & 400             \\
Beam Intensity per cycle   & 	\num{4.2e13}             \\
Cycle length [s]    & 	7.2             \\
Spill duration [s]    & 	1             \\
Expected r.m.s. spot size (H/V) [mm]    & 	6/6             \\
Avg. beam power on target [kW]	   & 355             \\
Avg. beam power on target during spill [kW] 	   & 2560             \\
Protons on target (PoT)/year	   &  \num{4e19}            \\
Total PoT in 5 year's data taking	   & \num{2e20}            \\
\hline
   \end{tabular}
   \label{shipP}
\end{table}

BDF is in the study phase and any planning is necessarily tentative.
The present accelerator schedule foresees a year's shutdown for the injector complex in 2025 (LS3).
The key BDF civil engineering package is the demolition of the junction cavern, and its reconstruction along with the 
first section of the new beam line leading to the new target complex.
This work package is planned to take at least 18 months and ideally would start at the end 2024.
This would imply extending the stop of the North Area into 2026.
The other major civil engineering work (target complex, experimental hall) can be executed with North Area beam-on.
Given commencement of work in 2024, first beam to BDF/SHiP could be expected at the start of 2027,
with preliminary beam-line commissioning at the end of 2026.

As part of the BDF study, a comprehensive cost estimate was performed. 
The two most important components of the total cost are the civil engineering, and the target complex.
Both packages employed the services of external specialists who either confirmed or produced cost estimates for major parts of the works.
Detailed breakdown of the other hardware and services was based on CERN expertise -- much of the required hardware is relatively standard CERN deployment.
The final cost estimate is detailed in the BDF Comprehensive Design Study~\cite{PBC:BDF}:
the preliminary estimate for the total material cost is of order 150 MCHF. 
It is important to note that the estimates have been performed at the study phase, and as such the uncertainties remain potentially significant.

\paragraph{SHiP}
\vspace{0.5cm}

The physics case and detector aspects of the SHiP program were documented in detail in, for example, \cite{Alekhin:2015byh} and \cite{2015arXiv150404956S} respectively, and updated in a presentation to the SPSC \cite{SHIPPROTOCDS} anticipating the publication of a Comprehensive Design Report in 2019. SHiP targets being the first dedicated high intensity beam dump experiment in the modern era, aiming at \num{2e20} protons on target with zero background.

\subparagraph{Physics motivation:} 
New physics in the MeV to GeV region is suggested by theoretical as well as phenomenological arguments.
On the theory side this region arises from a loop factor suppression of the electroweak scale. An explicit example for such a mechanism 
at work could be a supersymmetric dark sector that couples to the Standard Model only via loop interactions, and which relies on its supersymmetry breaking only from these interactions with the SM. 
Perhaps more intriguingly the same mass region also seems appealing as a messenger to dark matter. In particular in connection with suggestive astrophysical excesses such as the one observed by the Pamela satellite and/or the 511 keV line indicating excesses in the number of positrons. 
An additional strong indication for this mass region is the possibility to explain the baryon asymmetry of the Universe via leptogenesis 
with sterile neutrinos in this mass range in a minimal model that extends the SM only by right handed neutrinos for each of the SM ones. 

Taking into account the smallness of the relevant interactions, a high intensity experiment is required to obtain an observable number of events. This together with the targeted mass range makes a beam dump experiment with SPS protons ($\sqrt{s}=28\,{\rm GeV}$ for proton proton collisions) ideally suited for this task. 

\subparagraph{Detector implementation:}
The SHiP detector is being designed to  explore  Hidden  Sector  models  with  vector,  scalar,  neutrino, axion-like SM portals  (and  others) in the MeV-GeV range \cite{Alekhin:2015byh}, for which the key channels have a signature based on 2 tracks pointing to a hypothetical decay position. This results in an overall detector layout similar to NA62: a large vacuum decay vessel followed by a high precision tracker spectrometer. There are, however, two notable differences: the transverse dimension of the decay vessel is much larger than in NA62 in order to reach the charm and beauty mass decay acceptance domain, and SHiP will operate downstream of a beam-dump in a much quieter environment than NA62 in its standard operation mode. In addition to the mainstream spectrometer, SHiP is planned to be equipped with a high precision emulsion spectrometer located immediately upstream of the decay vessel. This subdetector will increase the discovery reach by providing sensitivity to re-interactions of long lived particles produced in the dump, and will collect a first high statistics sample of $\tau$-neutrino interactions to test lepton universality. It is made possible by the remarkable progress in emulsion fast automatic processing and analysis performed by the former OPERA groups, now members of SHiP. 

The two spectrometers have complementary physics programs and make SHiP a kind of dual platform. They are based on well established technologies which raise no particular concern. The key experimental issue is background control. The main SM particles expected within SHiP come from interactions of neutrinos produced in the beamdump and from punch-through of charged particles. The impact of the latter is minimized thanks to a sophisticated active $\mu$-shield located upstream of the decay vessel to sweep out muons exiting the dump. Dedicated beam tests have been performed to check the absolute rate and spectrum of muons produced in the dump and to further optimize the $\mu$-shield. Technological issues for its realization are being addressed with prototypes. Current estimates comfort the 0-background assumption of both neutrino and combinatorial contributions for the full life time of SHiP. More studies are ongoing to consolidate the predictions. 

The SHiP project is supported by a large collaboration currently gathering more than 50 Institutes from about 20 countries. The detector total cost is estimated to $\approx60$ MCHF. Construction and installation target a start of operation after LS3. 

\paragraph{TauFV}
\vspace{0.5cm}
\subparagraph{Physics motivation:} 

Violation of lepton flavor in the charged sector would be a clear signal of physics beyond the SM. Experiments are currently foreseen to significantly improve the sensitivity for $\mu\to e\gamma$ and $\mu\to eee$. The current limits on the decay of $\tau$ are significantly weaker. On the other hand theory suggests that any effects in the heavier $\tau$ would be significantly enhanced. Therefore the opportunity to search for $\tau\to \mu\mu\mu$ and other rare tau decays on the $10^{-10}$ level seems very attractive. 
In light of the currently existing anomalies in B decays as well as (small) anomalies in tau decays to neutrinos this seems particularly timely.  

\subparagraph{Detector implementation:}
TauFV plans to intercept $\approx2\%$ of the BDF proton beam upstream of the dump with a segmented W target, to look for ultra-rare $\tau$ decays. The signature for the priority decay channel in 3 muons consists in 3 $\mu$-tracks at the $\tau$ mass pointing to a common vertex. The detector overall layout is in an early design stage and is similar to LHCb: a vertex locator followed by a dipole, high precision trackers and calorimeters. Combinatorial background suppression requires very demanding performances of all detector components, including high granularity and timing better than 100 ps. Another issue is radiation hardness to withstand the harsh BDF environment. The starting R\&D for TauFV can benefit from its strong synergies with the LHCb upgrades for HL-LHC. Very preliminary estimates of backgrounds indicate a sensitivity to the $3\mu$ decay channel at a $10^{-10}$ level.

TauFV is being launched by a small motivated collaboration of several UK and Swiss Institutes. A possible implementation plan would include operation of a pilot project parallel to SHiP after LS3, followed by the full size experiment after LS4. This would pave the way to a long term upgradable facility devoted to ultra-rare lepton and meson decays, as well as precision measurements of e.g. CP violation.

\subsubsection{eSPS: a primary electron beam facility at CERN}
\vspace{0.5cm}

The SPS, which once accelerated electrons up to 22 GeV for injection into LEP, together with the R\&D invested into CLIC, offer CERN an opportunity to create a new high intensity primary electron beam facility.
The eSPS collaboration have recently submitted an expression of interest to the SPSC \cite{PBC:eSPS} with a description of the facility including its physics motivation, layout and initial civil engineering sketches for a possible implementation at the SPS. 
One potential application is a search for hidden particles in the MeV-GeV range with the missing momentum method, as proposed by the LDMX experiment. 
In addition such a facility could serve a beam dump experiment searching for dark sector mediators decaying to Standard Model particles.

\paragraph{BEAM FACILITY}
\vspace{0.5cm}   

The main stages of the facility are as follows.
\begin{itemize}
\item An approximately 70 m long X-band based linac based on CLIC technology in the existing TT4-5 transfer lines of the SPS would accelerate electrons to 3.5 GeV. 
TT4 is connected to the SPS via an existing tunnel TT61 which leads to the SPS Long Straight Section 6, LSS6. 
\item The SPS would be filled in 1 to 2 s with bunches 5 ns apart via TT60. 
The proposed SPS injection scheme features a fast bunch to bucket injection of the 200 ns trains with a 100 Hz repetition rate directly into the SPS 200 MHz RF-system.
\item Acceleration to 16 GeV in the SPS. Acceleration of electrons to 16 GeV requires a total RF voltage around 10 MV.
Existing 200 MHz 1 MV standing-wave cavities from the LEP era can be re-installed in an available location of the SPS lattice. 
Twelve of these cavities are in storage and in good condition. 
\item Slow extraction down the TT10 transfer line in 10 s using a resonant scheme similar to the one already used for
extracting proton beams from the SPS. The proposed scheme requires installation of a new electrostatic septum, four small slow bumper magnets
and two new thin magnetic septa, all in available spaces.
\item The beam is delivered to an experimental hall by bringing beam back on Meyrin site using the existing TT10 line.
This beam-line is capable of transporting the electron beam without modification. 
A new, short beam-line would branch from TT10 to the site that has been identified on the Meyrin site for the location of the experimental hall. 
\end{itemize}

The material costs of the facility are estimated to $\approx80$ MCHF \cite{PBC:eSPS}. Construction could be executed within the 5 years following a positive decision. 
A strong case is made for the possible use of the linac for accelerator R\&D and other studies.

\paragraph{LDMX-like EXPERIMENT}
\vspace{0.5cm}
\subparagraph{Physics motivation:} 
An eSPS beam would be of particular interest to investigate the hidden sector in the invisible decay channels as presently done by NA64.
The lower energy compared to the NA64 beam allows better constraints on the kinematics and better particle identification. 
Yearly integrated luminosities of up to $\approx 10^{16}$ EoTs, i.e. 4 orders of magnitude higher than collected by NA64, are anticipated if 1/3 of the SPS operation time is devoted to running with electrons.

\subparagraph{Detector implementation:}
The detector design is currently under study by the LDMX collaboration \cite{LDMXSLAC} for a possible implementation at SLAC or CERN. 
The LDMX concept is based on state-of-the art high precision tracking from the HPS experiment at JLab, and on fast high granularity calorimetry as developed for HL-LHC. 
The eSPS beam flexibility will allow to tune the bunch structure and intensity as function of the capabilities of the detector to disentangle events from each other, which is required by the missing momentum method.
The LDMX  detector is still in its R\&D and design phase and is supported by a collaboration of 7 US Institutes, together with CERN physicists for the eSPS option. The detector cost is expected to fit within a $\approx10$ MCHF envelope.  

\subsubsection{AWAKE++}
\vspace{0.5cm}
Following the successful demonstration of the AWAKE principle \cite{AWAKE}, the collaboration is considering to extend the set-up after LS2 towards a high intensity,
medium energy pulsed electron beam facility usable for investigation of the hidden sector \cite{PBC:AWAKE}.

\paragraph{BEAM FACILITY}
\vspace{0.5cm}

The first phase of AWAKE consisted of a 10 m long plasma cell taking 400 GeV proton bunches from the SPS. 
On-site are laser and electron beam production systems and a sophisticated array of instrumentation.
A proton bunch from the SPS co-propagates with a laser pulse, which creates a plasma in a column of rubidium vapour and seeds the modulation of the proton bunch into microbunches. 
The 400 GeV/c protons have a bunch length of 6 to 8 ~cm and the bunch is focused to a transverse size of approximately $200\,\mu{\rm m}$ 
at the entrance of the vapour source, with the proton bunch population varying shot-to-shot in the range $Np\approx(2.5-3.1)\times 10^{11}$ protons per bunch.

The collaboration is well supported and good progress has been made over the last years.  Notable milestones include demonstration of the self modulation of the proton bunches into microbunches, and more recently the successful acceleration of electrons in the wakefield of the proton microbunches \cite{AWAKE}.  
Demonstration of acceleration of electrons was the goal of phase 1 of the experiment and technical design of phase 2 is now ongoing. 
The goal after LS2 is to accelerate an electron beam to 5 -- 10 GeV in a 10 -- 20 m plasma cell to demonstrate the scalability of the AWAKE concept.

Anticipating the success of phase 2 the collaboration is in the process of developing potential applications of its concept \cite{PBC:AWAKE}. 
The most well-developed option is an electron beam dump experiment served by using 16 to 320 SPS proton bunches per 40 seconds with \num{3.e11} ppp, to produce an electron beam of $\mathcal{O}$(50 GeV).

\paragraph{DARK SECTOR EXPERIMENT}
\vspace{0.5cm}
\subparagraph{Physics motivation:} 
The proposed AWAKE electron beam would be of interest to investigate the hidden sector as presently done by NA64 at the North Area. The missing energy method, which requires tagging of individual events, cannot be used with the highly pulsed structure of an AWAKE beam. Searches should be performed in the appearance mode e.g. by looking for dark photon decays into electron or muon pairs. The integrated luminosity is anticipated to reach more than $10^{15}$ EoT, 3 orders of magnitude larger than with the current NA64 electron beam.
\subparagraph{Detector implementation:}
Adequate space for a detector and its services is available downstream of AWAKE in the former CNGS target hall and decay tunnel. These areas are still activated and would require $\approx1$ year to be cleaned out.  The detector set-up for a dark photon search in the electron decay channel could be similar to NA64 in its appearance configuration. Such a project is pending on the successful implementation of an operational AWAKE beam and could be installed at the earliest during LS3.

\subsection{LHC hosted projects}
\vspace{0.5cm}

Two domains connected to the LHC are of interest for the PBC studies. On one side, there is growing momentum within some LHC collaborations, notably LHCb, to implement internal targets allowing Fixed Target physics with the LHC beams. 
The potential of such programs has been fully investigated within the PBC mandate. In addition several collaborations are currently proposing to implement new detectors located far from the LHC collision points, which will search for new Long Lived Particles (LLP) in a parameter space which overlaps with that of the PBC projects. These experiments do physics with LHC colliding beams and therefore do not fit within the PBC mandate, but it was felt important to include them in the sensitivity studies of chapter 4 to compare their reach to that of the PBC proposals.     

\subsubsection{LHC Fixed Target}
\vspace{0.5cm}

Three types of fixed target implementations are being considered by the LHC FT working group \cite{PBC:LHCFT}: 

{\bf Crystal Channelling for Beam Splitting}:
A bent crystal with bending angle of $\approx$ 0.15 mrad, similar to the ones presently used for hadron-beam collimation at the LHC, separates from the circulating beam a fraction of the halo that then impinges on a target, safely retracted from the beam core envelope. A specific implementation of this scheme was proposed \cite{CRYSTAL-IDEA} for an experiment at the LHCb interaction point, to study the magnetic moments of short-lived heavy particles. In this scheme, called the ``double crystal" setup, the target is followed by a second crystal with a much larger bending angle of $\approx$ 15 mrad. Promising results on 6.5 TeV beam halo deflection were obtained at the LHC and the double crystal scheme had a first successful test at the SPS within the UA9 experiment.

{\bf Solid Targets}:
The crystals described above can also be used in a single-crystal setup where the deflected halo interacts
with a solid target internal to the beam pipe close to an existing detector. More conventional solid target setups where
the beam halo directly interacts with the in-vacuum target are also being envisioned.

{\bf Gaseous Targets}:
A Fixed Target physics programme has been initiated in LHCb by exploiting the forward
geometry of the detector and the availability of SMOG, a gas injection system originally implemented for measuring the luminosity by beam-gas imaging of the colliding beams. 
This has triggered new proposals for internal gas targets at the LHC, with or without polarization.
One feature of these proposals is the use of a storage cell, a cylindrical open-ended tube located around
the beam, which allows one to enhance the target thickness as compared to an equivalent flux of gas injected directly in the beam vacuum. The current understanding of the beam conditions \cite{PBC:LHCFT} indicates that instantaneous luminosities of a few times $10^{32}~{\rm cm^{-2}s^{-1}}$ (depending on the gas type) would be compatible with LHC operation.

Although not an exhaustive list of what could be done with the LHC beams, these three options are currently the most actively studied
by several groups, in particular the double-crystal setup and the gaseous target proposal. 
Possible implementations at different LHC locations are being elaborated by the proponents, working group’s members
and the relevant machine experts.

\subparagraph{Physics motivation:}
Fixed Target collisions at the LHC energy open a new kinematic domain, intermediate between SPS Fixed Target and RHIC collisions, to both high precision hadron and heavy ion measurements \cite{AFTER,LHCBFT}. Unpolarized targets can provide a copious yield of DY events and heavy flavor particles, allowing breakthroughs in the determination of the high-x gluon, quark and heavy quark content of the proton, which is important for the full exploitation of the high-luminosity phase of the LHC. Hadron yield measurements of interest for astrophysics would also be feasible. Polarized targets would provide a comprehensive set of Transverse Spin Asymmetry measurements including determination of quark- and gluon-induced TMD effects. Collision of ion beams on various target ions would extend the exploration of the QCD phase transition diagram with new signatures including e.g. the sequential suppression of $J/$$\psi$, $\psi$' and $\chi_c$ meson production.      

In addition the double crystal configuration provides the opportunity to measure the magnetic and electric dipole moments of short lived heavy baryons and of the $\tau$-lepton. 
Measurements of the magnetic dipole moments of heavy hadrons can provide valuable benchmarks for non-perturbative QCD calculations in this region. 
Electric dipole moments would be a sign of CP violation (cf. the discussion for the proton and nucleon below). Detailed simulations of the expected sensitivities within the LHCb detector \cite{LHCBEDM} indicate that precisions of a few \% of $\mu_N$ on the magnetic moments and of $10^{-17}$ e.cm on the electric dipole moments could be achievable. The moderate precision on the EDM would provide sensitivity only to new physics models with a strong enhancement of heavy flavor contributions to CP violation.

\subparagraph{Interface to LHC experiments:}
Performing Fixed Target measurements with internal targets located inside existing experiments offers the advantage to minimize the required investments while relying on well supported collaborations and detectors. The method was already pioneered by LHCb with SMOG, and first measurements of high-x charm and of p-He yield of anti-protons were recently released. LHCb is presently designing a storage cell, which could be installed during LS2 upstream of the VELO, improving the gas density by up to 2 orders of magnitude and providing more flexibility to the gas choice. The design of a polarized target system by the LHCSpin collaboration is also ongoing, benefiting from the expertise of similar devices built for the HERMES experiment at HERA. Such a target would be more difficult to insert within an existing experiment and could be installed at earliest during LS3. 

Studies are also ongoing for the implementation of an internal target at the ALICE interaction point, for which 3 positions are considered. Fixed Target measurements with ALICE would benefit from the forward arm including its high precision tracking upgrade. The ALICE central barrel with precise tracking, as well as charged hadron and electron identification, would extend the acceptance to very backward rapidities not accessible to LHCb. Because of its TPC ALICE would have a lower data taking rate than LHCb in fixed target mode, but the different focus of the collaboration could offer the possibility to dedicate specific periods of LHC proton running to Fixed Target operation. 

The costs for implementation of internal targets within LHCb or ALICE are small compared to those of other upgrades being considered for future LHC operation. The main issues are the prioritisation of the various physics programs within the collaborations, and the possible interference of Fixed Target instrumentation and operation with the collider-mode performance of the experiments and of the LHC.

\subsubsection{LHC-LLP experiments}
\vspace{0.5cm}

The long lifetime experiments FASER, MATHUSLA, CODEX-b and milliQan wish to search for new particles produced in LHC collisions with a too long lifetime or a too small electric charge to be detected in the main detectors. They propose to install new detectors away from the interaction points to detect long range decays or low signals from these putative particles.

\subparagraph{Physics motivation:}
Nature features many examples of unstable particles with lifetimes much longer than one would naively expect by estimating the decay rate from the mass of the decaying particle $\Gamma\sim m$.
There are several different reasons (often working together) that can provide parametrically longer lifetimes. The decaying particle could be protected by an approximate symmetry, the interactions mediating the decay could be weak or the decay could have very little available phase space.
For example, B-mesons have a liftetime that is 12 orders of magnitude larger than that of the $\rho$-meson despite being approximately 7 times heavier. Here symmetry enforces that the decay can only occur via the weak interactions. Similarly the beta decay of the neutron has only a relatively small phase space $\sim$~MeV compared to its mass $\sim$~GeV. Moreover it can only occur via the weak interactions that are suppressed by the electroweak scale.
Hence it is very conceivable that such particles could also exist beyond the Standard Model and many theoretical examples have been proposed.

Phenomenologically lifetimes greater than $\sim 10^{8}$~s and shorter than a few minutes are particularly interesting because they are less constrained by the LHC experiments and big bang nucleosynthesis, respectively. For a detailed discussion see, in particular, the physics case studies for FASER~\cite{Ariga:2018uku} and MATHUSLA~\cite{Curtin:2018mvb}.

\subparagraph{Detectors implementation:}
the four dedicated LHC-LLP proposed experiments FASER, MATHUSLA, CODEX-B and milliQan are briefly described below.

\vspace{0.5cm}
{\bf FASER} is a small and inexpensive experiment designed to search for light, weakly interacting particles at the LHC. Such particles are dominantly produced along the beam collision axis and may be long-lived, travelling hundreds of meters before decaying. To exploit both of
these properties, FASER is to be located along the beam collision axis, 480 m downstream from the ATLAS interaction point, in the unused tunnel TI12.

Given its small scale, its close relationship to the LHC machine, and the aim to install a phase 1 detector during LS2, an accelerator side PBC study group was formed to support the FASER proposal. Initial investigations involving civil engineering, the FLUKA team, integration, survey, and beam instrumentation revealed no obvious show-stoppers. In particular a study from the CERN STI group using the FLUKA simulation program was completed to assess backgrounds and the radiation level in the FASER location.
The results show that muons are the only high-energy (> 100 GeV) particles entering FASER from the IP, with an expected rate of 70 Hz (for the expected Run 3 conditions with a peak
luminosity of 2$\times$10\textsuperscript{34} cm\textsuperscript{-2}s\textsuperscript{-1}). No high-energy particles are expected to enter FASER from proton showers in the dispersion suppressor or from beam-gas interactions. The radiation level expected at the FASER location is very low due to the value of the dispersion function in the LHC cell closest to FASER.

The FASER project is scheduled in two phases: it is intended to operate a small detector during Run 3 (FASER phase 1) and to install a larger detector for HL-LHC operations (FASER phase 2). The collaboration submitted a letter of intent to the LHCC in September 2018, 
and has submitted a Technical Proposal \cite{FASERLOI} following the positive feedback of the Committee.  

\vspace{0.5cm}
{\bf MATHUSLA} is envisioned to be a large, relatively simple surface detector that can robustly reconstruct displaced vertices with good timing resolution.
The main component of the detector is an approximately 5-m thick tracker array situated above an air-filled
decay volume that is 20 m tall. The configuration used for the sensitivity estimates of chapter 4 assumes a detector of 200 m$\times$200 m in area located 100m above an LHC high-luminosity interaction point, with a 100m horizontal shift. The detector size and location are currently being optimized to take into account land constraints and opportunities, with the hope to be able to reduce the size while keeping similar sensitivity. The detector design is modular for a staged implementation. The total cost will be driven by civil engineering and the large area tracking detectors. The collaboration is investigating low-cost solutions with the challenging goal to keep the overall cost of the full size detector below 100 MCHF.

MATHUSLA submitted a Letter of Intent \cite{MATHUSLALOI} to the LHCC in September 2018. Considering the size of the project and its physics overlap with other PBC programs, the LHCC recommended to discuss it further within the PBC and strategy processes.

\vspace{0.5cm}
{\bf CODEX-b} proposes to take advantage of a large shielded space in the LHCb cavern that is expected to become available
after LS2 upgrades, to construct a Compact Detector for Exotics at LHCb (``CODEX-b").
With additional shielding from the primary interaction point, CODEX-b can operate in a low background environment,
eliminating the triggering challenges associated with ATLAS and CMS. 
The modest size of CODEX-b is anticipated to translate to relatively low construction and maintenance costs and a relatively short construction time-scale with proven, off-the-shelf components. The possibility to accumulate the 300~fb\textsuperscript{-1} integrated luminosity needed to reach the targeted sensitivity is however pending on approval of the LHCb Phase II upgrade for the HL-LHC era.   

\vspace{0.5cm}
{\bf milliQan} is proposed~\cite{2016arXiv160704669B} to be installed in the PX56 drainage gallery located above the CMS experiment cavern.
The proposed detector is a \num{1  x 1  x 3 } \si{\metre} plastic scintillator array. 
The array will be oriented such that the long axis points at the nominal CMS interaction point. It is specifically aimed at particles with a very small electric charge.

\vspace{0.5cm}

\subsection{New facilities}
\vspace{0.5cm}

\subsubsection{EDM ring}
\vspace{0.5cm}

The existing EDM storage ring community, based mainly in  Germany (JEDI) and Korea (srEDM), have joined their efforts with a fledgling CERN effort under the PBC auspices to form a loose collaboration known as CPEDM, with the aim of converging towards a common design of an EDM storage ring \cite{PBC:EDM}.

\subparagraph{Physics motivation:}
CP violation is a necessary ingredient to generate the matter antimatter asymmetry observed in the Universe. The CP violation encoded in the CKM matrix is not sufficient for this task. Therefore searching for new sources of CP violation is one of the most pressing tasks in fundamental physics. Electric dipole moments are potentially one of the most sensitive probes. 
For example the current limit on the neutron EDM in the 10\textsuperscript{-26} e.cm range allows to probe SUSY particles well into the TeV range and has even greater reach for other extensions of the SM.
Hadronic and leptonic dipole moments give complementary information on the respective particle sectors. In the hadron sector efforts have mostly focussed on neutrons as well as certain advantageous nuclei with unpaired nucleons. 
The proton on the other hand with its unshielded monopole charge has been considered experimentally very challenging. 
However, it has been realized that storage rings may offer the prospect to measure proton and nucleon EDMs with  a sensitivity level of 10\textsuperscript{-29} e.cm, an attractive target at one or two orders of magnitude better precision than that of hadronic EDM experiments based on other methods. 

In addition such an experiment also has the potential to search for oscillating EDMs that are a direct probe of dark matter consisting of axions or axion-like particles.

\vspace{0.5cm}

\subparagraph{Technical implementation:}

A proposal was published in 2011 by the storage ring EDM collaboration (srEDM). 
They proposed a method to search for and measure the electric dipole moment of the proton by using polarized protons at the so-called
magic momentum of 0.7 GeV/c in an all-electric storage ring with radius of $\sim$ 40 m and an E-field of $\sim$ 10 MV/m between plates separated by 3 cm. 
The proposal included the potential realization of such a storage ring at Brookhaven and included a relatively detailed costing.

The strength of the storage ring EDM method comes from the fact that it should be possible to store a large number of highly polarized particles for a
long time, achieve a long horizontal spin coherence time, and probe the transverse spin components as a function of time with a high sensitivity polarimeter.
At their magic momentum of 0.7 GeV/c, the proton spin and momentum vectors precess at the same rate in any transverse electric field. 
With the spin frozen along the momentum direction, the radial electric field acts on the assumed EDM vector and precesses the proton spin vertically for the duration of the storage time, building up its vertical component. The target storage time is of order 10\textsuperscript{3} s given by the estimated spin coherence time of the beam. 

Having considered various methods of freezing the spin along the momentum direction by applying a combination of magnetic and electric fields, 
the collaboration believes that the proton EDM method at the magic momentum using only electric fields is the simplest to implement.

Since the original proposal the Juelich EDM Investigations (JEDI) collaboration have exploited polarized deuterons in the COSY ring to make impressive progress in the manipulation and understanding of:
\begin{itemize}[noitemsep,topsep=0pt]
\item state of the art polarimeters and understanding of polarimeter systematic errors;
\item studies with polarized deuteron beams;
\item beam-based alignment, Rogowski coils, and the use of RF-Wien filters for deuteron EDM studies;
\item feedback stabilization to maintain frozen spin to better 1 ppb;
\item long spin coherence times in excess of 1000 s.
\end{itemize}

In parallel the srEDM collaboration has continued to develop understanding of systematic errors, and pioneered the potential application  of technology such as SQUIDs and magnetic shielding. 

Within CPEDM significant progress has been made in the understanding of the key systematic errors, and their potential limitation to the ultimate sensitivity of the storage ring approach has been quantified. 
The leading systematic is the potential of a radial magnetic field interacting with the magnetic moment to mimic the potential EDM signal. 
Magnetic shielding to reduce the background magnetic fields to the order of nT is foreseen.
Any residual radial magnetic field will lead to a vertical orbit split of counter-rotating beams on the order of picometres. 
Measurement of this split with SQUIDS has the potential to provide a handle on the radial magnetic field systematic.
Detailed analysis has shown that it is still a challenge for this measurement technique to allow the target sensitivity of 10\textsuperscript{-29} e.cm to be reached. 

Given the open questions and need to develop the measurement techniques, hardware components, and test the essential principles of operation,
a small prototype ring has been proposed as a necessary staging point on the path to a full ring.
The ring is designed to work either with only electric fields or with combined electric and magnetic fields.
The latter option will allow the spin to be frozen paving the way for proto-measurements of the proton EDM.
Given the existing expertise and infrastructure, the COSY site at Juelich is foreseen as a potential location for the prototype ring.

\subsubsection{Gamma Factory}
\vspace{0.5cm}

The Gamma Factory is a novel concept \cite{GAMMA-Witek} using the LHC to produce gamma ray beams with a breakthrough in intensity by up to 7 orders of magnitude, at very high $\gamma$-energies up to 400 MeV, compared to the present state of the art.

\subparagraph{Physics motivation:}

The initiative proposes  to produce, accelerate and store high-energy atomic beams 
in the CERN accelerator complex. Excitation of their atomic degrees of 
freedom by laser photons can form high-intensity primary beams of gamma rays and, in turn, 
secondary beams of polarised leptons, neutrinos, vector mesons, neutrons and radioactive ions.
The Gamma Factory target is to achieve a  leap,
by several orders of magnitude, in intensity and/or brightness with respect to existing facilities. 

This will open completely new territories and opportunities to precision measurements and \linebreak searches. Potential domains of interest include e.g. fundamental QED measurements, nuclear and neutrino physics with rare processes or exploration of the hidden sector. In addition, the storage of partially stripped ions in the LHC provides an effective high energy electron beam which could be used for DIS e-p physics by the LHC experiments. The Gamma Factory has therefore a rich potential provided the anticipated intensities are confirmed. The reach of fundamental physics highlights will be quantitatively estimated with improved simulation tools after the foreseen SPS proof of principle experiment.

\subparagraph{Key principles:}
Partially stripped ion (PSI) beams, composed of ions with a few electrons left, can be stored in the SPS and the LHC at very high energies over a large range of Lorentz factors: $ 30 < \gamma_L < 3000$, at high bunch intensities: $ 10^8 < N_{\rm bunch} <10^9 $, and at high bunch repetition rates of up to 20 MHz.  

Two phenomena contribute to the production of high intensity and high energy gamma rays. On one hand, the resonant excitation of the atomic levels thanks to the very large energy of the ions, provides a laser frequency Doppler boost of $2\gamma_L$. Moreover, the very large resonant absorption cross section, in the giga-barn range, and the high $\gamma_L$ factors in the SPS and LHC, would allow the excitation of deep-lying transitions with conventional lasers. Additionally, the consequent spontaneous atomic-transition emission of secondary photons produces a further $2\gamma_L$ boost in the photon frequency laboratory frame. Therefore, the processes of absorption and emission gives a total frequency boost of the initial laser of up to $4\gamma_L^2$, opening the possibility of producing at LHC gamma rays with energies above the muon pair production threshold. Last but not least, contrary to proton bunches, the PSI beams could be efficiently cooled increasing considerably the beam brightness. 

\subparagraph{Results and next steps:}
Investigations of PSI took place at CERN during 2017 and 2018 \cite{PBC:GAMMA} and included tests of strippers, PSI transfer and lifetime studies in the SPS. These culminated in a successful test in the LHC with lead nuclei with one remaining electron. The PSI were accelerated without issue to 6.5 TeV proton equivalent. At the top energy the lifetime of the beam was around 40 hours. 

This impressive result opens way for consideration of a proof of principle installation in the SPS. Here, the aim is to install a Laser and Fabry-P\'erot cavity in the SPS tunnel, and to investigate the production of X-rays from excited Lithium-like Pb+79.  The Gamma Factory group is preparing full documentation of the proposal for submission to the SPSC.

\subsubsection{nuSTORM}
\vspace{0.5cm}

A detailed study of nuSTORM that includes consideration of the implementation of the facility at FNAL was published in 2014.
The decision was taken by the P5 evaluation not to pursue the proposal at the time. 
In these studies the facility was optimised for the search for a light sterile neutrino. 
More recent considerations include the re-optimisation of the facility for the study of neutrino-nucleus scattering with concomitant adjustment
of the parameters for pion production and the energy range of the muon storage ring.
An Expression of Interest had been presented to the CERN SPSC by the collaboration in 2013 \cite{NUSTORMLOI}.

Based on the work performed for the FNAL proposal and more recent input from the nuSTORM collaboration,
an exploratory study \cite{PBC:NUSTORM} of the possible implementation of nuSTORM at CERN has been conducted under the auspices of PBC. 
It should be emphasised that this is a cursory look at the possibility given the limited availability of resources.
Nonetheless, a small team examined:

\begin{itemize}
\item The potential of SPS to deliver the required beam
\item The possibility for fast extraction and siting of facility
\item Initial civil engineering sketches
\item Consideration of target, horn, target complex and related beam absorbers
\item Preliminary consideration of radiation protection issues
\end{itemize}

The studies show that the SPS has the capability to produce the required beams, as was proven in the past by the delivery 
over 5 years of $\approx$ \num{2e20} PoT to CNGS at a higher beam energy on target.
The potential for competition for time in the SPS's duty cycle should be noted.

After examining the options, a potential extraction point from the SPS in LSS6 was chosen as a baseline leading to a potential siting on an unused area on CERN unfenced territory near the Meyrin site. 
LSS6 is presently equipped with fast extraction capabilities used to serve HiRadMat and the LHC.
The target and storage ring would be situated approximately 20 m underground in the geologically favourable molasse.
Based on depth, the previous CENF analysis, and wide-ranging experience, radiation protection would appear at first sight to be not a show-stopper but necessitating, of course, serious attention.

\subsection{Non-accelerator options}
\vspace{0.5cm}

The PBC technology group has investigated the potential synergies between CERN technological expertise and non-accelerator experiments possibly located outside of CERN \cite{PBC:TECHNO}. Two highlights include the IAXO and JURA projects proposed as successors to the CAST and OSQAR/ALPSII experiments, respectively.

\subsubsection{IAXO}
\vspace{0.5cm}
The IAXO project was initially presented to the SPSC in \cite{IAXOLOI}. 

\subparagraph{Physics motivation:} 
Following CAST, IAXO plans to look for axions by pointing a large\linebreak bore/high field magnet at the sun. Within the IAXO fiducial volume, axions emanating from the sun would convert into X-rays and be focused by specific X-ray mirrors to low noise X-ray detectors. 
The method is independent of DM assumptions and provides the current best limits on standard axions on a broad mass range reaching up to $\approx$ 1 eV, and is starting to exclude the QCD axion in this mass range. 
The full IAXO helioscope would gain a factor $\approx$20 on the CAST sensitivity, and a staged option ("BabyIAXO") under consideration could already exclude the axion interpretation of several astrophysical observations.

\subparagraph{Detector implementation:}
A large worldwide IAXO collaboration has recently been founded, extending the contours of the CAST collaboration. IAXO gathers the relevant expertise on X-ray optics and low noise detectors and can use the CAST set-up as a R\&D platform for IAXO pathfinder detectors. Within this framework CERN brings its ATLAS expertise in large toroids and contributes to the design of the BabyIAXO magnet which will have a peak field of 4.1T. Contrary to CAST which was using an LHC magnet, IAXO has no strong reason to be sited at CERN, and DESY has expressed interest to be the host institute. A longer term CERN contribution to the design and exploitation of this flagship axion program would however naturally extend the past strong support to CAST. 

\subsubsection{JURA}
\vspace{0.5cm}

\subparagraph{Physics motivation:} 
Light Shining through a Wall (LSW) experiments exploit one of the few model-independent laboratory methods to search for axions.
Axions are postulated to be produced by interaction of laser photons with a high magnetic field, and similarly converted back to photons in the detection region after crossing a photon proof wall.
The first generation of LSW experiments had limited sensitivities, but new set-ups involving laser amplification with Fabry-P\'erot cavities, longer path lengths and higher B-fields are expected to be competitive with axion helioscopes.

\subparagraph{Detector implementation:}
Following the completion of the first generation experiments at \linebreak CERN (OSQAR) and DESY (ALPS), the second generation project ALPS II is currently equipping 20 HERA magnets with optical cavities to increase the sensitivity by several orders of magnitude. First ALPS II results are expected in the early 2020's. JURA anticipates the ALPS II success by proposing to prepare a third generation LSW experiment combining the state of the art ALPS II optics with future very high field magnets developed for the FCC. 
A long term collaboration between DESY and CERN is proposed in this respect.      

\subsubsection{Others}
\vspace{0.5cm}
Beyond IAXO and JURA a variety of other initiatives could benefit from CERN facilities and expertise. The PBC technology working group has identified 5 key CERN technologies of interest for non-accelerator experiments \cite{PBC:TECHNO}: Magnets, Radio Frequency cavities, Optics, Vacuum and Cryogenics. 
Of particular importance is the CERN expertise on combining these technologies within operational experimental setups. Collaborating with external institutes will also allow CERN to benefit from the latest outside developments. Some of the physics projects considered for these technologies are briefly mentioned here:  
\begin{itemize}
\item{} STAX is a microwave cavity LSW experiment that can search for axion-like particles and low mass dark photons with a resonant technique. It can benefit from both the available magnet and cavity technologies as well as cryogenics.
\item More generally RF cavities are of crucial interest for low mass axion searches. CERN has a particular expertise in operating such cavities in high field magnets, as well as with low noise RF detection techniques at very low temperature.
\item Optical techniques such as Fabry-P\'erot cavities are most important for searches of new low mass particles coupled to photons or for investigation of new QED effects. It is proposed to set up an "Optics Technology Hub" to connect CERN to other expert laboratories in the field including Gravitational Wave facilities, in order to share and develop state-of-the art methods in this domain.
\item{} Based on the above optical R\&D and high field magnets developed for LHC, HL-LHC and FCC, a new experiment is proposed at CERN to search for vacuum magnetic birefringence (VMB@CERN), with a sensitivity which should allow detection of this long-standing QED prediction.
\item{} New physics beyond the Standard Model could also lead to new forces acting on matter. Examples are chameleon models of dark energy, but also scalar and vector dark matter models. These models could be explored with new highly sensitive force sensors such as suggested by the aKWISP proposal. This could benefit from surface coating techniques and test facilities available at CERN as well as cryogenics.
\item{} Searches for WIMP dark matter require large scale detectors with multi-tonne scale amounts of active target material. DARKSIDE aims at ultimate sensitivity close to the neutrino floor, using a large volume of depleted liquid argon (LAr). DARKSIDE has already a good collaboration with CERN for the development of LAr depletion methods, and is designing its detection techniques and cryogenics in close collaboration with CERN.
\end{itemize}

\newpage
\subsection{Summary tables}
\vspace{0.5cm}

The beam requirements of the proposed experiments and facilities are shown in tables \ref{tab:summarySPS} and \ref{tab:summaryFac1}.
An overview of the status and prospects of the individual implementation of the projects is given in table \ref{tab:summaryprojects}.
The positioning of the projects in the overall landscape is discussed in the next section.

\begin{table}[h]
\begin{center}
\caption{Summary of proposed experiments at the SPS and PS}
\label{tab:summarySPS}
\begin{tabular}{llclcc}
\hline
Experiment & Primary & PoT/year & Secondary & Location & Beam line \\
\hline
NA61++ & ions/p scan &  & ions/hadrons < 120 GeV & ENH1 & H2 \\
NA61++ & 400 GeV p &  & hadrons < 10 GeV (tertiary) & ENH1 & H2 \\
NA64++(e) & 400 GeV p & \num{1e18} & 100 GeV e- & ENH1 & H4 \\
 &  &  &  &  &  \\
COMPASS+ & 400 GeV p &  & muons, pions & EHN2 & M2 \\
COMPASS++ & 400 GeV p &  & 100 GeV K; $\bar{p}$; RF-separated & EHN2 & M2 \\
MUonE & 400 GeV p &  & 150 GeV muons & ENH2 & M2 \\
NA64++(h) & 400 GeV p &  & 20-50 GeV $\pi$;  $K^-$; 200 GeV p & ENH1 & H4 \\
NA64++($\mu$) & 400 GeV p &  & 100 GeV muons & ENH2 & M2 \\
 &  &  &  &  &  \\
NA62++ & 400 GeV p & \num{1e18} & Hidden sector & ECN3 & K12 \\
KLEVER & 400 GeV p & \num{1e19} & $K_L$ & ECN3 & P42/K12  \\
NA60++ & ions/p scan &  &  & ECN3 &  \\
DIRAC++ & 400 GeV p &  & pions;  kaons & ECN3 &   \\
 &  &  &  &  &  \\
SHiP & 400 GeV p & \num{4e19} & Hidden sector & BDF & TT21 \\
TauFV & 400 GeV p & \num{8e17} &  & BDF & TT21 \\
&  &  &  &  &  \\
REDTOP & 2-3 GeV p & \num{1e18} & & PS & \\
\hline\hline
\end{tabular}
\end{center}
\end{table}

\begin{table}[h]
\begin{center}
\caption{Overview of proposed new beam lines and facilities}
\label{tab:summaryFac1}
\begin{tabular}{lllll}
\hline
Facilty & Type & (Primary) Beam & PoT/year & Status \\
\hline
BDF@SPS & Beam dump & p 400 GeV & \num{4e19} & Design study \\
AWAKE++@SPS &  PDWA/target & p 400 GeV & & Exploratory \\
eSPS & Linac/synchrotron/target & e 16 GeV & & EoI \\
EDM & Storage ring & polarized p  & & Feasibility \\
nuSTORM@SPS & Target/storage ring & p 100 GeV & \num{4e19} & Feasibility \\
$\gamma$-Factory@LHC & Storage ring & PSI & & Feasibility \\
\hline\hline
\end{tabular}
\end{center}
\end{table}

\newpage

\begin{table}[h]
\begin{center}
\caption{Implementation status and prospects of the proposed PBC beam experiments. COMPASS+ corresponds to the part of the COMPASS++ program feasible with existing beams. LHC-FT++ identifies the LHC-FT programs associated to more demanding set-ups such as polarized targets or double crystals (considered as part of the detector in the table), as opposed to more conventional storage cells. The LHC-FT collaborations refer to the communities associated to LHCb and ALICE which are motivated by LHC-FT physics. For the $\gamma$-Factory the "Beam" column refers to the facility, and "Collaboration","Detector" to the expected user communities and their experiments. The $\gamma$-Factory costing will strongly depend on the application.}
\label{tab:summaryprojects}
\begin{tabular}{lcccccc}
\hline\hline
   & A &  ready  &  ready  &  adequate  &  < 10 M\texteuro  &  Run 3  \\
Quote: & B & need upgrade & under design   & to strengthen  & 10-50 M\texteuro  &    Run 4 \\
   &  C  &  to be built  & need R\&D & to be built   & > 50 M\texteuro   &  Run 5 \\ 
\hline\hline
\textbf{Project}  & \textbf{Physics} & \textbf{Beam}
& \textbf{Detector} & \textbf{Collaboration} & \textbf{Cost} & \textbf{Earliest} \\
& \textbf{highlight} & \textbf{requirement} & \textbf{maturity} &  & \textbf{beam+det} & \textbf{operation} \\ 
\hline\hline
NA61++  & QGP Charm    & B &  B & A & A & A \\
COMPASS+ &  $R_p$ \& QCD       & A &  B & A & A & A \\
COMPASS++ &  QCD       & B &  B & B & B & B \\
MUonE   &  HVP(g-2)$_\mu$    & A &  B & B & A & A \\
LHC-FT  &  QCD         & A &  B & B & A & A \\
LHC-FT++  &  spin/MM/EDM & A &  C & B & A & B \\
NA60++  &  QGP phase   & C &  B & C & B & B \\
DIRAC++ &  chiral QCD  & C &  B & C & B & B \\
NA62++  &  dark sector & B &  A & A & A & A \\
KLEVER  &  $K^0\rightarrow\pi^0\nu\bar{\nu}$   & B & C & B & B & B \\
NA64++  &  dark photon & A &  B & A & A & A \\
SHiP    &  dark sector \& $\nu_{\tau}$ & C &  B & A & C & B \\
TauFV   &  $\tau\rightarrow3\mu$   &  C & C & B & C & C \\
REDTOP  &  $\eta$ decays  & B  &  C &  B  & B & B \\
EDM ring  &  p EDM     &  C  &  C &  B &  C &  C \\
eSPS    & dark photon  &  C  &  B &  B &  C &  B \\
AWAKE++ & dark photon  &  C  &  B &  A &  B &  B \\
nuSTORM & $\sigma(\nu)$  & C  & C &  B &  C &  B \\
$\gamma$-Factory  &  high rate $\gamma$  &  C &  C &  C & - & C \\
\hline\hline
\end{tabular}
\end{center}
\end{table}

\newpage
\section{PBC projects in the global context}

\vspace{0.5cm}
The physics reach and implementation issues of the proposed PBC projects are discussed within the CERN and worldwide contexts, summarizing the main findings of the detailed studies reported in the QCD \cite{PBC:QCD} and BSM \cite{PBC:BSM} PBC working group reports. These documents also provide all references to the experiment results and prospects reported here. 

\subsection{QCD-oriented projects}
\vspace{0.5cm}

\subsubsection{Hadron Vacuum Polarization input for $(g-2)_{\mu}$}
\vspace{0.5cm}

\begin{figure}[!htb]

   \centering
   \includegraphics*[width=500pt]{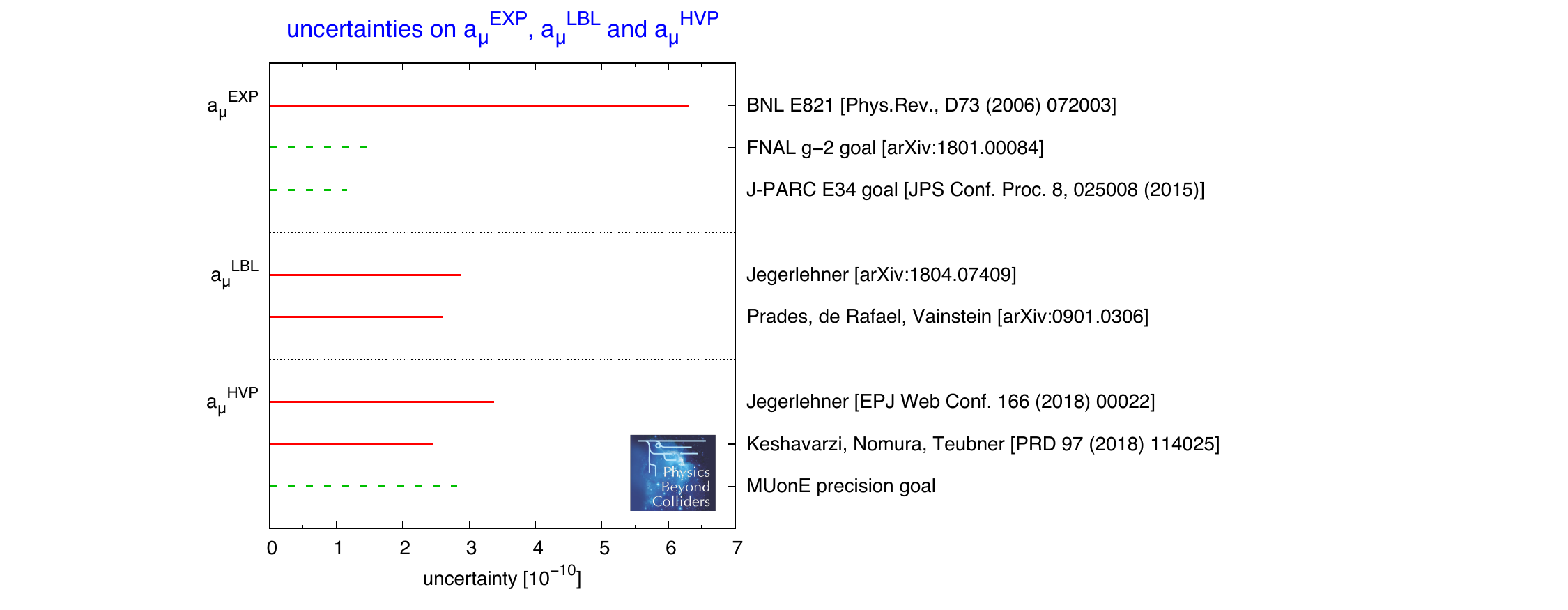}
   \caption{Overview of the (g-2)$_\mu$ landscape. Top: current (red) and projected (dashed green) experimental uncertainties of past and future measurements. Middle and bottom: dominant sources of theory uncertainty. Light by Light contribution (middle), as estimated from theoretical calculations, and hadron vacuum polarization (bottom), extracted from $e^+e^-$ data (red) and anticipated by MUonE (dashed green). The MUonE goal assumes a systematic uncertainty equal to the statistical uncertainty and added in quadrature. (See~\cite{PBC:QCD} for details.)}
   \label{fig:HVP}

\end{figure}

MUonE comes as a timely project in the context of the improved (g-2)$_\mu$ experiments currently starting in the US and in preparation in Japan. The experimental and theoretical landscape is summarized in figure~\ref{fig:HVP}. 

The expected improvement of the experimental uncertainty is a factor 4. A full exploitation of the improved measurements as a test for the Standard Model and a tool to search for new physics requires a commensurate reduction and consolidation of the theoretical uncertainties.
Lattice computations of the HVP will not yet be competitive in a foreseeable future. The current method based on e+/e- low energy data combines many heterogeneous data samples, raising specific issues in the estimation of the systematic errors. Incremental improvement of the precision is regularly reported based on improved procedures and new data sets, but the method may hit a precision ceiling in the future. As seen in figure \ref{fig:HVP} the initial goal of MUonE is not to provide a breakthrough in the achieved precision, but to complement and consolidate the available estimations using a completely independent method with orthogonal systematics. Reaching the current precision of the e+/e- method will already be an experimental challenge. However MUonE is based on a uniquely homogeneous data sample and presents no known intrinsic limitation, paving the way to possible improvements in the future.

\subsubsection{Proton Radius}
\vspace{0.5cm}

\begin{figure}[!htb]

   \centering
   \includegraphics*[width=500pt]{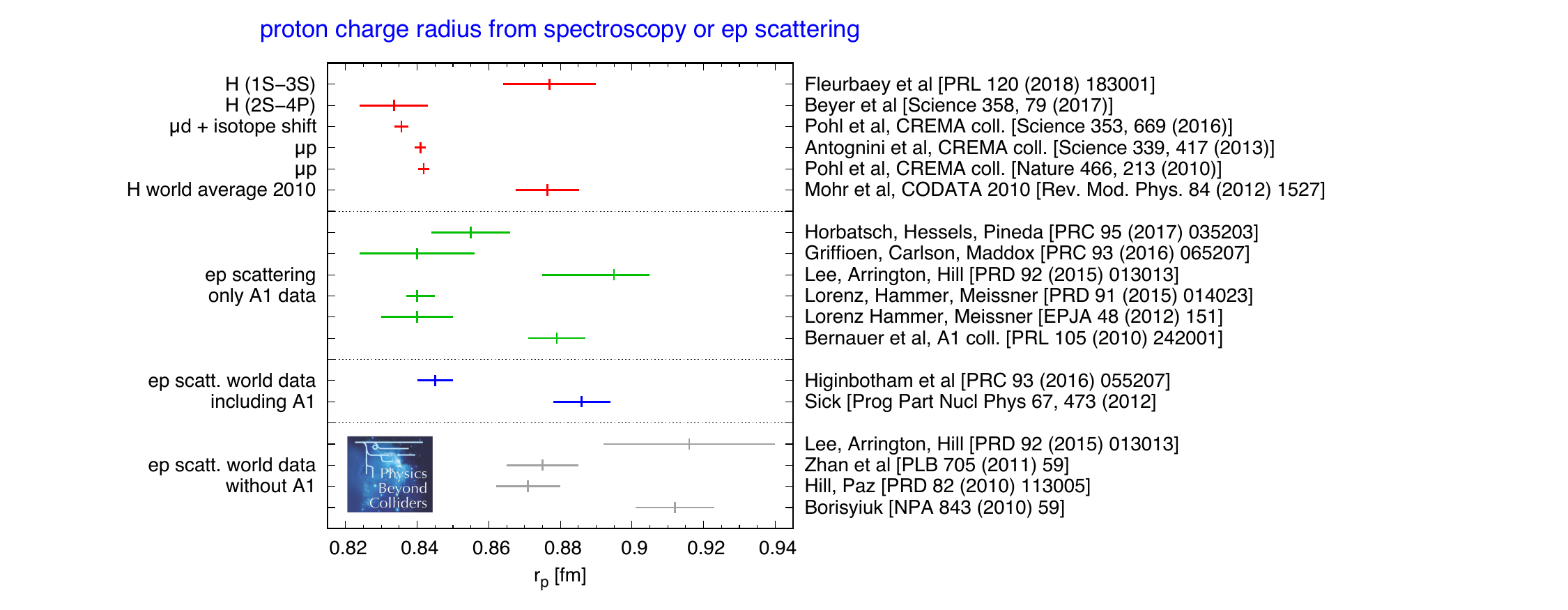}
   \caption{Overview of proton radius measurements with hydrogen spectroscopy, muonic atom spectroscopy and recent analyses of electron-proton scattering. (See~\cite{PBC:QCD} for details.)}
   \label{fig:RP}

\end{figure}

The COMPASS++ proposal to measure the proton radius in $\mu$p elastic scattering is motivated by the current confused situation of this field. As illustrated in figure \ref{fig:RP}, the spectroscopy and ep scattering methods currently available 
provide different values when compared to each other or from one group to another. The intrinsic motivation to solve the puzzle is not the absolute determination of the proton radius, but to clarify whether the interactions of the electron and the muon to the proton differ, which would indicate lepton universality violation.

The MUSE experiment is scheduled at PSI in 2019-20 to address the puzzle: MUSE will provide a direct comparison of $\mu$p and ep low energy scattering in the same experiment. Both electric charges of the beams will be used to control 2-photon effects. An issue in the comparison arises from higher order QED interaction terms which are higher in ep than in $\mu$p scattering. COMPASS++ has the advantage that these higher order QED corrections are lower at high energy than at the low energy used by MUSE. The COMPASS++ measurement may therefore be a useful independent contribution to the resolution of the puzzle.

\subsubsection{Low Energy QCD: Chiral dynamics and spectroscopy}
\vspace{0.5cm}

Several measurements proposed by the PBC projects allow to test QCD in the low energy domain. 

Of particular interest for chiral perturbative QCD are the measurements of the $K$-polarizability by COMPASS++ and of the $\pi$$K$ scattering length by DIRAC++, since they both test the theory in the full SU(3) sector. There is little competition expected to DIRAC++ since the extraction of the scattering length from meson decays, as was done by NA48 for the $\pi\pi$ system, seems difficult for the $\pi$$K$ system because of many parasitic resonances.

Another highlight of COMPASS++ is the proposed high-statistics revisit of strange spectroscopy. A potential long term competition may come from the 10 GeV charged kaon beam foreseen in the possible future extension of the J-PARC Hadron Experimental Facility, but the COMPASS++ high energy measurements will benefit from a cleaner separation of the strange system from the outgoing baryon. Other COMPASS++ spectroscopy measurements will face the competition of the E50 experiment on the new 20 GeV J-PARC hadron beam in construction, and of PANDA at FAIR for spectroscopy with antiproton beams. 

Finally new insights in the understanding of heavy-light systems may come from the proposed measurement of the magnetic moments of short-lived baryons with a double crystal system at LHC. The expected precision of a few \% should allow to disantangle the theoretical calculations which have a wide spread of predictions \cite{PBC:QCD}. The method can also provide a competitive measurement of the $\tau$ magnetic moment.

\subsubsection{Structures and Spin}
\vspace{0.5cm}

\begin{figure}[!t]
   \centering
   \hspace*{1.1cm}\includegraphics*[width=500pt]{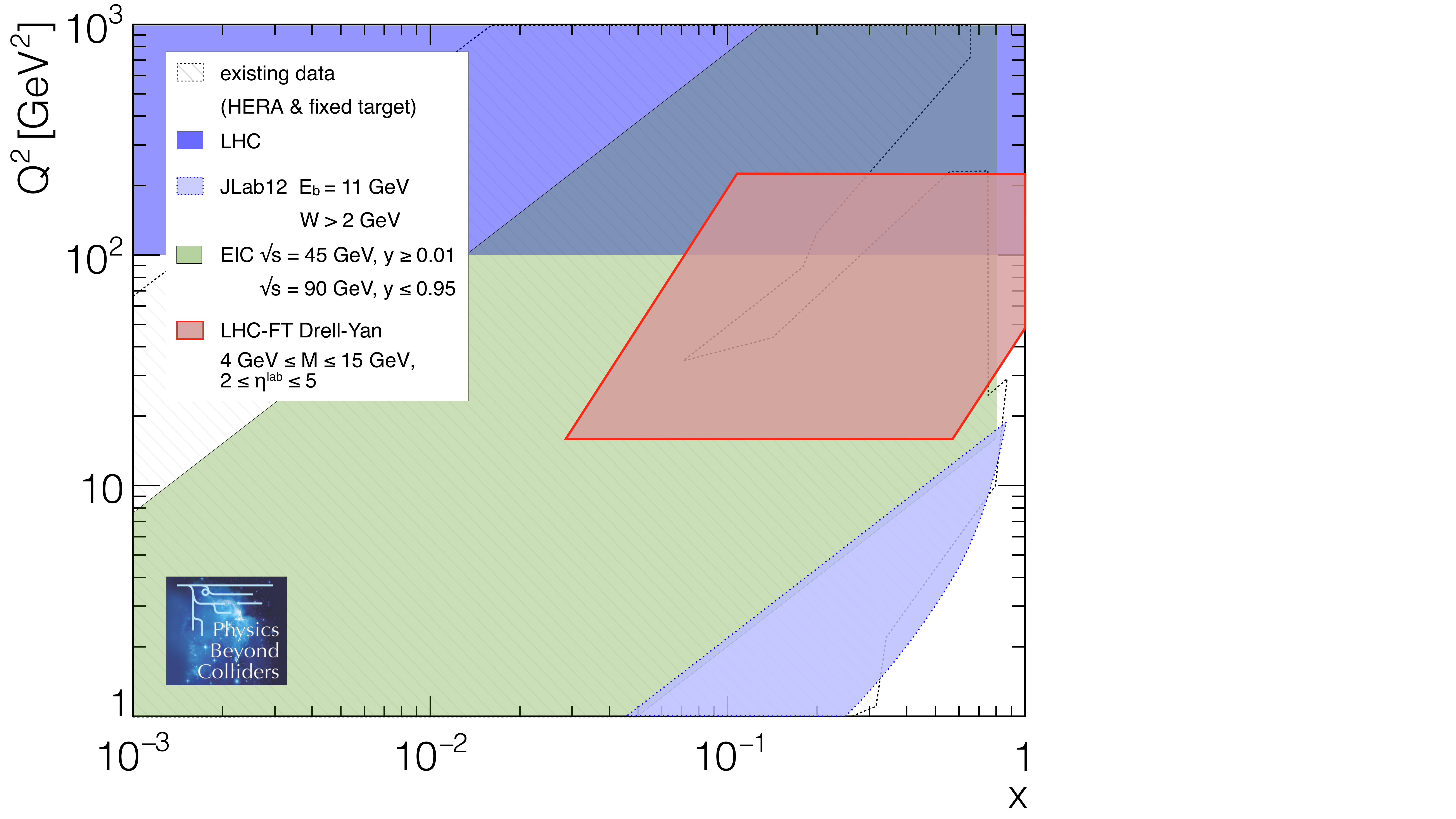}
   \caption{Kinematic coverage of past, present and future programs for the proton PDFs measurement. (See~\cite{PBC:QCD} for details and references.)}
   \label{fig:pdf}

\end{figure}

Investigation of the hadron structure and spin is foreseen both by COMPASS and the fixed target program at LHC. The kinematic coverage of the programs is compared to the worldwide landscape in figure~\ref{fig:pdf} for the proton structure. The messages are the same for the nuclear structure and spin asymmetry measurements: COMPASS lies in the intermediate domain between JLab and the former ep collider HERA, whereas LHC-FT covers a high-x high-$Q^2$ domain, already explored by HERA and LHC but with a limited statistical precision.

The COMPASS++ strategy to focus on hadron beams in the long term seems meaningful: it provides an approach complementary to the flagship program of the field, JLab, as well as to the future electron-ion collider EIC, which both use e-hadron collisions. The COMPASS++ program for structure functions and spin measurements needs, however, prioritisation, with a focus on measurements which make best use of the CERN unique kinematical domain at high energy. One highlight based on existing beams would be a high statistics $\pi^+$ DY sample for pion structure and sea extraction, which would be complementary to the $\pi$ structure extractions foreseen at JLab12 and EIC.

\begin{figure}[!t]

   \centering
   \includegraphics*[width=200pt]{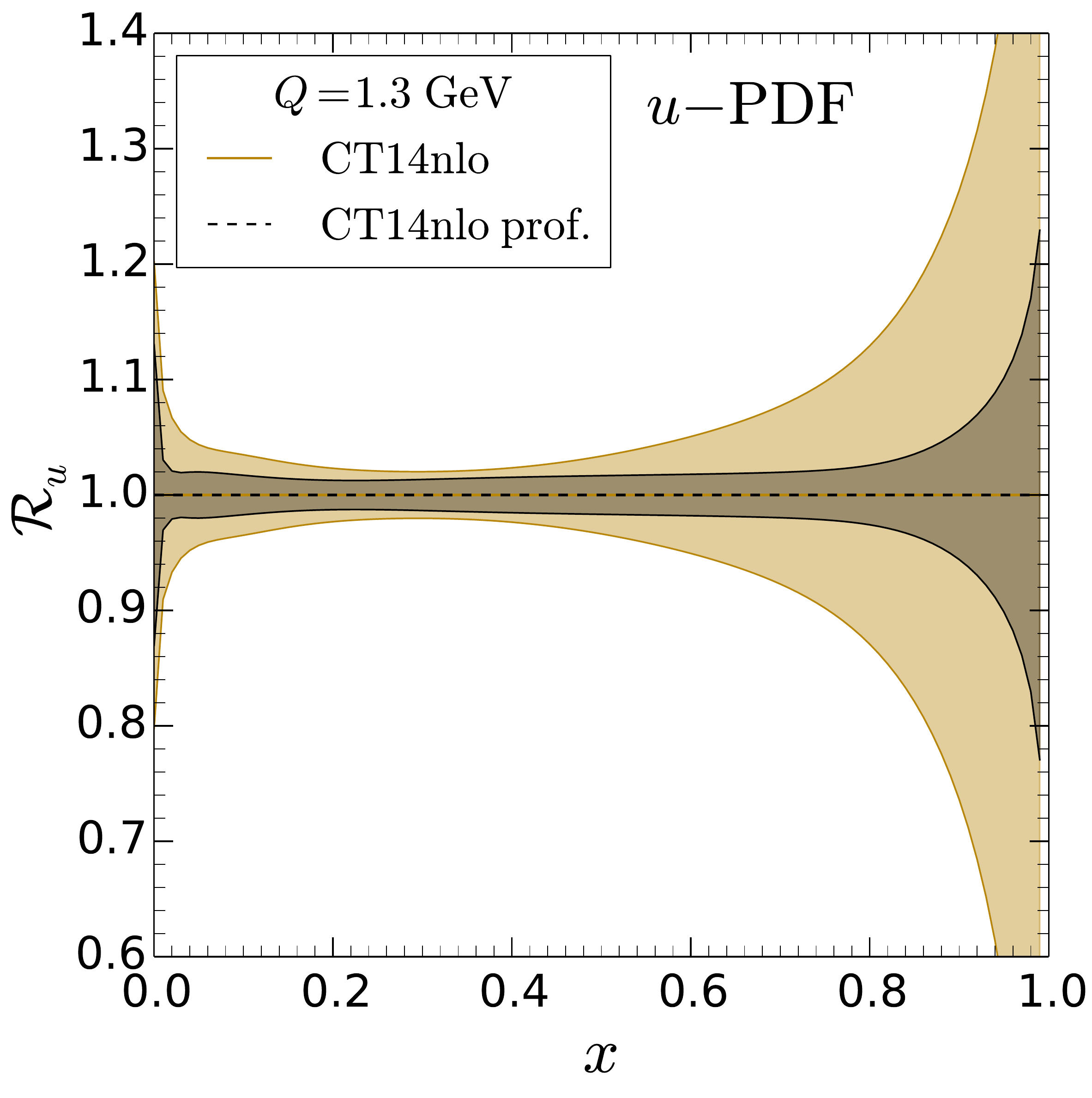}
   \caption{Improvement in the precision of the proton u-quark distribution expected with 10$fb^{-1}$ of DY data collected in pp collisions by LHCb-FT (taken from~\cite{AFTER}). Such an integrated luminosity should be compatible with LHC operation conditions (section 3.3.1).}
   \label{fig:updf}

\end{figure}

LHC-FT has a potential broad physics reach for both structure and spin measurements, making use of specific capabilities of LHCb and ALICE. The program could be an important ingredient of the full exploitation of the LHC potential in the long term. One highlight is the prospect for high-statistics measurements in the high-x high-$Q^2$ domain. As an example, figure \ref{fig:updf} shows the improvements on the proton high-x u-quark distribution expected with 10 fb$^{-1}$ of DY data accumulated with LHCb-FT. This would be an important contribution to improve the future sensitivity of HL-LHC searches at high mass. The high performance of LHCb for heavy flavor measurements will also allow to constrain the high-x gluon distribution. These considerations also apply to the measurements of spin asymmetries with polarized targets.

\subsubsection{Quark-gluon Plasma}
\vspace{0.5cm}

\begin{figure}[!t]

   \centering
   \includegraphics*[width=300pt]{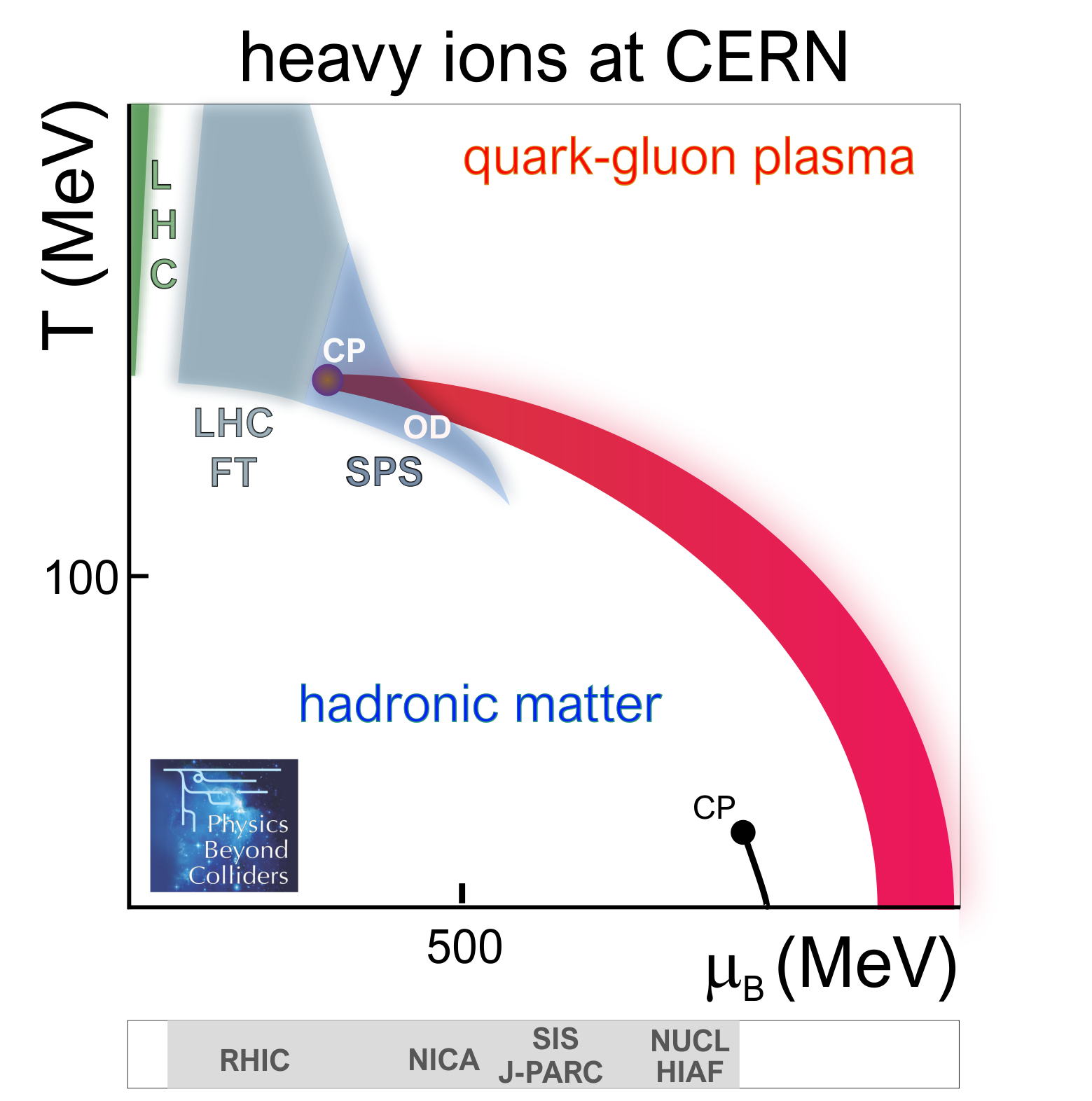}
   \caption{Phase diagram of QCD with regions covered by running and planned experiments. The lower boundary of the grey and blue shaded area follows the chemical freeze-out. The upper boundary relates to the parameters at the early stage of the collisions. The potential critical end point is labeled with CP, the onset of deconfinement with OD. The black line at small temperatures and high densities corresponds to the nuclear liquid-gas transition, also ending in a critical end point CP. The temperature-density ranges of LHC, LHC-FT and SPS experiments are indicated with the shaded green, grey and blue areas, respectively. The density ranges of other experiments are indicated in the bar below the figure. They include RHIC at BNL, NICA at JINR, SIS100 at FAIR, J-PARC-HI at J-PARC, the Nuclotron at JINR (NUCL), and HIAF at HIRFL. (Figure and caption as in~\cite{PBC:QCD}, see text there for more details.)}
   \label{fig:qgp}

\end{figure}

As illustrated in figure \ref{fig:qgp}, both the SPS and LHC-FT access a specific domain of the QCD phase transition, close to the expected QCD critical point (SPS) or between SPS and RHIC/LHC (LHC-FT). In particular a wide coverage of the crossover region is important to consolidate extrapolations to the critical point region from measurements and lattice computations at low $\mu_{B}$.

At SPS there is a clear case to revisit the phase transition with open charm as proposed by NA61++, since this observable was not accessible to the pioneering SPS experiments of the 90's, and is complementary to charmonium. NA60++ also offers a unique opportunity to better characterize the QCD phase diagram with high statistics low-E dimuons by a scan of the expected critical point region.

At LHC, Fixed Target operation opens a new unique kinematical domain, intermediate between SPS and RHIC, to the precise characterization of the QGP with high statistics measurements. RHIC will explore the same region with its energy scan but the statistical precision of its measurements will be limited by the achievable luminosity in collider mode. 

\newpage
\subsection{BSM-oriented projects}
\vspace{0.5cm}

Beyond the wealth of BSM models developed by theorists, the relevant experimental search strategies result from one common constraint: the new phenomena should not yet be excluded by existing experiments. Most of the existing models achieve this by making the newly introduced particles either heavy or very weakly coupled, or a combination of the two. The sizes of the masses and of the relevant effective couplings determine the search strategies to explore the different types of new physics: higher mass requires higher energy to produce the new particles, whereas the production (and subsequent detection) of very weakly coupled particles requires a combination of high intensities, high precision and low backgrounds. In some cases indirect measurements of subtle effects (typically linked to the breaking of symmetries) allow to substitute precision for energy, i.e. a highly precise low energy experiment can indirectly explore particles of very high mass.

\begin{figure}[!t]
   \centering
   \includegraphics*[width=13cm]{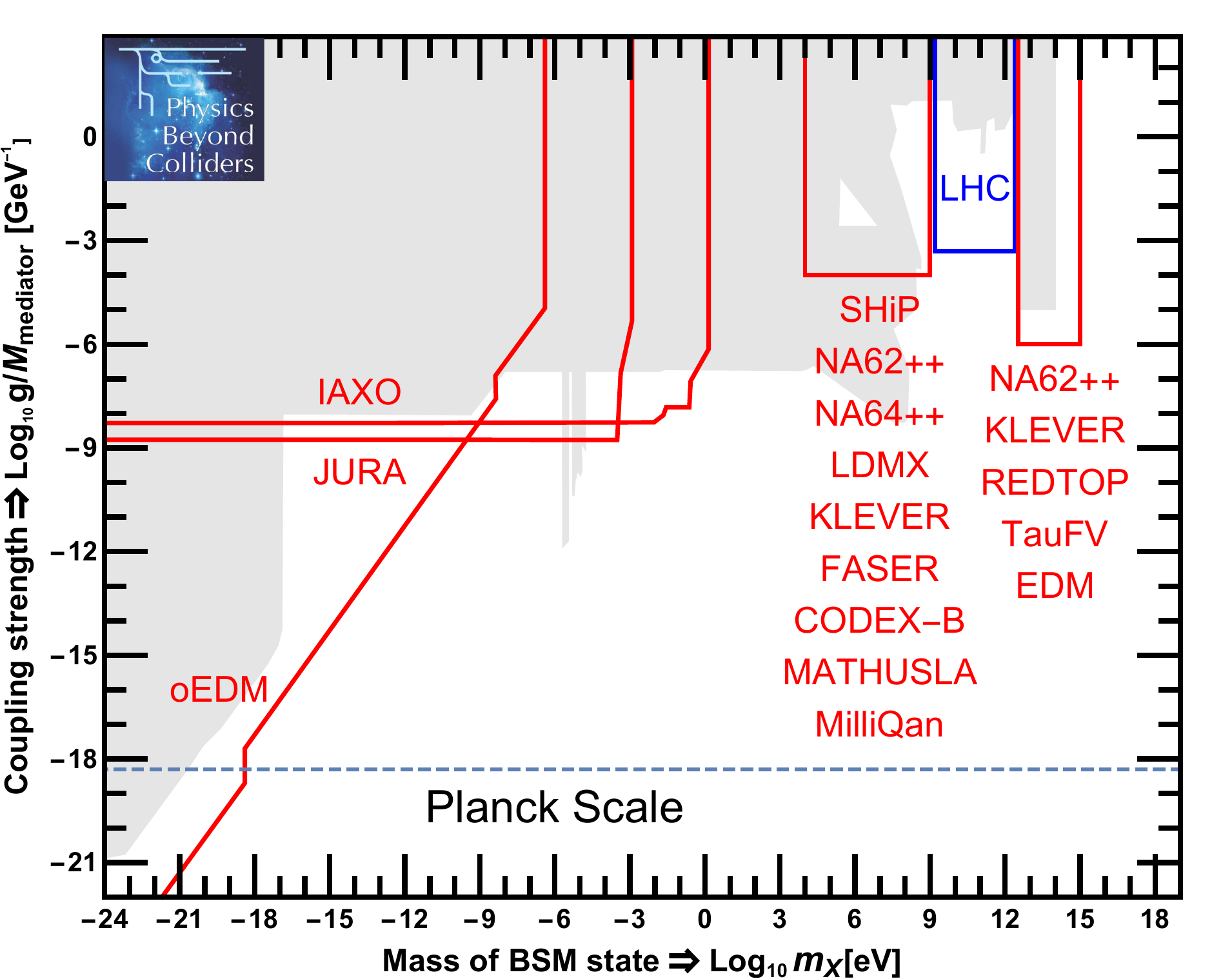}
   \caption{Schematic overview of the BSM landscape, based on a selection of specific models, with a rough outline of the areas targeted by the experiments considered in the PBC sensitivity studies. The x-axis corresponds to the mass $m_{X}$ of the lightest BSM state, and the y-axis to the scale of the effective new interaction where $M_{mediator}$ is the mass of a heavy mediator and $g$ its dimensionless coupling constant to the Standard Model. For indirect probes that do not produce a new particle the relation $m_{X}=M_{mediator}$ is assumed. It is important to note that the indicated areas are only schematic. The true sensitivities of individual experiments can vary significantly as function of the model. The LHC area includes the sensitivity at high mass of the LHC-LLP dedicated proposals, corresponding mostly to models different from those targeted by the PBC projects in the GeV region.
The grey shaded area outlines the currently excluded regions for a class of models corresponding to the benchmarks BC9 and BC11 (cf. e.g.~\cite{Abel:2017rtm,Jaeckel:2010ni,Irastorza:2018dyq}) plus a rough indication of flavor and CP constraints. (As in~\cite{PBC:BSM}.)}
   \label{fig:comprehensive}
\end{figure}

A rough outline of this picture can be seen in figure~\ref{fig:comprehensive}, which schematically indicates the existing limits as well as the levels of improvement possible with the PBC experiments on some exemplary models.

From this very schematic outline a couple of important conclusions can already be drawn:
\begin{itemize}
\item The PBC projects offer significant discovery potential over a wide range of masses and couplings.
\item Very sensitive low energy experiments target the sub-eV mass area. The extremely weak interactions explored correspond to very high underlying scales of new physics. Crucial motivation and hints that are explored include very light dark matter, the strong CP problem, as well as a variety of astrophysical hints.  
\item SPS Fixed Target beam-dump-like experiments and long lived particle searches at LHC have unique capabilities to target the MeV-GeV domain, in between lower energy experiments and LHC direct searches. Particles in this area can be messengers to dark matter, but also dark matter itself. Direct searches at accelerators provide crucial insights complementarity to large scale WIMP dark matter searches, that become less sensitive in the mass range below 10 GeV and are also less sensitive to velocity suppressed interactions.

\item The precision tests of flavor violation (lepton and quark), as well as of CP violation, probe new particles in a mass range exceeding LHC direct searches.

\end{itemize}

Beyond this qualitative picture the specific and quantitative potential offered by the various experiments is discussed below, highlighting their individual capabilities and their place within the worldwide context. 

\subsubsection{Searches for new sources of symmetry violation}
\vspace{0.5cm}

\subsubsection*{Rare meson decays}
\vspace{0.5cm}

Within PBC rare meson decays are explored by NA62, KLEVER and REDTOP.
They probe new sources of quark flavor violation (NA62, KLEVER) and CP violation (REDTOP).

Rare $K$ and $B$ decays cover complementary domains of the flavor sector for BSM searches. In the strange sector the ultra-rare decays $K^{+}\to\pi^{+}\nu\bar{\nu}$ and $K^{0}_{L}\to \pi^{0}\nu\bar{\nu}$ are again complementary in several respects since the two branching ratios have different sensitivities to most BSM models.

NA62 currently offers a unique opportunity for a first precise measurement of the $K^{+}\to\pi^{+}\nu\bar{\nu}$ decay. This measurement has a high priority in the context of the current $B$ anomalies. For the $K^{0}_{L}\to \pi^{0}\nu\bar{\nu}$ decay KLEVER faces the competition from KOTO at J-PARC. The present KOTO results are 2 orders of magnitude away from the SM branching ratio, but the experiment is a long term project improving regularly: extrapolating the current results taking into account the planned detector upgrades and new data, the KOTO projected sensitivity reaches the SM value in the coming decade, and there are prospects for further improvements with the higher intensity beam-line foreseen in the J-PARC Hadron Experimental Facility extension in discussion. The competitiveness of the KLEVER project will therefore have to be monitored as function of the future evolution of KOTO. 

REDTOP is currently the only project worldwide aiming at a very high statistics study of the $\eta$ meson rare decays. The peculiarities of this meson offer unique prospects but present the drawback that BSM physics has to compete with fast SM decay processes, therefore requiring very high statistics to be detected. The competitiveness of REDTOP for a discovery relies on its capacity to accumulate the necessary large $\eta$ meson sample.

\subsubsection*{Rare lepton decays}
\vspace{0.5cm}

Most current experiments dedicated to charged lepton flavor violation, such as MEG and Mu3e at PSI, Mu2e at FNAL and COMET at J-PARC, deal with the first two lepton generations. TauFV addresses similar channels in the third generation sector. Leading results in the field originate from the $B$ factories: in the coming decade BELLE II expects to improve the current limit on $\tau \rightarrow 3\mu$ by more than one order of magnitude, reaching a sensitivity of a few $10^{-10}$ on the branching ratio. TauFV has the potential to further improve by about an order of magnitude, and its sensitivity will not be limited by the $\tau$ production rate on the BDF beam, which leaves room for future improvements. 

The TauFV sensitivity remains far from the $10^{-15}$ to $10^{-16}$ levels expected to be reached in the coming years by Mu3e@PSI for the $\mu \rightarrow 3e$ process, but the third generation sector provides unique insights in a domain where many BSM models predict larger effects than in the first two generations.

\subsubsection*{Electric Dipole Moments}
\vspace{0.5cm}

\begin{figure}[!htb]

   \centering
   \includegraphics*[width=450pt]{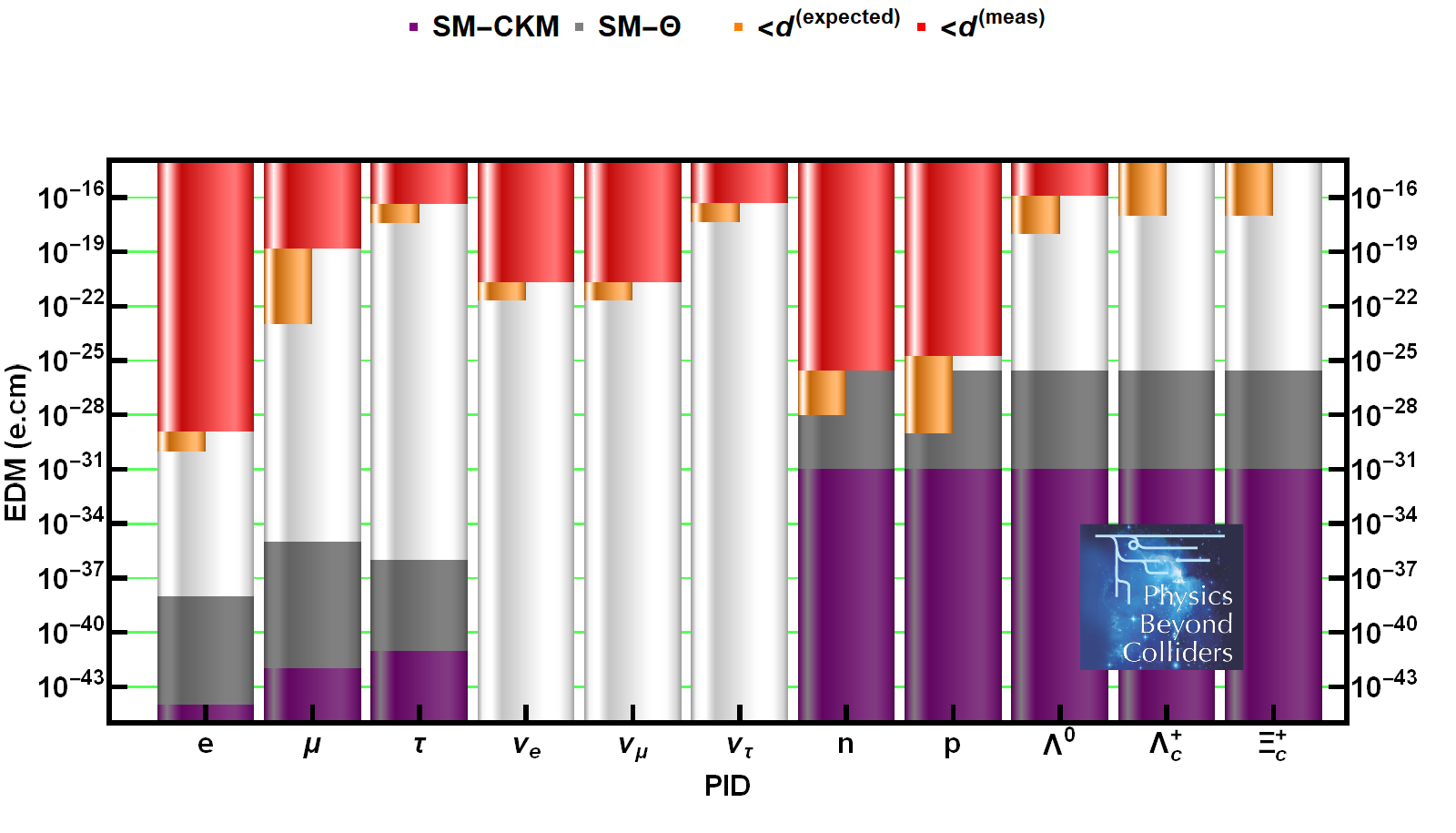}
   \caption{Status and prospects of lepton and hadron EDM measurements. The yellow bars indicate the improvements expected in the next decade versus the current limits shown in red. The purple and grey bars show the SM expectations based on CKM CP-violation and QCD with the maximal $\theta$ term allowed by the limit on the neutron EDM, respectively. The former gives an impression when searches will become SM limited, the latter shows one option of how, a so far undiscovered, SM parameter can manifest itself. However, it could also be interpreted as signal of explicit symmetry breaking effects in models with an axion, giving information of the underlying model and perhaps even its connection to gravity. (See~\cite{PBC:BSM} for details and references.)}
   \label{fig:edm}

\end{figure}

The current status and prospects of EDM measurements are summarized in figure \ref{fig:edm}.

The nucleon EDM is sensitive not only to BSM physics, as lepton EDMs, but also to the QCD $\theta$-term. The neutron and proton EDMs provide complementary information to disantangle between the two sources of CP violation. Figure \ref{fig:edm} indicates that sensitivities at the $10^{-28}$ to $10^{-29}$ e.cm level will cover a significant domain of possible BSM contributions. In this context the moderate precision expected from the short-lived baryon EDM extraction with an LHC-FT double crystal setup may have a limited discovery potential, but the measurements would explore an uncharted territory. 

A proton EDM measurement will significantly contribute to the global picture only if its precision is similar or better than that of the neutron EDM. For neutrons there are several set-ups existing in the world which regularly improve the experimental methods: the community aims to reach a precision close to $10^{-28}$ e.cm in the coming decade. New ideas are coming up for a larger next generation experiment which could federate part of the community and offer long term prospects. 

At present proton EDM measurements lie behind neutrons in precision. There are new projects, e.g. the molecular EDM experiment CeNTREX in the US, which are expected to improve the current limit of $2\times 10^{-25}$ e.cm by more than one order of magnitude in the coming years. The EDM storage ring discussed within PBC could take the proton EDM precision significantly beyond that of the neutron. However, reaching this level of precision represents a step of $\approx$10 orders of magnitude compared to previous EDM storage ring experiments. 

\newpage
\subsubsection{Searches for Hidden Sector particles}
\vspace{0.5cm}

An important class of searches deals with new relatively low mass but very weakly interacting particles. PBC experiments can contribute to this domain both in the sub-eV and the MeV-GeV ranges.

\begin{figure}[!t]
   \centering
   \includegraphics*[width=388pt]{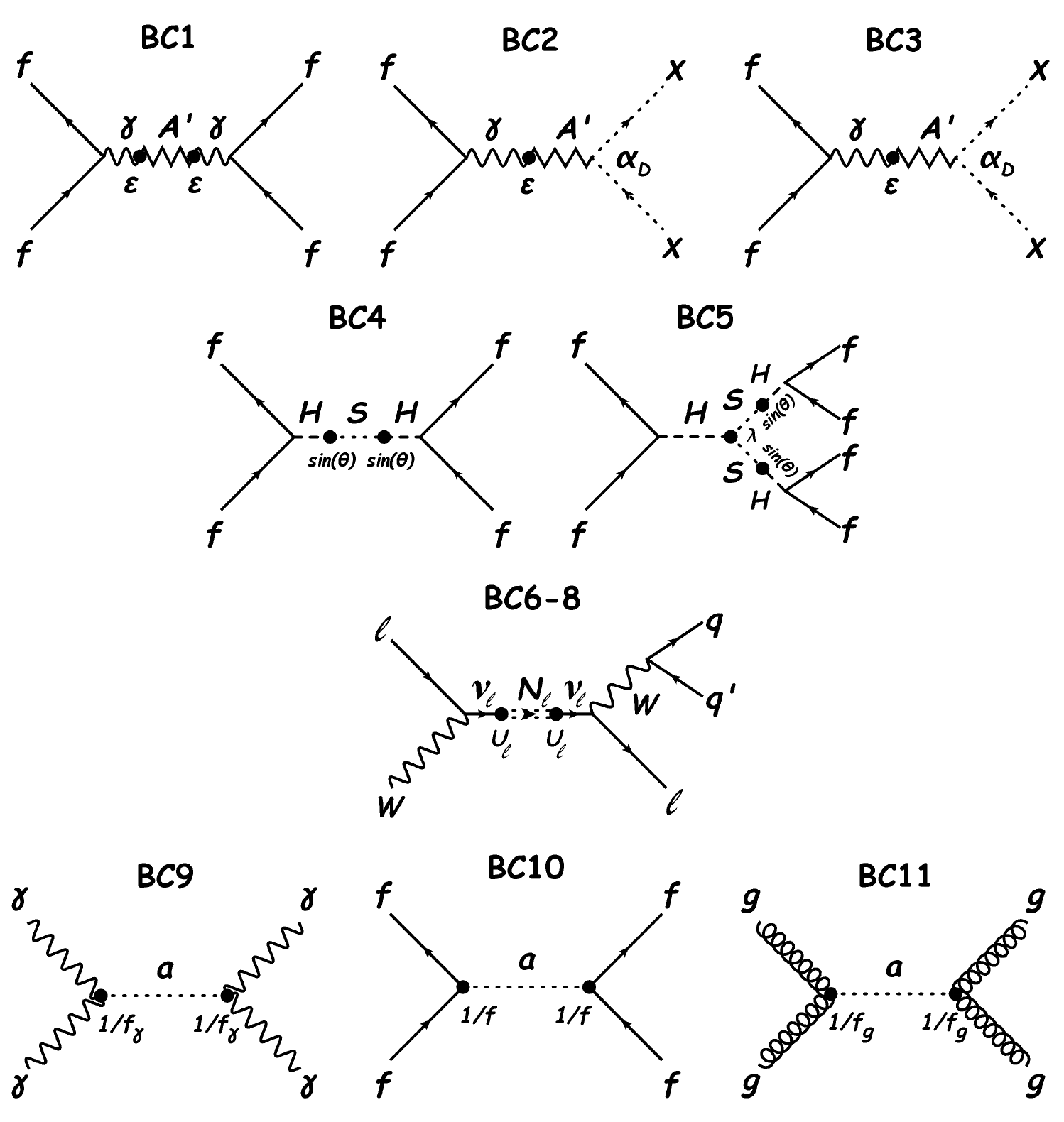}
   \caption{Typical interaction processes of the hidden sector with Standard Model particles for the various benchmark models. The letters $f$, $\ell$, $\nu$ and $q$ denote SM fermions, leptons, neutrinos and quarks, respectively. The free parameters of the models are the couplings indicated at the vertices and the masses of the hidden sector particles: dark photon $A'$, hidden sector matter particle $X$, new singlet scalar $S$, heavy neutral lepton $N_{\ell}$ and axion-like particle $a$. BC3 is the limit of BC2 for $m_{A'}=0$. 
   In experiments these fundamental interaction processes are realized in different ways and lead to a variety of signatures. In the PBC experiments production of the new particles proceeds mainly via Bremsstrahlung and Primakoff-like processes as well as meson decays. Signatures include the appearance of SM particles ``out of nothing'' by the decay of a long lived neutral particle (e.g. beamdumps, LLP searches), missing energy/momentum features if the particle does not decay or decays into other weakly coupled particles (e.g. NA64++, LDMX, NA62++, KLEVER), and the weak scattering of the weakly interacting particles themselves (e.g., SHiP emulsion target, milliQan).
   }
   \label{fig:diagrams}
   \end{figure}

\subsubsection*{Benchmark models}
\vspace{0.5cm}

In order to compare the physics reach of the projects, the PBC BSM working group has defined a set of 11 benchmark theoretical models of the hidden sector ~\cite{PBC:BSM}. Their goal is 2-fold: 
\begin{itemize}
\item achieve a broad coverage of possible experimental signatures;
\item provide a common simple theoretical framework to the experiments for estimation and comparison of their sensitivities.
\end{itemize}
Figure~\ref{fig:diagrams} shows the typical interaction processes of the new hidden particles with the Standard Model for each benchmark case (BC1-11). The BC models cover the simplest fields which can be added to the SM lagrangian to explain open questions such as dark matter, neutrino masses and baryogenesis. If the new particles carry no direct charges under the SM gauge groups, i.e. if they belong to a so-called ``hidden sector'', they can only interact via a small number of ``portals'' with the SM particles relevant for their experimental detection. On the theoretical side these benchmark models represent the range of dominant (lowest dimensional) gauge invariant interactions of hidden sector particles with the SM. They are given by renormalizable (BC1-8) or dimension 5 (BC9-11) operators. 
\vspace{0.5cm}

More specifically the BC models include:
\begin{itemize}
\item{\bf Hidden vectors} interacting via a small kinetic mixing with the hypercharge U(1) gauge group of the SM. Models of this type have been intensely discussed in the context of dark matter. \\
{\bf BC1} features only a massive vector which consequently decays into SM particles.\\
{\bf BC2} in addition contains hidden sector (``dark matter'') particles. The massive vector therefore decays dominantly into these very weakly interacting final states.\\
{\bf BC3} is similar to BC2 but the dark photon is massless and the hidden sector matter particles appear with an effective small electric ``millicharge''.
\item{\bf Hidden scalars} interacting via the Higgs portal. They have been studied in the context of dark matter but also in the context of attempts to solve the hierarchy problem.\\
{\bf BC4} has the scalar  interacting dominantly via mixing with the Higgs. All production and decay proceeds via this mixing and is proportional to the SM couplings of the Higgs to the relevant particles.\\
{\bf BC5} includes in addition the possibility for production via interactions of the Higgs boson with two hidden scalars.
\item{\bf Heavy neutral leptons (HNL)}, also known as right handed neutrinos, interacting via a coupling to the Higgs and left handed SM neutrinos. This is one of the simplest extensions of the SM accounting for neutrino masses and mixings, baryogenesis and potentially also dark matter.\\ {\bf BC6}, {\bf BC7} and {\bf BC8} correspond to a HNL interacting exclusively with the e-, $\mu$- and $\tau$-neutrinos, respectively.
\item{\bf Axion-like particles (ALP)} are scalar or pseudo-scalar particles, present in many extensions of the SM, and the name-giving axion has been introduced to solve the strong CP problem. They are also good dark matter candidates and for higher masses messengers to a dark matter sector.\\ {\bf BC9}, {\bf BC10} and {\bf BC11} have dominant interactions with 2 photons, 2 SM fermions (derivative interaction) and 2 gluons (QCD axion case), respectively.
\end{itemize}

\subsubsection*{Sensitivity overview and comparison}
\vspace{0.5cm}
Using the benchmark models the BSM group~\cite{PBC:BSM} performed a comprehensive and detailed comparison of the sensitivities of the full range of experiments proposed within PBC and put them into the worldwide context of existing and future experiments. 

At this stage a word of caution is necessary: when comparing the project sensitivities it should be kept in mind that there are still significant differences in the level of maturity of the various estimations. Table~\ref{tab:maturity} shows that some of the projects take into account all experimental effects and background sources with detailed simulations and, in some case, real data, whereas others have straightforward assumptions which still need to be proven. 

\begin{table}[h]
\begin{center}
\caption{Level of maturity of the estimations of the projects sensitivities to the hidden sector}
\label{tab:maturity}
\begin{tabular}{cccc}
\hline\hline
Project & Background & Efficiency & Inputs \\
\hline\hline
NA62++ & 0-BG assumed & partly included & $10^{16}$ PoT run in BD mode  \\
KLEVER & partly included & included & fast simulation  \\
REDTOP & included & included & full simulation  \\
NA64++(e) & included & included & real data  \\
NA64++($\mu$) & 0-BG assumed & 100 \% assumed & M2 $\mu$ beamtest  \\
eSPS/LDMX & included & included & full simulation at 4 GeV  \\
AWAKE++ & 0-BG assumed & 100 \% assumed  & toy model  \\
SHiP & 0-BG assumed & included & full simulation \\
CODEX-b & 0-BG assumed & included & full simulation \\
FASER & 0-BG assumed & 100 \% assumed & BG simulations \& in situ measurements  \\
MATHUSLA200 & 0-BG assumed & 100 \% assumed  & cosmic \& LHC BG fluxes  \\
milliQan & included & included & full simulation \\
\hline\hline
\end{tabular}
\end{center}
\end{table}

Results for a selection of the benchmark models covering the main experimental signatures are discussed below. The full set of sensitivity plots is available in~\cite{PBC:BSM}.

\bigskip

\begin{figure}[!t]
   \centering
   \hspace*{1.3cm}\includegraphics*[width=388pt]{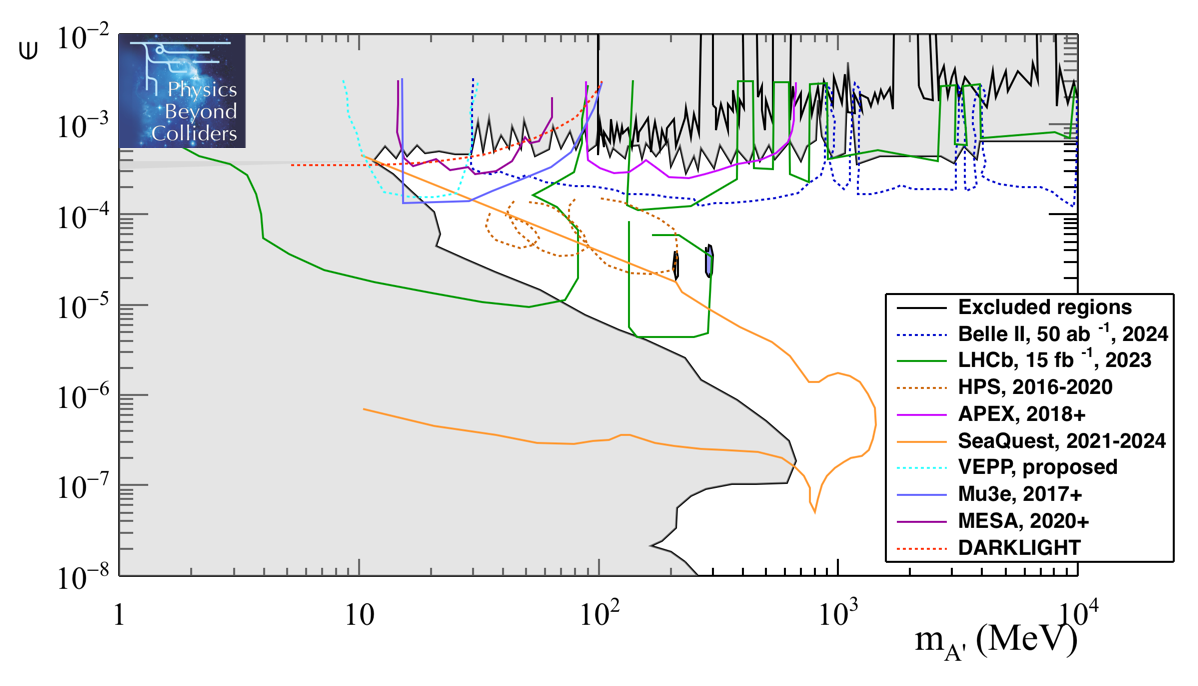}
   \caption{Projected sensitivities to the dark photon visible mode (BC1) of worldwide experiments ongoing or in discussion. The filled area corresponds to already excluded regions. (See~\cite{PBC:BSM} for details and references.)}
   \label{fig:BC1a}
   \centering
   \includegraphics*[width=400pt]{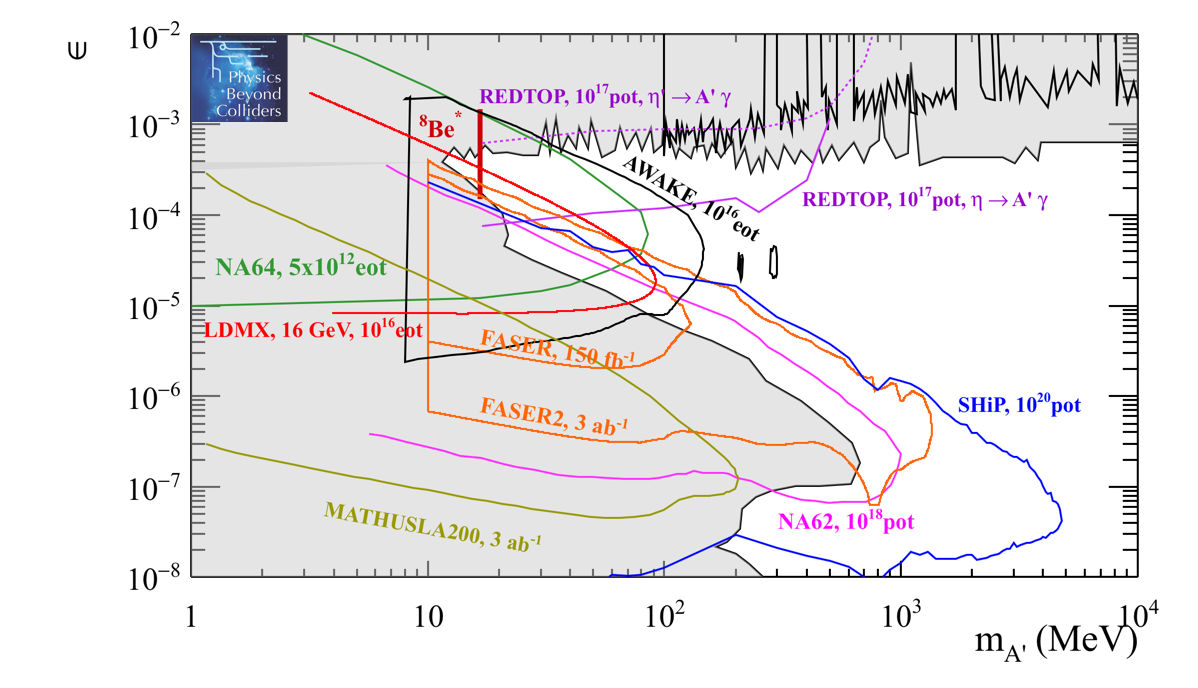}
   \caption{Projected sensitivities to the dark photon visible mode (BC1). The filled area corresponds to already excluded regions. (See~\cite{PBC:BSM} for details and references.)}
   \label{fig:BC1b}
\end{figure}

The Hidden Vector model of BC1 is an intensely studied model. This can be seen in figure~\ref{fig:BC1a} which shows the existing limits together with sensitivities of worldwide experiments for the next 10 years. This is to be compared to the sensitivities of the experiments considered by PBC, shown in figure~\ref{fig:BC1b}. Here we note the excellent opportunities to compete for discovery by the relatively small scale experiments and upgrades NA64++(e), NA62++ and FASER. On the longer term SHiP provides significant discovery potential beyond any other proposal.

\begin{figure}[!t]
   \centering
   \includegraphics*[width=400pt]{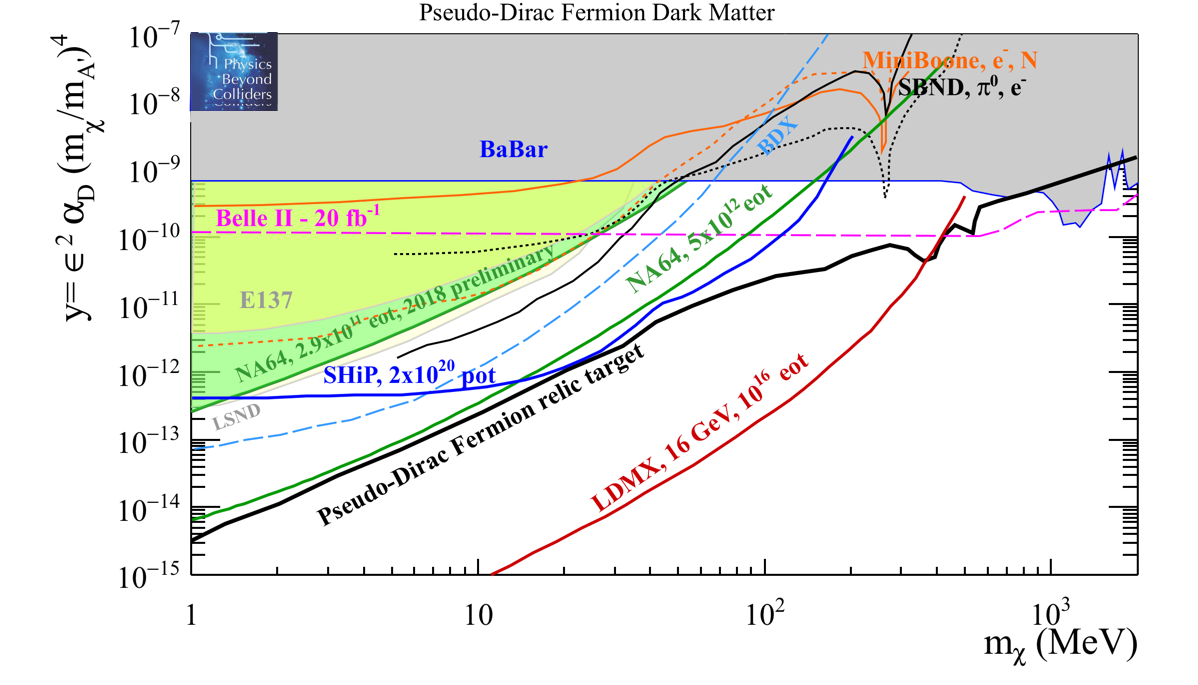}
   \caption{Projected sensitivities to the dark photon invisible mode (BC2). The filled areas correspond to already excluded regions. Sensitivities are shown in a parameter space where a relic particle responsible for the full DM content would lie on a single line (pseudo-DIRAC fermion case shown). Exclusion limits are derived as function of the A' mass $m_{A'}$ and its coupling $\epsilon$ to SM particles. The mass of the relic DM particle and its A' coupling are fixed to $m_{\chi}$ = 1/3 $m_{A'}$ and $\alpha_D$ = 0.1,  respectively. (See~\cite{PBC:BSM} for details and references.)}
   \label{fig:BC2b}
\end{figure}

For particles with dominantly invisible decays (BC2), an exciting physics opportunity is provided by the possibility that the invisible decay products can also be a candidate for the dark matter in the Universe. This is clearly one of the underlying motivations for these invisible searches. Figure~\ref{fig:BC2b} shows that missing energy and momentum search strategies pursued by NA64++, and in the longer term future by an LDMX-like experiment, provide significant sensitivity. NA64++(e) can explore part of the relevant area in the near future, with a sensitivity close to the relic density target line, which could provide a case for further upgrades. On the long-term an LDMX-like experiment extends significantly the coverage of this target space.

Comparing figures~\ref{fig:BC1b} and \ref{fig:BC2b} gives a good overview of the complementarity between the different search strategies employed by the PBC experiments. 
The possibility of both visible and invisible decays favours pursuing dual detection strategies. An example being the search for very weak scatterings of dark particles in the emulsion target of SHiP that provides complementary sensitivity compared to searches for visible decay products.

\begin{figure}[!t]
   \centering
   \includegraphics*[width=400pt]{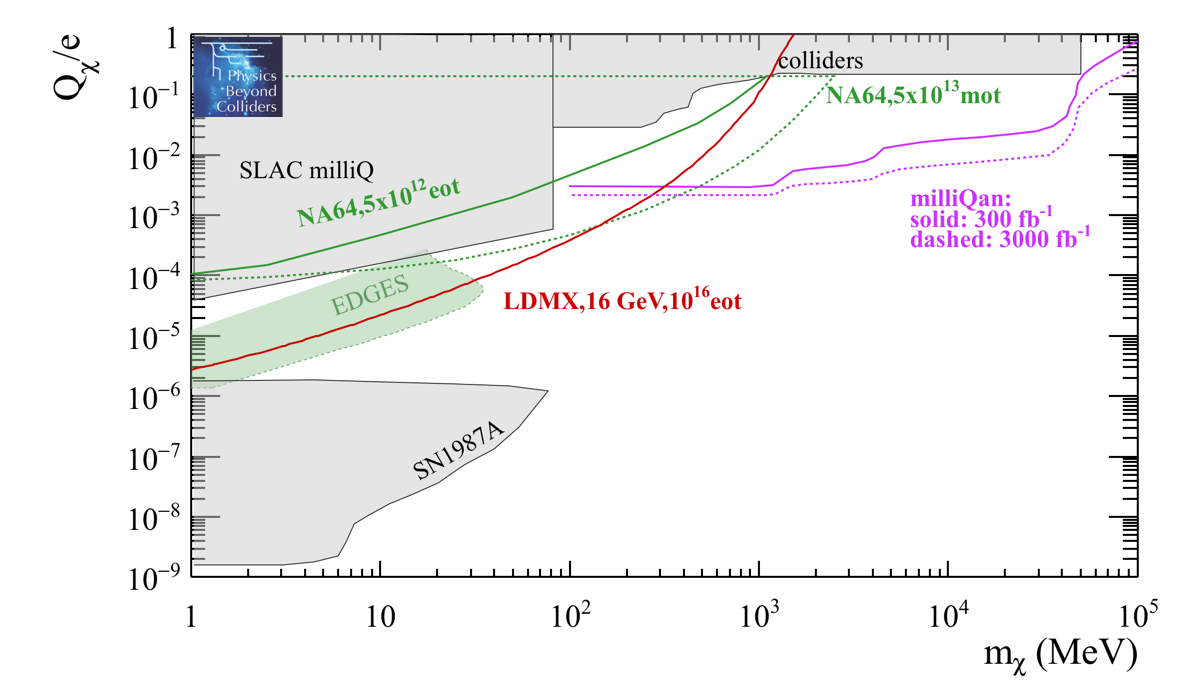}
   \caption{Projected sensitivities to millicharged particles (BC3). The filled areas correspond to already excluded regions. (See~\cite{PBC:BSM} for details and references.)}
   \label{fig:BC3}
\end{figure}

One physics goal of NA64++($\mu$) is not covered by the benchmark models, but is of particular interest: it is the possibility to explain the long-standing deviation of the $(g-2)_{\mu}$ from its Standard Model value with a very weakly coupled vector boson dominantly coupled to muons.
The Phase I short run of NA64++($\mu$) with a muon beam provides a unique opportunity to test this. 

A longer run of NA64++($\mu$) would provide sensitivity to millicharged particles (BC3) through missing energy. This is compared to the reach of other experiments in figure~\ref{fig:BC3}. The motivation of an extended Phase II run of NA64++($\mu$) will depend on the results of milliQan.

\begin{figure}[!t]
   \centering
   \includegraphics*[width=400pt]{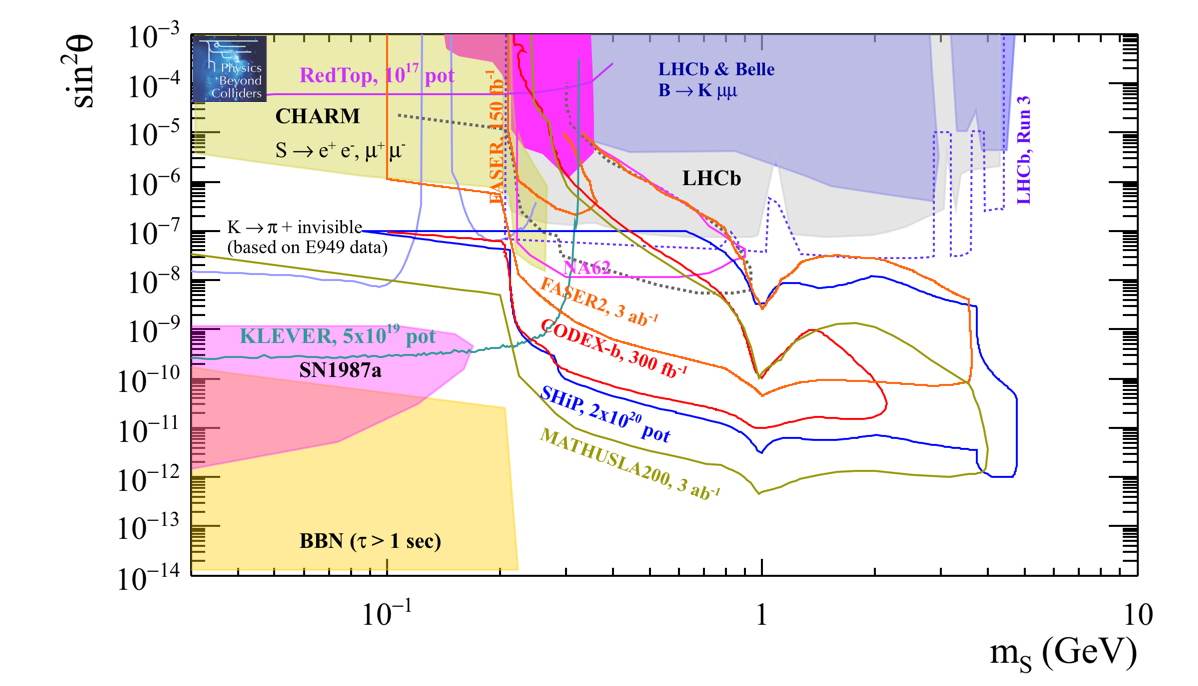}
   \caption{Projected sensitivities to dark scalars (BC4). The filled areas correspond to already excluded regions. (See~\cite{PBC:BSM} for details and references.)}
   \label{fig:BC4}
\end{figure}

\begin{figure}[!t]
   \centering
   \includegraphics*[width=400pt]{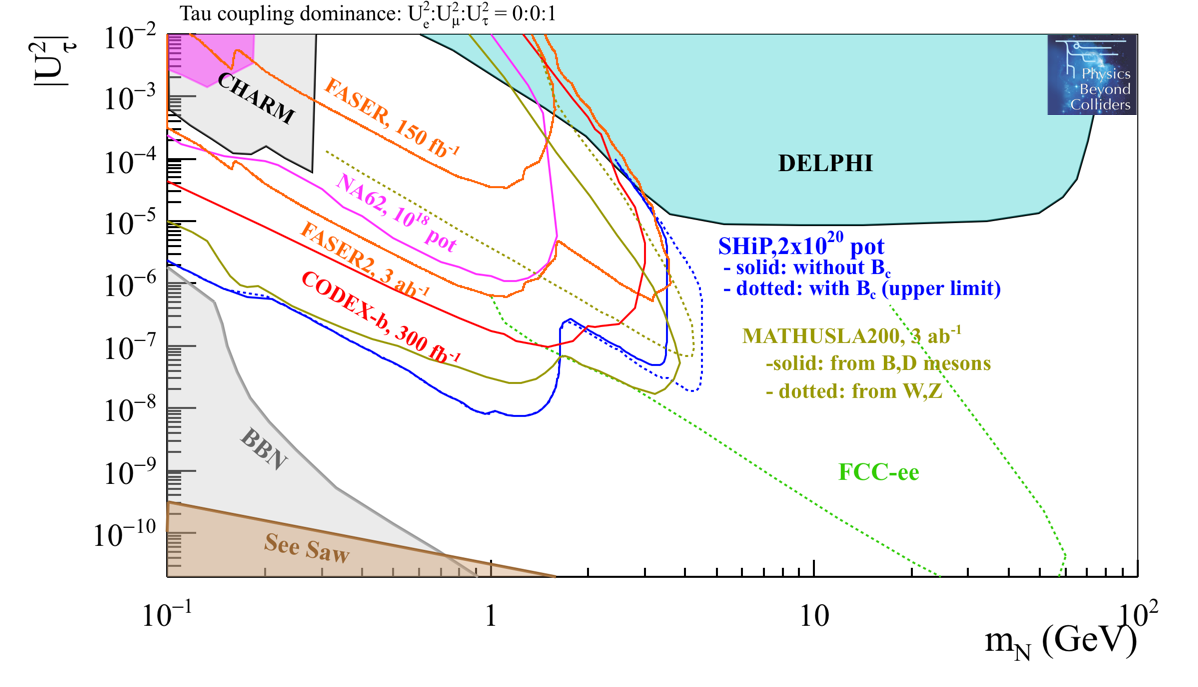}
   \caption{Projected sensitivities to HNLs coupled to the $\tau$ (BC8). The filled areas correspond to already excluded regions. The two SHiP exclusion curves correspond to 2 extreme assumptions on the poorly known $B_c$ production cross section at the BDF energy. (See~\cite{PBC:BSM} for details and references.)}
   \label{fig:BC8}
\end{figure}

Figures~\ref{fig:BC4} and \ref{fig:BC8} complement the landscape with hidden scalars and heavy neutral leptons. The benchmark cases BC4 with scalar-Higgs mixing, and BC8 of an HNL interacting with $\tau$-neutrinos, are shown as typical examples. 
In both cases significant uncharted territories can be covered, with the main players being SHiP and the LHC-LLP searches. They feature a similar mass reach due to the production via meson decays. For BC4 they have complementary coupling reach, and a significant new range at lower mass can be explored by KLEVER, but also NA62 (not shown on the plot). For heavy neutral leptons (BC8) SHiP has the broadest coverage. Early discovery  potential is provided by NA62++ and FASER.

\begin{figure}[!t]
   \centering
   \includegraphics*[width=350pt]{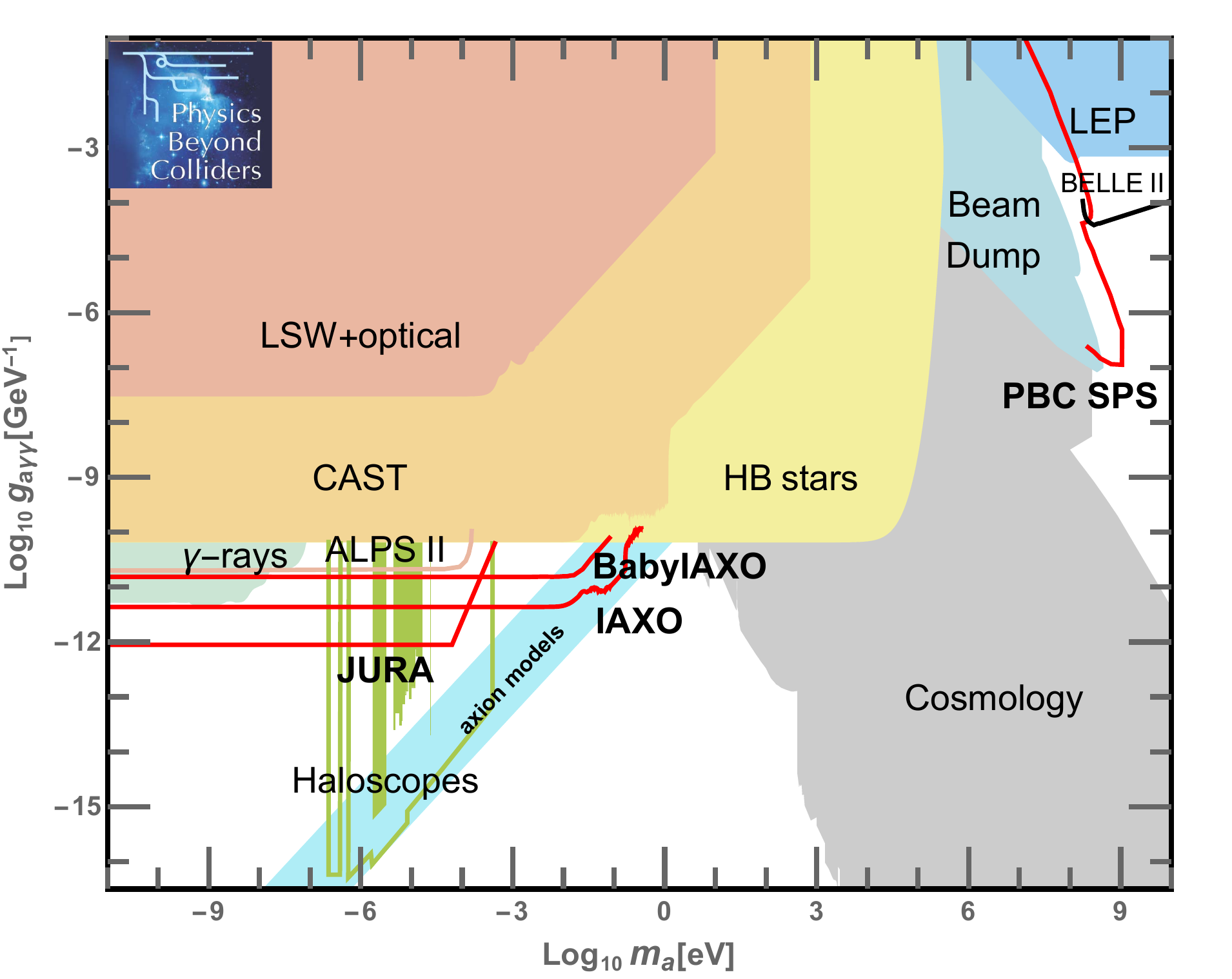}
   \caption{Projected sensitivities to axions coupled to photons (BC9) (see~\cite{PBC:BSM} for details and references). The filled areas correspond to already excluded regions (cf. e.g.~\cite{Jaeckel:2010ni,Irastorza:2018dyq}).}
   \label{fig:axions}
\end{figure}

The full complementarity of the PBC projects can be explicitly seen in the case of axion-like particles shown in figure~\ref{fig:axions}. In this figure a much wider range of masses from $10^{-11}$~eV to 10 GeV is shown. 
In an explicit realization of the schematic figure~\ref{fig:comprehensive}, IAXO and JURA contribute to searches in the sub-eV region, while fixed target and LHC-LLP searches target the MeV to GeV region. A detailed zoom on the MeV-GeV domain can be found in~\cite{PBC:BSM}: since the experimental signatures are similar to those of dark photons in the visible mode, the projects sensitivities compare qualitatively as previously discussed for BC1 (figure~\ref{fig:BC1b}). 

In the sub-eV area, IAXO can provide world-leading sensitivity for a wide range of masses. Baby-IAXO already has the ability to test a significant part of the parameter range suggested by a variety of astrophysical hints. The full IAXO implementation then improves by more than one order of magnitude over current limits from astrophysics as well as the successful precursor experiment CAST. Importantly it is also a unique experiment able to test multi-meV mass QCD axions, complementing searches of axion dark matter in the $(1-100)\mu$eV range.
In the very long term the $3^{rd}$ generation LSW experiment JURA has the potential to improve beyond IAXO.

A similar picture as for photon-coupled axion-like particles can also be drawn for gluon couplings (BC11). Here the sub-eV range can be probed by the EDM ring experiment searching for an oscillating EDM that is expected if ultralight axion-like particles constitute the dark matter in the Universe. This can potentially even reach sensitivity to underlying physics at the Planck scale. An indication of the corresponding sensitivity is shown as the low mass oEDM curve in the schematic overview figure~\ref{fig:comprehensive}. 

\newpage
\subsection{Implementation issues}
\vspace{0.5cm}

The implementation of some of the PBC QCD-oriented projects raises issues associated to their expected siting in the CERN experimental halls.

There is no strong competition on the NA61 beam-line beyond the regular test-beam users. NA61 is also a competitive platform for measurements of hadron production in replica targets of the future neutrino beams to better control their fluxes, as well as for cosmic ray experiments. A prolongation of NA61 after LS2 does not raise any particular operational issue.

The situation is different in the EHN2 hall presently used by COMPASS, where the new projects MUonE and NA64++($\mu$) Phase I wish to use the M2 $\mu$-beam after LS2. They would compete with the foreseen COMPASS++ $R_p$ measurement. The high energy / high intensity M2 $\mu$-beam is unique worldwide. For an optimal exploitation of the beam at its maximal intensity, it is important to investigate how the three measurements could fit together within Run 3 as regards both detector siting in the hall and possible parallel operation of the experiments. The initial studies of the Conventional Beams \cite{PBC:CONV} and QCD \cite{PBC:QCD} working groups are promising and should be strengthened. In the longer term competition in the EHN2 hall may decrease as a function of the evolution of the MUonE and NA64++($\mu$) programs. 

DIRAC++ and NA60++ require an underground hall for radioprotection issues. The only practical option is the ECN3 hall which hosts NA62. It was checked that the DIRAC++ and NA60++ beams could fit together in the hall \cite{PBC:CONV}, but their implementation can be envisioned only after NA62 frees the hall.

On the LHC, LHCb is already well engaged in Fixed Target physics with SMOG and the SMOG2 proposal. In their Fixed Target operation, LHCb and ALICE may focus on different physics signals due to different acceptances, data acquisition rates and operation modes. For both experiments, the physics reach will highly depend on the feasibility of simultaneous Fixed Target and collision operation of the LHC. For ALICE there is a possibility of dedicated Fixed Target operation during LHC proton running.

\vspace{0.5cm}

As for QCD-oriented projects, the implementation of a few BSM-oriented projects is constrained by siting considerations. The NA64++($\mu$) case was already discussed. The timing of the KLEVER installation in the ECN3 hall is also an issue: the optimal phasing of NA62 and KLEVER may depend on the future NA62 results, the evolution of the current $B$-anomalies and the expected sensitivity of $K^+$ vs $K^0$ rare decays to the favored BSM explications of the anomalies, if confirmed.

The design of the BDF and of the SHiP detector are both mature for an implementation decision. 
An operation of NA62++ in beamdump mode on the short term would provide useful insights for the SHiP final design. Anticipating a small upstream experimental hall in the BDF baseline configuration would trigger the option of a long term facility dedicated to rare decays, along the lines of the TauFV proposal. This would significantly extend the physics reach of the BDF.

There is an opportunity for a short-term implementation of LDMX on LCLS-II at SLAC, pending approval of the LCLS-II extracted beam-line. Beyond the LDMX experiment, eSPS and AWAKE have an interesting potential for dark photon searches in the medium to long term. The implementation of such new e-beams at CERN for this purpose should be considered in regard of the opportunities offered by the DESY XFEL upgrades.

The CERN beams are not optimal for REDTOP and would have difficulties to deliver the required integrated luminosity necessary for sensitivity to new physics. The detector also requires significant R\&D and design before construction can start. An implementation at CERN could be a staged option but the above arguments suggest that a direct implementation at FNAL would be more efficient.

The study of the proton EDM storage ring has raised open questions which require a prototype ring to be answered, before construction of the full size ring can be pursued. The prototype ring could be sited in a Laboratory associated to the CPEDM collaboration. Good options to validate the required technologies include the COSY infrastructure to which part of the CPEDM collaboration is associated.

\newpage

\section{References}
\vspace{0.3cm}

\printbibliography

\newpage
\section{Appendix 1: definition of acronyms}

\textbf{}

\textbf{AD}: Antiproton Decelerator

\textbf{ALP}: Axion-Like Particle

\textbf{BDF}: Beam Dump Facility

\textbf{BSM}: Beyond Standard Model

\textbf{CLIC}: Compact LInear Collider

\textbf{CNGS}: CERN Neutrino beam to Gran Sasso

\textbf{CP}: Charge Parity

\textbf{CW}: Continuous Wave

\textbf{DIS}: Deep Inelastic Scattering

\textbf{DVCS}: Deeply Virtual Compton Scattering

\textbf{DY}: Drell-Yan

\textbf{EDM}: Electric Dipole Moment

\textbf{EIC}: Electron Ion Collider

\textbf{EoT}: Electron on Target

\textbf{FCC}: Future Circular Collider

\textbf{FT}: Fixed Target

\textbf{GPD}: Generalized Parton Distribution 

\textbf{HL-LHC}: High Luminosity LHC

\textbf{HNL}: Heavy Neutral Lepton

\textbf{HVP}: Hadronic Vacuum Polarization

\textbf{ILC}: International Linear Collider

\textbf{LBL}: Long Base Line

\textbf{LFV}: Lepton Flavor Violation

\textbf{LHC}: Large Hadron Collider

\textbf{LHCC}: LHC Committee

\textbf{LIU}: LHC Injectors Upgrade

\textbf{LLP}: Long Lived Particle

\textbf{LS}: Long Shutdown

\textbf{LSW}: Light Shining through a Wall

\textbf{MM}: Magnetic Moment

\textbf{NA}: North Area

\textbf{PBC}: Physics Beyond Colliders

\textbf{PDF}: Parton Distribution Function

\textbf{PDWA}: Proton Driven Wakefield Acceleration

\textbf{PoT}: Proton on Target

\textbf{ppp}: proton per pulse

\textbf{PS}: Proton Synchrotron

\textbf{PSI}: Partially Stripped Ion

\textbf{QCD}: Quantum Chromo Dynamics

\textbf{QED}: Quantum Electro Dynamics

\textbf{QGP}: Quark Gluon Plasma

\textbf{RICH}: Ring Imaging CHerenkov

\textbf{RF}: Radio Frequency

\textbf{SBL}: Short Base Line

\textbf{SIDIS}: Semi Inclusive Deep Inelastic Scattering

\textbf{SM}: Standard Model

\textbf{SPS}: Super Proton Synchrotron

\textbf{SPSC}: SPS Committee

\textbf{TMD}: Transverse Momentum Dependent

\textbf{TPC}: Time Projection Chamber

\textbf{WIMP}: Weakly Interacting Massive Particle

\newpage
\section{Appendix 2: brief overview of worldwide beam facilities}

\begin{table}[!htb]
\begin{center}
\caption{Beam facilities worldwide (* indicates study)}
\label{tab:facww}
\begin{tabular}{lllll}
\hline\hline
\textbf{Lab}  & \textbf{Facility}           & \textbf{Beam}  & \textbf{Energy} & \textbf{Current/POT} \\
\hline
Cornell & Synchrotron & e+ & 5.3 GeV & $\sim$2.3 nA\\
&cBeta (ERL)& e- & 90 -- 270 MeV & 100 uA extr. \\
\hline
Fermilab & FAST & e- & 150 MeV &     \\
 & & protons & 2.5 MeV  &   \\
 & MTEST &  protons & 120 GeV  & \num{5e5}  \\
&&pi&1 -- 66 GeV  & \num{7e4} -- \num{1e6} \\
 & MCenter & p,pi &  $\geq$ 200 MeV  &   \\
& Booster & p for $\mu$  &  8 GeV &  \num{16e12} POT per cycle  \\
&  Booster      & p for $\nu$ &  8 GeV  &    \num{4.3e12} ppp  \\
& Main Injector & p for $\nu$ &  120 GeV  &   \num{49e12} ppp \\
\hline
Triumf &BL1A & p & 180-500 MeV  & 50 -- 75 kW  \\
\hline
KEK & SuperKEKB & e+/e-  & 4 on 7 GeV   &  \num{8e35} cm$^{-2}$s$^{-1}$ \\
&PF photon factory & e- & 2.5 GeV &  450 mA \\
& PF-AR & e- & 6.5 GeV  & up to 60 mA  \\
& LINAC & e+/e-  & 10 GeV  &   \\
 & ATF & e- & 1.3 GeV  &  \num{1e10} e/bunch \\
\hline 
 J-PARC       &   LINAC &  H-  &   400 MeV  &  $\sim$45 mA \\
    &  RCS MLF  & p for n & 3 GeV & $\sim$1 MW  \\
  &  RCS MLF & p for $\mu$  &  3 GeV &   \\
  &   MR  FX  &  p for $\nu$  (T2K) &     30 GeV  & 490 kW \num{2.5e14} ppp  \\
  &   MR  SX  &  p for hadrons  &     30 GeV  &  51 kW \\
  &  MR &   p for $\mu$        &    8 GeV  &  3.2 kW (phase II:  56 kW)  \\

\hline
Novosibirsk &  VEPP-2000 &  e+e- collider   & 2 GeV CoM  &  \num{1e32} cm$^{-2}$s$^{-1}$ \\
            &   Super c-tau factory*  &e+e- collider   &   2 -- 5 GeV CoM  &  \num{1e35} cm$^{-2}$s$^{-1}$  \\
\hline
SLAC   &  ESTB   &   e-    &    2-15 GeV  &  1 to 10$^9$ e-/pulse \\
        & LCLS II Sector 30*   &   e-  &    4 GeV  & low current, quasi-CW  \\
\hline        
Mainz  &  MAMI   &e-  &  1600 MeV  &  140 $\mu$A  \\
       &   MESA  &  e-   &  155 MeV  & 150 $\mu$A (1mA in ER mode)  \\
\hline       
JLab  & Hall-A  & e-  &  1-11 GeV  & 1 - 120 $\mu$A  \\
  & Hall-B  & e-  &  1-11 GeV  &  1 - 160 nA \\
  & Hall-C  & e-  &  1-11 GeV  &  2.5 - 150 $\mu$A \\
      & Hall-D  & e-  &  12 GeV   & 1-2 $\mu$A   \\
\hline      
Frascati &  Linac/BTF  & e+/e-  &  25 to 750 MeV   & 1 - 10$^{10}$ per pulse  \\
         &  DAFNE   & e+e- collider  &  1.02 GeV CoM  &   $\sim$1 A\\
\hline         
DESY   & Petra III & e-   & 6 GeV &  100 mA \\
& European XFEL & e- & 17.5 GeV  &  5 mA      \\
\hline
PSI   & Cyclotron      &   p for $\pi$ and $\mu$    &   590 MeV     &  2.4 mA (1.4 MW)      \\
  & Cyclotron      &   p for n  (SINQ)  &   590 MeV     &  2.4 mA (1.4 MW)      \\
  & Cyclotron      &   p for ultracold n   &   590 MeV     & 2-3\% duty factor full power   \\
  \hline\hline
\end{tabular}
\end{center}
\end{table}

\newpage
\section{Appendix 3: PBC Working Group membership}

\vspace{0.5cm}

The PBC working groups include accelerator experts, independent physicists and proponents of the projects. Their core members are listed below with the names of the conveners underlined. Other contributors are visible on the author lists of the working group reports. 

\subparagraph{ACCELERATOR WORKING GROUPS}

\subparagraph{Beam Dump Facility:}
\underline{Mike Lamont} (CERN), \underline{Brennan Goddard} (CERN),\linebreak Lau Gatignon (CERN), \underline{Marco Calviani} (CERN), Simone Gilardoni (CERN), Claudia Ahdida (CERN), Simon Marsh (CERN), John Osborne (CERN),  \underline{Richard Jacobsson} (SHiP), Francisco Galan Sanchez (CERN), Jonathan Gall (CERN).

\subparagraph{EDM ring:}
Mike Lamont (CERN), Gianluigi Arduini (CERN), \underline{Christian Carli} (CERN), Mei Bai (Jülich/GSI), Klaus Jungmann (Gröningen), \underline{Yannis Semertzidis} (electrostatic ring proponent), Jörg Pretz, Achim Stahl, Ed Stephenson, \underline{Hans Ströher} and Paolo Lenisa (JEDI proponents).

\subparagraph{Conventional beams:}
\underline{Lau Gatignon} (CERN), \underline{Markus Brugger} (CERN), Johannes Bernhard (CERN), Nikos Charitonidis (CERN, Alexander Gerbershagen (CERN), Simone Gilardoni (CERN), Evgueni Goudzowski (NA62++),  Matthew Moulson (KLEVER), Paolo Crivelli (NA64++), Clara Matteuzzi (MUonE), Oleg Denisov (COMPASS++), Antoni Aduszkiewicz (NA61++), Enrico Scomparin (NA60++), Leonid Nemenov (DIRAC++), Corrado Gatto (REDTOP).

\subparagraph{LHC Fixed Target:}
Brennan Goddard (CERN), \underline{Stefano Redaelli} (CERN), Johannes Bernhard (CERN), Simone Gilardoni (CERN), Cynthia Hadjidakis (AFTER), Walter Scandale (UA9), \linebreak \underline{Massi Ferro-Luzzi} (LHCb), Andrea Dainese (ALICE).

\subparagraph{Technology:}
\underline{Andrzej Siemko} (CERN), Dimitri Delikaris (CERN), \underline{Babette Döbrich} (CERN), Jörn Schaffran (DESY), Herman ten Kate (IAXO), Pierre Pugnat (LSW), Giovanni Cantatore (KWISP), Livio Mapelli (DARKSIDE), Guido Zavattini (VMB), Antonio Polosa (Nanotubes), Paolo Spagnolo (GHz LSW).

\subparagraph{Complex performance study:}
\underline{Giovanni Rumolo} (CERN), \underline{Hannes Bartosik} (CERN) in collaboration with LIU, 
Eirini Koukovini Platia (CERN).

\subparagraph{AWAKE++ study:}
\underline{Edda Gschwendtner} (CERN) and \underline{Matthew Wing} (AWAKE)

\subparagraph{nuSTORM study:}
Mike Lamont (CERN), \underline{Ken Long} (nuSTORM proponent), \underline{Jonathan Gall} (CERN),
Wolfgang Bartmann (CERN), Claudia Ahdida (CERN), Marco Calviani (CERN), Simone Gilardoni (CERN).

\subparagraph{Gamma Factory study:}
Frank Zimmermann (CERN), \underline{Reyes Alemany Fernandez} (CERN), \linebreak Mike Lamont (CERN), John Jowett (CERN), \underline{Witold Krasny} (gamma factory), \underline{Brennan Goddard} (CERN).

\subparagraph{eSPS study:}
Lyn Evans (CERN), Steinar Stapnes (CERN)

\subparagraph{FASER study:}
\underline{Jamie Boyd}, Mike Lamont, Rhodri Jones, Jonathan Gall, John Osborne, Francesco Cerutti

\vspace{1.0cm}
\subparagraph{PHYSICS WORKING GROUPS}

\subparagraph{BSM:}
Maxim Pospelov (Victoria Univ., Canada), \underline{Clare Burrage} (Nottingham Univ., UK), \linebreak \underline{Sasha Rozanov} (CPPM Marseille, France), \underline{Giuseppe Ruoso} (INFN-Legnaro, Italy), Gaia Lanfranchi (INFN-Frascati, Italy), Felix Kling (Univ. California, USA), Klaus Jungmann (VSI-Groningen, Netherlands), Klaus Kirch (ETH/PSI, Switzerland), Tommaso Spadaro (NA62++), Cristina Lazzeroni (KLEVER), Kostas Petridis (SHiP), Guy Wilkinson (TauFV), Sergei Gninenko (NA64++), Igor Irastorza (IAXO), Axel Lindner (LSW), Yannis Semertzidis (EDM), Fernando Martinez Vidal (EDM with crystals), Philip Schuster (LDMX/eSPS), Anthony Hartin (AWAKE++), David Curtin (MATHUSLA), Albert de Roeck (milliQan), Michele Papucci (CODEX-b), Jonathan L. Feng (FASER), Isabel Pedraza (REDTOP).

\subparagraph{QCD:}
\underline{Markus Diehl} (DESY, Germany), \underline{Jan Pawlowski} (Heidelberg Univ., Germany), \linebreak \underline{Gunar Schnell}  (UPV/EHU \& Ikerbasque, Spain), Alain Magnon (CERN),
Graziano Venanzoni (MUonE), Gerhard Mallot (COMPASS++), Jean-Philippe Lansberg (AFTER), Achille Stocchi (crystal), \linebreak Juerg Schacher (DIRAC++), Gianluca Usai (NA60++), Szymon Pulawski (NA61++), Giacomo Graziani (LHCb-FT), Andrea Dainese (ALICE-FT).

\end{document}